\documentclass[aps,prb,twocolumn,superscriptaddress]{revtex4-2}

\usepackage[colorlinks=true, allcolors=blue]{hyperref}
\usepackage{graphicx}
\usepackage{amsmath}
\usepackage{amssymb}
\usepackage{changes}
\usepackage{braket}
\usepackage{blkarray}
\usepackage{cancel}
\usepackage{soul}
\usepackage{xcolor}
\usepackage[flushleft]{threeparttable}

\begin{document}

\title{Downfolding from Ab Initio to Interacting Model Hamiltonians: \\ Comprehensive Analysis and Benchmarking of the DFT+cRPA Approach}

\author{Yueqing Chang}
\affiliation{Center for Materials Theory, Department of Physics \& Astronomy, Rutgers University, Piscataway, New Jersey 08854, USA}

\author{Erik G. C. P. van Loon}
\affiliation{NanoLund and Division of Mathematical Physics, Department of Physics, Lund University, Lund, Sweden}

\author{Brandon Eskridge}
\affiliation{Department of Physics, College of William and Mary, Williamsburg, Virginia 23187, USA}
\affiliation{Center for Computational Quantum Physics, Flatiron Institute, New York, NY 10010, USA}

\author{Brian Busemeyer}
\affiliation{Center for Computational Quantum Physics, Flatiron Institute, New York, NY 10010, USA}

\author{Miguel A. Morales}
\affiliation{Center for Computational Quantum Physics, Flatiron Institute, New York, NY 10010, USA}

\author{\\Cyrus Dreyer}
\affiliation{Department of Physics \& Astronomy, Stony Brook University, Stony Brook, NY 11794, USA}
\affiliation{Center for Computational Quantum Physics, Flatiron Institute, New York, NY 10010, USA}

\author{Andrew J. Millis}
\affiliation{Department of Physics, Columbia University, New York, NY 10027, USA}
\affiliation{Center for Computational Quantum Physics, Flatiron Institute, New York, NY 10010, USA}

\author{Shiwei Zhang}
\affiliation{Center for Computational Quantum Physics, Flatiron Institute, New York, NY 10010, USA}

\author{Tim O. Wehling}
\affiliation{I. Institute for Theoretical Physics, Universität Hamburg, Notkestraße 9-11, 22607 Hamburg, Germany}
\affiliation{The Hamburg Centre for Ultrafast Imaging, 22761 Hamburg, Germany}

\author{Lucas K. Wagner}
\email{lkwagner@illinois.edu}
\affiliation{Department of Physics, University of Illinois at Urbana-Champaign, Urbana, Illinois 61801, USA}

\author{Malte R\"osner}
\email{m.roesner@science.ru.nl}
\affiliation{Institute for Molecules and Materials, Radboud University, 6525 AJ Nijmegen, the Netherlands}

\begin{abstract}

    Model Hamiltonians are regularly derived from first-principles data to describe correlated matter. However, the standard methods for this contain a number of largely unexplored approximations. 
    For a strongly correlated impurity model system, here we carefully compare a standard downfolding technique with the best possible ground-truth estimates for charge-neutral excited state energies and wavefunctions using state-of-the-art first-principles many-body wave function approaches.
    To this end, we use the vanadocene molecule and analyze all downfolding aspects, including the Hamiltonian form, target basis, double counting correction, and Coulomb interaction screening models. 
    We find that the choice of target-space basis functions emerges as a key factor for the quality of the downfolded results, while orbital-dependent double counting correction diminishes the quality. Background screening to the Coulomb interaction matrix elements primarily affects crystal-field excitations. Our benchmark uncovers the relative importance of each downfolding step and offers insights into the potential accuracy of minimal downfolded model Hamiltonians.  

\end{abstract}

\date{\today}

\maketitle

\section{Introduction}

  The computational cost of solving electronic Hamiltonians increases rapidly with the size of the electronic Hilbert space, i.e., with the number of orbitals and electrons. This presents a substantial constraint on electronic structure calculations for molecules and an essential problem for the study of correlated solids. 
  To overcome this dilemma, it is desirable to construct models with fewer electronic degrees of freedom in a systematic and controllable way. For this purpose, one needs to select a smaller \emph{target} Hilbert space and determine the structure of the target Hamiltonian and its matrix elements. 
  In practice, typically, no more than a handful of orbitals per unit cell are kept in the target space.
  Altogether, this procedure is known as downfolding~\cite{Aryasetiawan2022} and serves as a bridge connecting \textit{ab initio} methods such as density functional or $GW$ theory with higher-level many-body treatments including (extended) dynamical mean field~\cite{Georges96} or dual theories~\cite{Rubtsov08, Rubtsov12}, \textit{ab initio} D$\Gamma$A \cite{Galler19}, self-energy embedding~\cite{Zgid15} or Eliashberg theory~\cite{lee2023electronphonon}, which cannot be performed on the full Hilbert space. There have been various suggestions for these downfolding schemes, ranging from constrained density functional perturbation theory for phononic degrees of freedom~\cite{falter_renormalization_1981,falter_unifying_1988,nomura_ab_2015} to the constrained random phase approximation (cRPA)~\cite{Aryasetiawan2004}, constrained density functional theory~\cite{PhysRevB.38.6650},
  constrained $GW$~\cite{Hirayama13} and constrained functional renormalization group~\cite{Honerkamp12} approaches for purely electronic Hamiltonians, next to embedding schemes, e.g., starting from coupled-cluster references~\cite{RevModPhys.79.291, lyakh_multireference_2012}
  or via projection schemes~\cite{PhysRevLett.131.200601}. 
  Our focus here is on interacting target space Hamiltonians of the form $H=\hat{t}_{\alpha\beta}c^\dagger_\alpha c^{\phantom{\dagger}}_\beta+\hat{U}_{\alpha\beta\gamma\delta} c^\dagger_\alpha c^\dagger_\beta c^{\phantom{\dagger}}_\gamma c^{\phantom{\dagger}}_\delta$, where $\hat{U}$ is a static Coulomb interaction in the target space, $\hat{t}$ defines the single-particle energies and the Greek indices run over the electronic target space.
  Generalized Hubbard model Hamiltonians of this form are the standard in ``DFT++''~\cite{lichtenstein_ab_1998} (or density functional theory based embedding) approaches to describe correlation effects in solids for which various solvers have been developed and implemented.
  
  Although the idea of downfolding is conceptually simple, and reductions of the Hilbert space are ubiquitous in quantum physics, the question has remained on \emph{how} one should do this in a systematic and practical way. By definition, the original Hilbert space is too large to do exact calculations, so one needs to use approximate methods whose errors are usually not controllable. At the same time, one needs to avoid double counting of correlation effects. This arises because of the interplay between the two-body $\hat{U}$ terms and the single-particle term $\hat{t}$ in the Hamiltonian. The latter is frequently extracted from a density-functional theory (DFT) calculation that already contains some interaction and correlation effects. 
  Several single-particle double counting corrections to $\hat{t}$ exist in the literature~\cite{anisimov_first-principles_1997,muechler_quantum_2022,sheng_greens_2022,haule_exact_2015}, with some physical arguments supporting their use, but there is no general consensus on how to tackle this problem.
  In fact, the uncertainty surrounding these single-particle double counting corrections can be a limiting factor for the accuracy of DFT++ approaches, see, e.g., Refs.~\cite{kristanovski_role_2018,kotliar_electronic_2006} for further discussions.

  In terms of the interaction in the target space, downfolding \textit{a priori} generates many-fermion interactions of arbitrary order~\cite{Honerkamp12} and not just one- and two-particle terms. Furthermore, these interactions might be non-instantaneous (frequency-dependent)~\cite{Aryasetiawan2004}, requiring an action rather than a Hamiltonian formalism. 
  Even if one accepts the restriction to a static two-particle Coulomb interaction tensor, its strength should be different from the corresponding Coulomb interaction in the full space to account for the screening by electrons outside of the target space, while screening by the other electrons inside the target space should not be included. The latter is to avoid a double counting of screening processes to the Coulomb interaction $\hat{U}$ within the target space. A variety of constrained methods~\cite{falter_renormalization_1981,falter_unifying_1988,nomura_ab_2015,Dederichs1984,McMahan1988,Zhang91,Aryasetiawan2004,Hirayama13} have been developed to avoid this two-particle double counting issue, e.g., via partial (constrained) screening approaches.

  Evaluating the performance of downfolding techniques has proven challenging in practice, as the matrix elements of the downfolded model are not measurable quantities themselves and as the models are often solved using approximate solvers. This multi-step procedure makes it hard to establish how accurate the downfolding by itself is since deviations from full first-principles calculations or experimental data might also originate in approximations made while solving the low-energy model. 

  Given the widespread use of cRPA calculations based on DFT input to derive interacting model Hamiltonians and their promise of accurately describing correlation effects at low numerical costs, it is critical to quantify their accuracy. 
  To this end, we analyze here the sensitivity of this method at hand of the vanadocene molecule. The latter is a good proxy for correlated bulk materials or embedded correlated defects with well-defined interacting sub-spaces within gapped semiconducting backgrounds (refered to as rest spaces).
  Because vanadocene is a relatively small system, we are able to use state-of-the-art first-principles methods to derive reliable reference data, and the downfolded models are solved exactly, which allows us to focus on errors in the downfolding procedure.
  To avoid obtaining correct spectra with incorrect states, we assess both energies and the character of the states.
  We systematically assess how the model Hamiltonian form, the target space basis, the impact of double counting corrections, and the (constrained) screening models affect the ground state and excited state energies and wavefunctions.

  In the following, we first explain our benchmarking strategy and introduce the vanadocene test system. Afterwards, we discuss the first-principles reference data before we present our step-wise benchmark, showing how each downfolding step affects the comparison to the reference system. The impact of our findings is discussed at the end.

 \section{Results}

\subsection{Benchmarking Strategy}

    \begin{figure}[h]
        \centering
        \includegraphics[width=0.99\columnwidth]{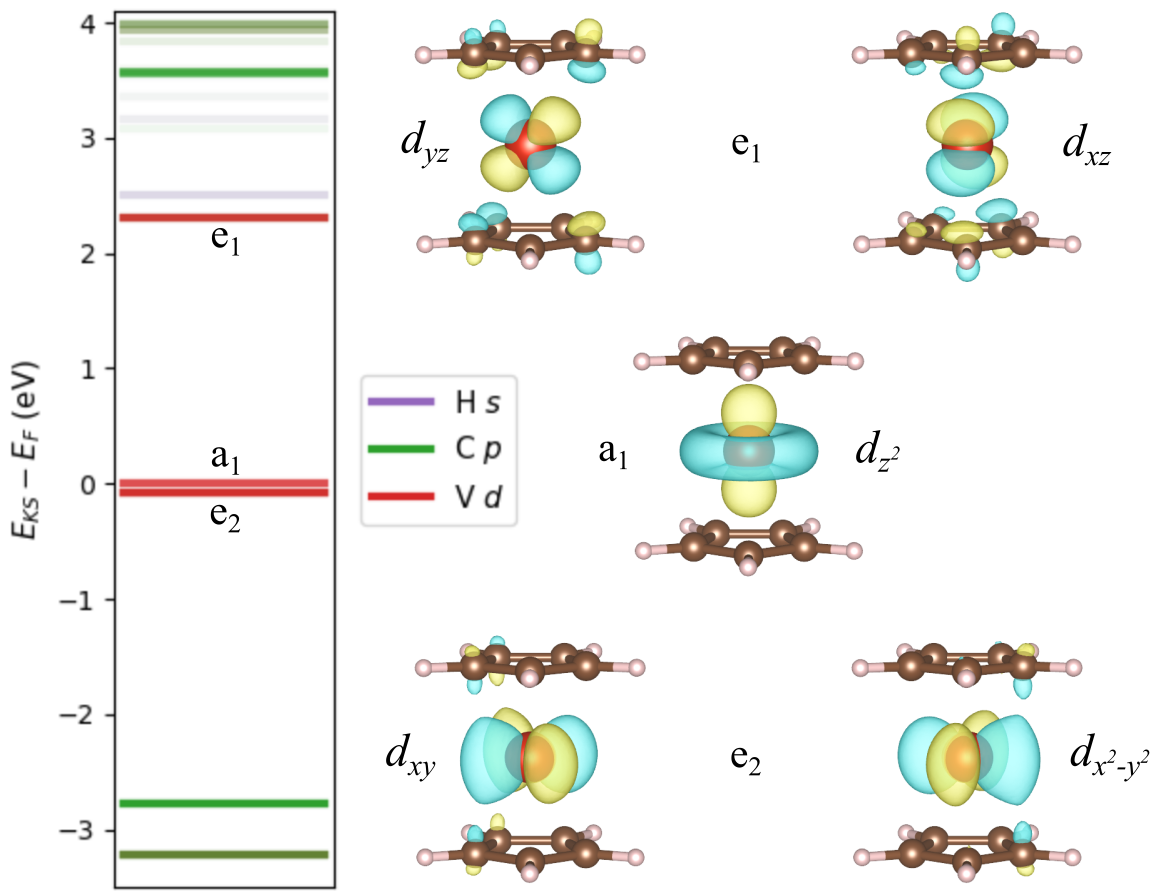}
        \caption{Single Particle Properties. Left: DFT Kohn-Sham energies. Purple, green, and red colors indicate the projected orbital weight of each Kohn-Sham state, being mostly H $s$, C $p$, or V $d$, respectively.
        Red lines mark single-particle energies resulting from diagonalization of the downfolded single-particle Hamiltonian, which perfectly overlap with the Kohn-Sham ones. Right: Maximally localized Wannier functions used within the downfolding procedure.}
        \label{fig:mlwf}
    \end{figure}

  To obtain accurate reference data, we use a combination of several state-of-the-art first-principles methods which each provides systematically improvable many-body wave functions to treat the electronic correlations at an attainable cost. Specifically, we apply equation of motion coupled cluster (EOM-CC)~\cite{koch_coupled_1990, stanton_equation_1993}, auxiliary field quantum Monte Carlo (AFQMC)~\cite{ZhangKrakauer2003}, and fixed-node diffusion Monte Carlo (DMC)~\cite{foulkes_quantum_2001},
  which have been shown to accurately describe the excited states for transition metal molecules~\cite{balabanov_ab_2000, pathak_excited_2021, Rudshteyn2020, shee_potentially_2023}.

  As the test system, we have chosen the unperturbed eclipsed vanadocene molecule VCp$_2$~\cite{prins_bonding_1968}, which consists of a V atom situated between two planar and parallel C$_5$H$_5$ rings as illustrated in Fig.~\ref{fig:mlwf}.
  The unperturbed molecule has $D_{5h}$ symmetry with a five-fold rotation axis, five two-fold rotational axes orthogonal to the five-fold rotational axis, and a mirror plane in the $xy$-plane. The vanadocene molecule may also exist in a staggered geometry with $D_{5d}$ symmetry. Previous DFT calculations predict a difference of only 13 meV in the ground state energy between the $D_{5h}$ and $D_{5d}$ structures with the $D_{5h}$ structure as the global ground state, consistent with experiment~\cite{xu2003,GARD1975181}. On the level of DFT, there are 5 V $3d$ dominated Kohn-Sham states forming the highest occupied and the lowest unoccupied molecular orbitals, which host 3 electrons in total. The $D_{5h}$ symmetry of the crystal field generated by the carbon rings leads to two doubly degenerate ($e_1$, $e_2$) and one non-degenerate ($a_1$) single-particle energies for these $d$-dominated molecular orbitals as depicted in Fig.~\ref{fig:mlwf}.
  This way, VCp$_2$ forms an optimal test bed for cRPA-based \textit{ab initio} downfolding approaches: it hosts a well-defined partially filled correlated target space spanned by the V $d$ orbitals, which is well disentangled from the carbon ring ``background'' electronic structure, which has a significant single-particle gap of more than $6\,$eV. The latter is important to guarantee that the RPA treatment of the background screening is adequate~\cite{PhysRevB.104.045134}.
  
  Our goal is to benchmark the effects of electronic correlations in this molecule. For this purpose, we use the same frozen positions of the nuclei in all methods and do not consider any electron-phonon coupling. Furthermore, all methods start from the same ccECP pseudopotentials~\cite{annaberdiyev_new_2018} and ignore spin-orbit coupling. 
  In this sense, our benchmark should be understood as a comparison of several computational methods for the interacting electron Hamiltonian defined by this pseudopotential and these atomic positions rather than as a benchmark to experimental data.

  For the downfolding, we start from conventional DFT calculations and project onto a set of localized orbitals. 
  These are subsequently used to perform constrained random-phase approximation~\cite{Aryasetiawan2004,Aryasetiawan2006,Aryasetiawan2022} (cRPA) calculations, which are widely used for deriving Coulomb interaction matrix elements for downfolded models~\cite{Nomura12,Biermann11,Werner15}. 
  The resulting downfolded models are evaluated using exact diagonalization to avoid any artifacts from approximate model solutions.

Using the reference data from accurate quantum chemistry calculations, we assess the effects of changing the downfolding procedure as a sort of sensitivity analysis, in which an independent variable (the choice in downfolding procedure) is varied to determine the sensitivity of the output (the eigenstates of the embedded problem). 
For each step in the DFT+cRPA procedure, we consider several reasonable and common choices and analyze the sensitivity of the results.

\subsection{First-principles Reference}

  Before we delve into the benchmark of the downfolded results, we first establish the first-principles references for the vanadocene molecule. In this study, we focus on the lowest four spin-flip and the lowest two crystal-field excitations, using three first-principles methods: EOM-CCSD, AFQMC, and DMC. They are among the most accurate quantum chemical methods for the investigation of excited states in correlated systems. Calculations have been systematically converged, and details of the methods can be found in Section~\ref{sec:methods}.
  Calculations have been performed without imposing specific point group symmetry properties on many-body wave functions.

  We found with all reference methods that the ground state is $^4A_{2}$ with $S=3/2$, consistent with previous experimental and theoretical studies~\cite{prins_bonding_2003,GARD1975181,xu2003,jackson_vanadocene_2012,Phung2012,nain_harnessing_2022}. Specifically, the DMC ansatzes were chosen to be of $S_z=1/2$ on the $S=3/2$ quartet. The configuration of the DMC ground state is determined to be $(e_{2})^2(a_{1})^1(e_{1})^0$, characterized by the one-body reduced density matrix of a natural orbital basis (see Section~\ref{sec:methods} for details). The symmetry of the many-body state computed using AFQMC is inherited from the trial wave function used. For the ground state, AFQMC calculations were performed using a CASSCF trial wave function with $(e_{2})^2(a_{1})^1(e_{1})^0$ $3d$-orbital occupancy.

    \begin{figure}[h]
        \centering
        \includegraphics[width=0.70\columnwidth]{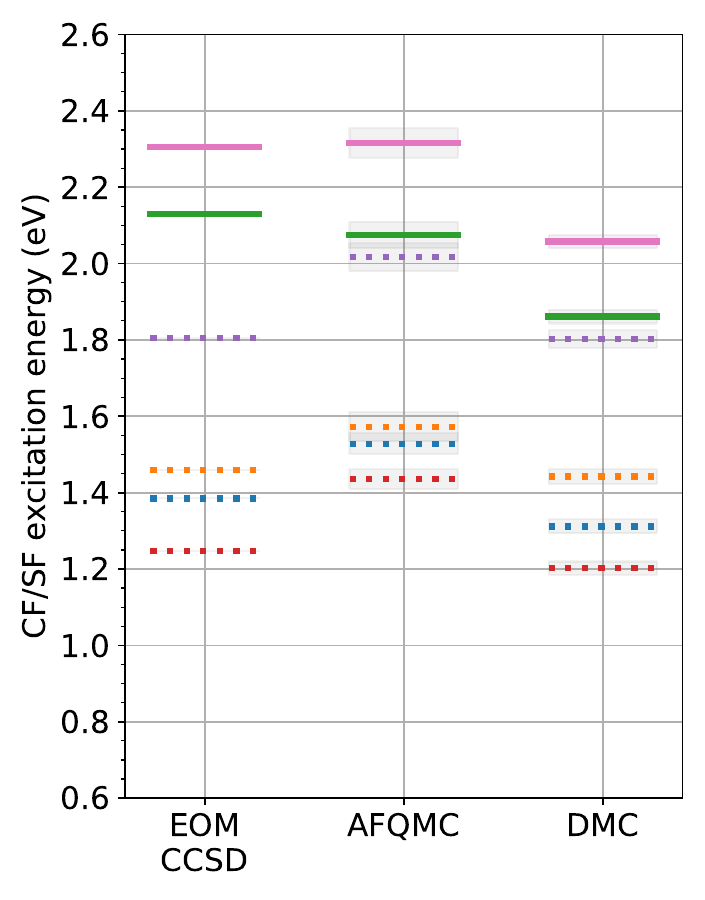}
        \caption{First-principles excitation energies obtained using EOM-CCSD, AFQMC, and DMC.
        The ground state energies obtained from all the methods have been shifted to align at zero.
        Dashed and solid lines represent spin-flip (SF) and crystal-field (CF) excitations, respectively, and the shaded regions around the energy levels show statistical uncertainties.}
        \label{fig:EX_FP}
    \end{figure}

  Fig.~\ref{fig:EX_FP} shows a summary of the charge-neutral excitation spectra obtained using these first-principles methods. 
  The ground state energies are aligned at zero, and the dashed (solid) lines represent spin-flip (crystal-field) excitations, identified by states' zero (finite) occupation on the high-energy $e_{1}$ manifold. 
  All three first-principles methods are in qualitative agreement regarding the order and degeneracy of all spin-flip and crystal-field excitation energies.
  Quantitatively, all methods agree to within about 200 meV.

  The energy difference between excitation energies of the same type (i.e. within either the set of spin-flip excitations or the set of crystal-field excitations) is remarkably similar across all methods.
  According to Fig.~\ref{fig:EX_FP}, the lowest three spin-flip excitation energies are found to be within windows of $<250\,$meV, while the fourth spin-flip excitation energy is higher than the lowest three spin-flip states by 350$\,$meV (DMC). The crystal-field excitations are identified to be higher than the lowest four spin-flip excitations.
  The two crystal-field excitations are found to be separated by about 200$\,$meV for each first-principles method.
  Overall, the first-principles methods obtain the same ordering and character of the many-body eigenstates of interest, but differ in energy by approximately 60$\,\sim\,$200$\,$meV.
  
  In the following sections, we proceed with the discussion of the six lowest charge neutral many-body excitations with $S_z=1/2$, which take place within the V $d$ shell and are thus describable within the minimal subspace of the downfolding calculations.

\subsection{Downfolding Results}

  Now that we are familiar with the test system and the first-principles reference data, we turn to the results obtained using our downfolding procedure. We aim to systematically test and scrutinize all relevant steps in the DFT+cRPA procedure, especially those for which multiple strategies exist in the literature such that choices need to be made. 
  Specifically, we will discuss the chosen form of the model Hamiltonian, the target space basis, the single-particle double counting procedure, and finally, the Coulomb screening model.
  The downfolding procedure is qualitatively assessed based on the predicted ground state properties, the ordering of the excitation energies, the many-body characteristics of the excited states, and the shape of their charge densities.
  To quantify the agreement of the (charge-neutral) excitation energies of the downfolded model with the first-principles reference data, we further calculate
  \begin{align}
      \chi^2(\{E_n\}) = \frac{1}{2\sigma^2} \sum_{i,n} \left( E_n^{\text{ref},i} - E_n \right)^2,
  \end{align}
  where $E_n$ are the charge-neutral excitation energies in the downfolded model and $E_n^{\text{ref},i}$ are the corresponding excitation energies according to the $i$th first-principles method with $i\in\{\text{EOM-CCSD,AFQMC,DMC}\}$. 
  $\chi^2$ is the log-likelihood from a Bayesian inference of the embedding model probability, based on the reference values. 
  It naturally accounts for the fact that some of the states are in close agreement among the reference sets, and some are not. 
  We determine $\sigma$ as the average over the variances of the $n$ excitations within the reference set (EOM-CCSD, AFQMC, DMC), which results in $\sigma \approx 0.1\,$eV.
  $\chi^2$ accounts for the fact that the reference data itself has errors, and so it represents our uncertainty about the exact result.

  The purpose of this quantifier is to summarize how consistent a given downfolding procedure is with the excitation spectrum of the reference data in a single number, given that there is uncertainty within the reference data set itself. 
  The smaller $\chi^2$, the better the match.
  
  Note that this procedure requires us to identify which excitations belong together (the label $n$, the colors of the energy levels in the figures). We achieve this based on the quantum numbers, the spatial structure of the spin-resolved charge densities $\rho_n^\sigma$, and the many-body characteristics of the corresponding excited wave functions for each state $n$ as explained in section~\ref{sec:methods}~B.

We consider several variations of the DFT+cRPA procedure. 
The choice of localized orbitals for the embedded space was either constructed using maximally localized Wannier functions~\cite{Marzari97} (MLWF) or first guess Wannier functions (FGWF). 
The single particle energies were constructed either using double counting corrections, no double counting corrections, and by explicitly modifying crystal field energies, which emulate potentially using different DFT functionals to create the one-particle energies. 
We then move to considering the construction of the interacting part of the downfolded Hamiltonian by considering monopole screening and frequency dependence.
 
      \begin{figure}[h]
        \centering
        \includegraphics[width=0.99\columnwidth]{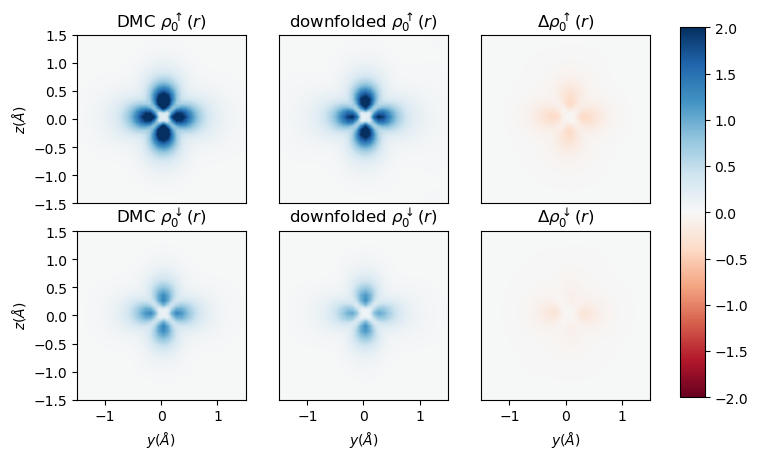}
        \caption{DMC and downfolded ground states charge densities. $\rho^{\uparrow/\downarrow}(r)$ from each method is normalized by the corresponding $\rho^{\downarrow}(r)$ from the same method.}
        \label{fig:GS}
    \end{figure}

\subsubsection{Reference DFT+cRPA procedure}

To limit the number of calculations, we start with a set of common choices that yield reasonable agreement with the reference data.
We use MLWF localized orbitals, single particle energies with no double counting corrections, and the full four index cRPA screened Coulomb matrix elements in the $\omega \rightarrow 0$ limit. 
As shown for the ground state in Fig.~\ref{fig:GS} and for all excited states in section~\ref{sec:methods}~B this model is in qualitative agreement with the reference data, with a loss function $\chi^2$ of 27.3. 
We will consider modifications to this strategy, and examine the changes in $\chi^2$; if $\chi^2$ varies upon a change in the choice, then the DFT+cRPA results are sensitive to the choice, and if it does not, then the DFT+cRPA procedure is not sensitive to the choice. 
In this way, we can identify what approximations deserve further analysis.

\subsubsection{Target Space Basis Set}

    \begin{figure}[h]
        \centering
        \includegraphics[width=0.70\columnwidth]{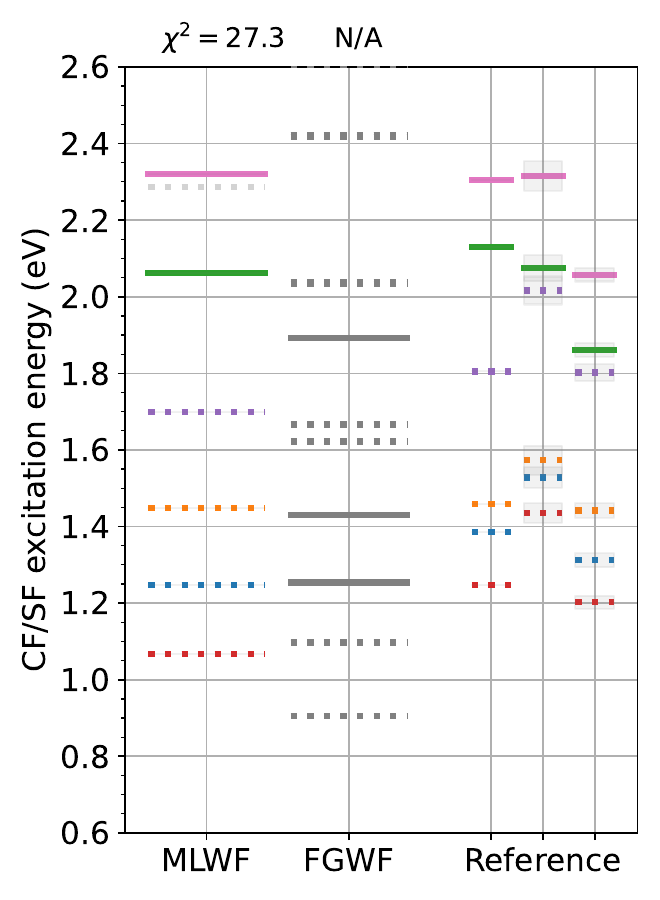}
        \caption{Basis set influence on downfolded excitation energies. Comparison between maximally localized and first guess Wannier function basis sets using static cRPA Coulomb interactions. Dashed and solid lines represent spin-flip (SF) and crystal-field (CF) excitations, respectively. (Light) Grey excitation energies could not be (have not been) characterized.}
        \label{fig:EX_Basis}
    \end{figure}

  We proceed with the discussion of the chosen basis sets. 
  For comparison to the conventional MLWF, which are constrained here to reproduce the DFT single-particle energies, we also use a projector-like~\cite{amadon_plane-wave_2008,haule_dynamical_2010,karolak_general_2011,schuler_charge_2018} first guess Wannier function (FGWF) basis set to calculate all model Hamiltonian matrix elements. 
  The FGWF Wannier function basis set differs in its construction by not applying any inner (or ``frozen'') energy window, such that the DFT single-particle energies are not reproduced, and by using a much larger outer energy window, such that the resulting wave functions are more localized. All details can be found in section~\ref{sec:methods}.
  In this case, both basis sets are capable of reproducing the correct ground state, so we proceed with the analysis of the charge-neutral excitation spectrum. 
  Fig.~\ref{fig:EX_Basis} shows all spin-flip and crystal-field excitations between $0.6$ and $2.6\,$eV. 
  For the FGWF basis, various double and four-fold degenerate spin-flip excitations appear in the energy window of interest, with significantly different characters, such that a 1:1 mapping to the spin-flip excitations from the other methods cannot be established and $\chi^2$ cannot be defined. Correspondingly, the FGWF excitation energies are colored gray in Fig.~\ref{fig:EX_Basis}.
  
  This comparison strikingly shows that projector-like first guess Wannier functions, which are not constrained to reproduce the DFT single-particle energies (see section~\ref{sec:methods} for details), can result in quantitatively and even qualitatively wrong many-body excitation properties. 

    \begin{table}
        \centering
        \caption{Bare and cRPA screened density-density ($v_{iijj}$) and Hund's exchange ($v_{ijij}$) Coulomb matrix elements in the Wannier basis of the MLWF in eV. \label{tab:CME}}
        \begin{ruledtabular}
            \begin{tabular}{r|ccc|ccc}
             & \multicolumn{3}{c}{$v_{iijj}^\text{bare}$}
             & \multicolumn{3}{c}{$v_{ijij}^\text{bare}$}\\
             & $e_2$ & $e_1$ & $a_1$ & $e_2$ & $e_1$ & $a_1$ \\
             \hline
             $e_2$ & 13.454 & 12.438 & 12.985 &   & 0.437 & 0.637   \\
             $e_1$ & 12.438 & 13.357 & 13.135 & 0.437 &   & 0.317   \\
             $a_1$ & 12.985 & 13.135 & 14.466 & 0.637 & 0.317 & \\
             \hline
             & \multicolumn{3}{c}{$U_{iijj}^\text{cRPA}$}
             & \multicolumn{3}{c}{$U_{ijij}^\text{cRPA}$}\\
             & $e_2$ & $e_1$ & $a_1$ & $e_2$ & $e_1$ & $a_1$ \\
             \hline
             $e_2$ & 5.757 & 4.998 & 5.153 &   &  0.396 &  0.570 \\
             $e_1$ & 4.998 & 5.900 & 5.379 & 0.396 &    &  0.302  \\
             $a_1$ & 5.153 & 5.379 & 6.247 & 0.570 &  0.302 & \\
             \hline
             & \multicolumn{6}{c}{$U^\text{cRPA}$ monopole screening only} \\
             & $e_2$ & $e_1$ & $a_1$ & $e_2$ & $e_1$ & $a_1$ \\
             \hline
             $e_2$ & 5.838 & 4.956 & 5.096 &  & 0.436 & 0.647  \\
             $e_1$ & 4.956 & 5.979 & 5.390 & 0.436 &   & 0.319  \\
             $a_1$ & 5.096 & 5.390 & 6.335 & 0.647 & 0.319 &    \\
            \end{tabular}
        \end{ruledtabular}
    \end{table}

  \begin{table}
        \centering
        \caption{Bare density-density ($v_{iijj}$) and Hund's exchange ($v_{ijij}$) Coulomb matrix elements in the Wannier basis of the FGWF in eV. \label{tab:CME-FGWF}}
        \begin{ruledtabular}
            \begin{tabular}{r|ccc|ccc}
             & \multicolumn{3}{c}{$v_{iijj}^\text{bare}$}
             & \multicolumn{3}{c}{$v_{ijij}^\text{bare}$}\\
             & $e_2$ & $e_1$ & $a_1$ & $e_2$ & $e_1$ & $a_1$ \\
             \hline
             $e_2$ & 18.450 & 15.058 & 16.733   & &  0.578 &  0.838   \\
             $e_1$ & 15.058 & 14.531 & 15.314   & 0.578 &&  0.407   \\
             $a_1$ & 16.733 & 15.314 & 18.127   & 0.838 &  0.407 & \\
             \hline
             & \multicolumn{3}{c}{$U_{iijj}^\text{cRPA}$}
             & \multicolumn{3}{c}{$U_{ijij}^\text{cRPA}$}\\
             & $e_2$ & $e_1$ & $a_1$ & $e_2$ & $e_1$ & $a_1$ \\
             \hline
              $e_2$ & 6.599  & 5.349  & 5.491   & &   0.508  & 0.307 \\
              $e_1$ & 5.349  & 6.146  & 5.809   & 0.508  & &   0.379 \\
              $a_1$ & 5.491  & 5.809  & 7.062   & 0.307  & 0.379  
            \end{tabular}
        \end{ruledtabular}
    \end{table} 
  
  As shown in Table~\ref{tab:KSEnergies}, the single-particle energies in the FGWF basis are not as different from MLWF basis functions as other modifications; however, the interactions as shown in Tables \ref{tab:CME} and \ref{tab:CME-FGWF} differ by up to 27\%. 
  Thus, at least in this case, the basis functions are important mainly because of their influence on the interactions rather than the single-particle energies.

\subsubsection{Single-Particle Energies}

  Although it is convenient to extract $\hat{t}$ from the DFT energies, there are several uncontrolled approximations in doing so. 
  First of all, the Kohn-Sham (KS) DFT energies are known to be auxiliary quantities that are not guaranteed to accurately represent the excitations of the system~\cite{Perdew83,Sham85}, even if the exact density functional were available. 
  Secondly, the DFT calculation already incorporates some electronic interaction effects, while the (ED) solution of the downfolded Hamiltonian will add further interaction effects. 
  Correcting for this double counting could be either achieved by starting from constrained $GW$ calculations~\cite{Hirayama13} or by using a so-called double counting correction term in the model Hamiltonian, which aims to remove any Coulomb interaction effects from the single-particle starting point.
  We consider two modifications to the single-particle energies: a Hartree double-counting correction,\cite{bockstedte_ab_2018,ma_quantum_2020,muechler_quantum_2022} which decreases the energy difference between $a_1$ and $e_1$ by more than $1\,$eV such that the crystal-field excitations shift significantly up in energy (see Tab.~\ref{tab:KSEnergies}), and rigid shifts of the energy difference between $a_1$ and $e_1$ levels by approx. $\pm$150\,meV. 
  We use the rigid shifts rather than varying the functional because varying the DFT functional also changes orbitals, which could result in conflated effects. For discussions on how to possibly vary the single-particle correction scheme for different DFT functionals, we refer the interested reader to Refs.~\cite{bockstedte_ab_2018,muechler_quantum_2022}.

    \begin{figure}[h]
        \centering
        \includegraphics[width=0.99\columnwidth]{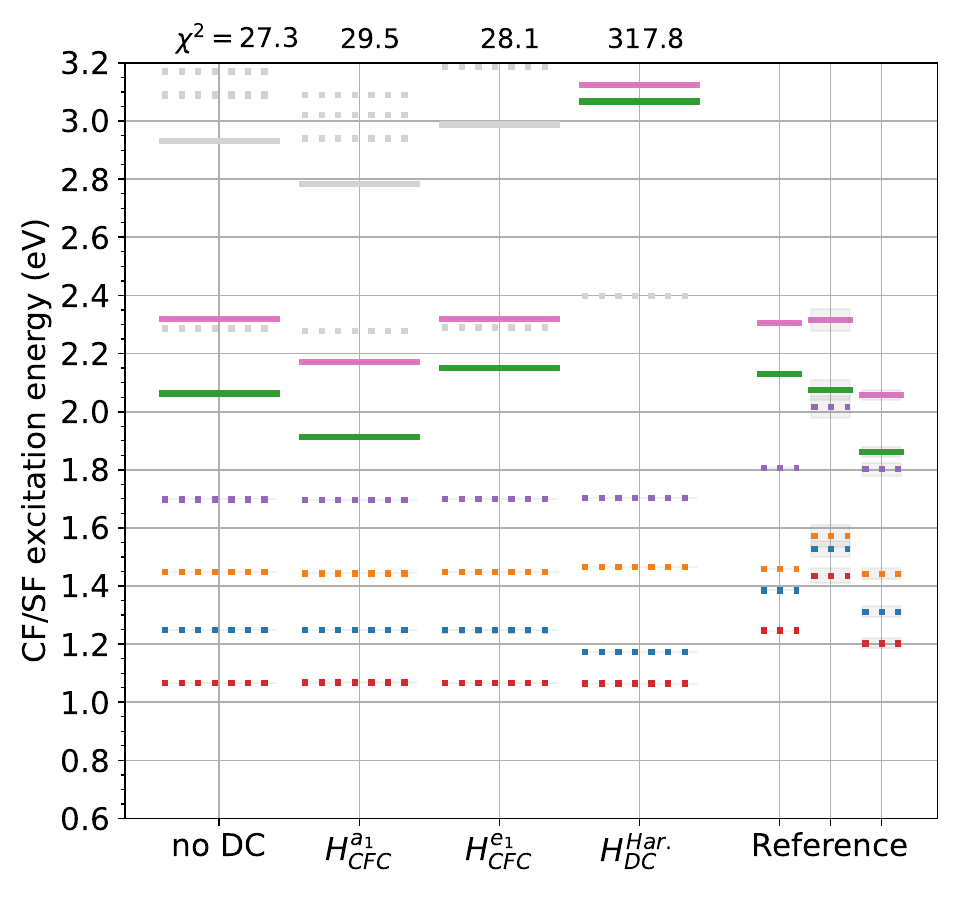}
        \caption{Single-particle double-counting correction influence on downfolded excitation energies using the MLWF basis and static cRPA Coulomb interactions. Dashed and solid lines represent spin-flip (SF) and crystal-field (CF) excitations, respectively. Light grey excitation energies have not been characterized.}
        \label{fig:EX_DC}
    \end{figure}
 
  Fig.~\ref{fig:EX_DC} shows how modifications to the single-particle terms affect the excitation energies in the downfolded model. 
  The spin-flip excitations are barely affected by single-particle energy changes, as they are mainly defined by exchange Coulomb matrix elements, which are unaffected by single-particle double counting corrections. 
  The crystal-field excitations can, however, change significantly: Increasing the energy difference between $a_1$ and $e_1$ with $H_{\text{CFC}}^{a_1}$ by about $150\,$meV lowers the crystal-field excitation energies, while decreasing the difference between $a_1$ and $e_1$ by about $150\,$meV using $H_{\text{CFC}}^{e_1}$ shifts the crystal-field excitations up in energy.
  This trend also holds using the Hartree double counting correction, which decreases the energy difference between $a_1$ and $e_1$ by more than $1\,$eV such that the crystal-field excitations shift significantly up in energy (see Tab.~\ref{tab:KSEnergies} for further details).
  
  In all of these cases, we can exactly map the excited many-body states to the reference ones, allowing us to calculate $\chi^2$. We see that $H_{\text{CFC}}^{a_1}$ and $H_{\text{CFC}}^{e_1}$ only mildly affect $\chi^2$. 
  The Hartree single-particle double counting correction, however, significantly increases $\chi^2$, yielding a poor agreement with the reference data.
  From this, we can conclude that the Hartree double counting is not a favorable correction scheme for charge-neutral local excitations in our vanadocene test system. 
  A similar effect from Hartree double counting corrections has already been observed for Fe impurities in AlN~\cite{muechler_quantum_2022}. 
  In fact, we find the best agreement with the first-principles reference without applying a double counting correction at all. 
  This is in line with conventional double counting corrections, as regularly applied in DFT++ schemes~\cite{solovyev_corrected_1994,czyzyk_local-density_1994,anisimov_band_1991,liechtenstein_density-functional_1995}, which do not have an orbital dependence within the correlated space. 

  The effects of the single-particle energies on the accuracy of the model are relatively straightforward compared to the other terms. 
  As long as the energies are not too much in error, the ground state and spin excitations are robust to their precise value. 
  The crystal field excitations are shifted roughly linearly with the single-particle energies. 

    \begin{figure}[h]
        \centering
        \includegraphics[width=0.81\columnwidth]{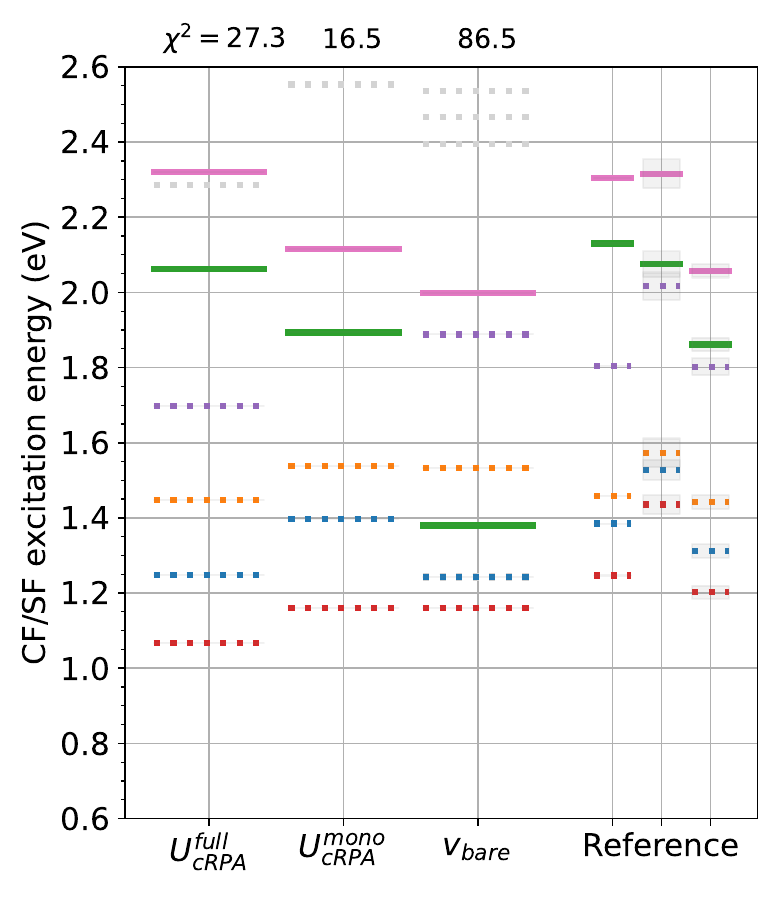}
        \caption{Static screening model influence on downfolded excitation energies using the MLWF basis and no double counting correction. Dashed and solid lines represent spin-flip (SF) and crystal-field (CF) excitations, respectively. Light grey excitation energies have not been characterized.}
        \label{fig:EX_Eps}
    \end{figure}

\subsubsection{Static Screening Model}
  
  We now investigate how different screening models affect the charge-neutral excitation spectrum. 
  To this end, we show in Fig.~\ref{fig:EX_Eps} the original data as obtained from using the conventional screened cRPA $U_{ijkl}$ tensor (referred to as ``full''), together with results obtained from a cRPA tensor which is predominantly screened in the monopole channel (defined here as the leading eigenvalue of the screening, as explained in section~\ref{sec:methods}), and results from using the bare (unscreened) Coulomb interaction tensor. 
  The ``monopole'' cRPA model is inspired by Scott and Booth~\cite{scott_rigorous_2023}, and further details can be found in section~\ref{sec:methods}.

  The full cRPA results that have already been discussed yield a $\chi^2 = 27.3$. 
  Restricting the cRPA screening predominantly to the monopole channel, such that predominately density-density Coulomb matrix elements are screened, yields a better agreement with $\chi^2=16.5$. 
  This additional constraint on the cRPA screening shifts the crystal-field excitation energies down by about 200$\,$meV, and clusters the spin-flip excitations into 3:1 pairs, with the highest spin-flip excitations being closer to the lowest crystal-field excitations (in Fig.~\ref{fig:EX_Eps} they actually now overlap). 
  This results in a better agreement with the reference data as quantified by the reduced $\chi^2$ value. 
  When screening is completely neglected, i.e., when we use the bare $v_{ijkl}$ Coulomb tensor, the agreement with the reference data as measured by $\chi^2$ becomes, however, unsatisfactory compared to the (monopole) cRPA cases.
  In the bare case, the crystal-field excitation energies are strongly underestimated, which mixes the overall ordering of the spin-flip and crystal-field excitations, resulting in a large error,  $\chi^2=86.5$. 
  The spin-flip excitations, however, also show the 3:1 clustering similar to the monopole screening case.
 
  From this study, we learn that the cRPA screening model is indeed better than bare Coulomb interactions. 
  However, the screening model is not perfectly accurate; a modification of only screening the monopole channel, motivated by the underlying RPA screening by long-range charge fluctuations~\cite{scott_rigorous_2023}, results in better agreement of the model eigenstates with the quantum chemistry reference results, at least for vanadocene.

    \begin{figure}[h]
        \centering
        \includegraphics[width=0.99\columnwidth]{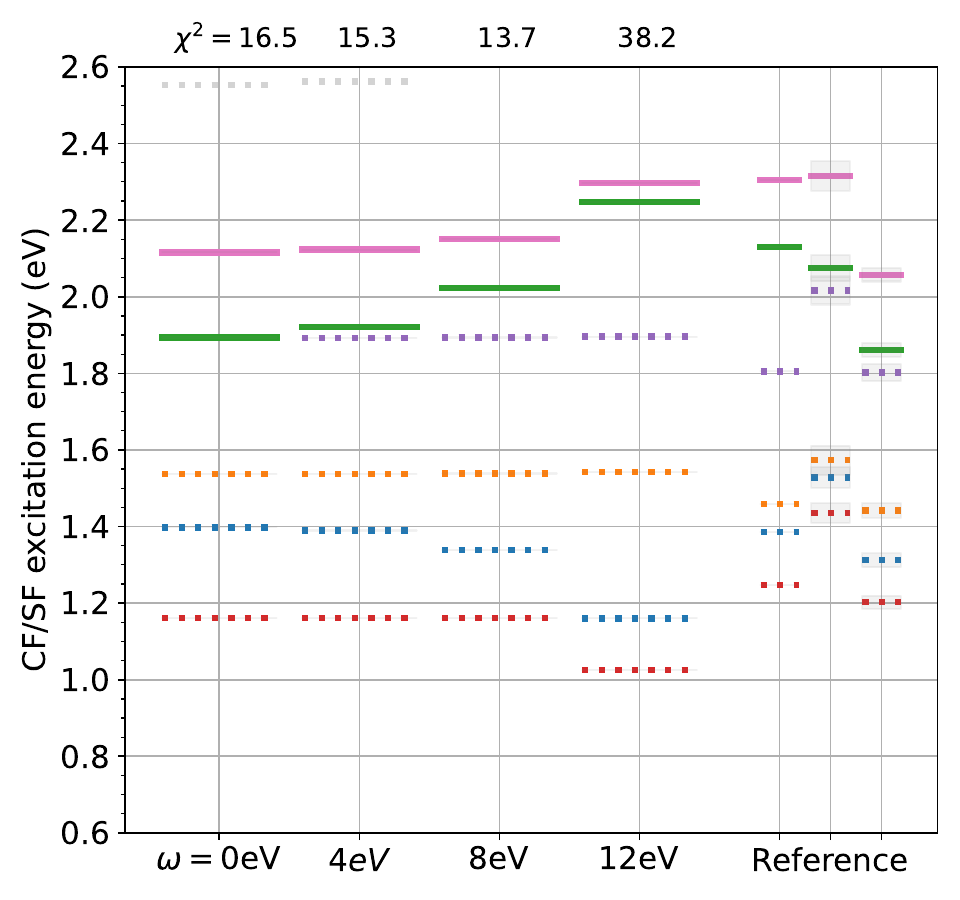}
        \caption{cRPA screening frequency influence on downfolded excitation energies within the monopole channel using the MLWF basis and no double counting correction. Dashed and solid lines represent spin-flip (SF) and crystal-field (CF) excitations, respectively. Light grey excitation energies have not been characterized.}
        \label{fig:EX_Freq}
    \end{figure}

\subsubsection{Finite Frequency Screening}

The Coulomb tensor in the cRPA approximation is frequency-dependent~\cite{Aryasetiawan2004}, i.e., $\hat{U}_\text{cRPA}(\omega)$. 
To obtain a static Hamiltonian which is easier to solve, it is conventional to evaluate $\hat{U}_\text{cRPA}(\omega=0)$, based on the idea that the low energy target space physics is slow compared to the screening processes. 
To assess this approximation, we study how the results change when the cRPA Coulomb matrix elements are evaluated at finite frequency. 
Here, we perform this analysis based on a simple model of the frequency dependence, the plasmon pole model (see Methods for details).

  Fig.~\ref{fig:EX_Freq} shows the excitation energies using a Coulomb tensor $\hat{U}(\omega)$ evaluated at various $\omega$, which are all below the estimated background screening plasmon pole frequency of approximately $16.5\,$eV. 
  Note that these are all calculated using the monopole screening model of Fig.~\ref{fig:EX_Eps}, which corresponds to $\omega=0$ here. 
  For $\omega=4\,$eV and $\omega=8\,$eV, we find an increase in the crystal-field excitation energies and a concomitant small improvement in the match with the reference data as quantified by $\chi^2$. 
  The spin excitation energies barely change for these $\omega$ values.  
  $\omega=12\,$eV is close to the plasmon pole frequency of $16.5\,$eV, so the Coulomb matrix elements change more rapidly, and the excitation energies change more as a result. 
  At this point, the agreement with the reference data deteriorates substantially. 
  We should note that the theoretical justification of cRPA relies on the screening frequency $\omega$ being smaller than the gap in the rest space~\cite{PhysRevB.104.045134}, which is roughly $6\,$eV in vanadocene (Fig.~\ref{fig:mlwf}). 
  I.e., for $\omega > 6\,$eV, the dynamics of particle-hole excitations start to become relevant, which we cannot capture in the static Hamiltonian formalism.
  
  With regard to the frequency choice in the cRPA background screening, we thus find slightly better agreement with the reference data upon using $\omega>0$ in comparison to $\omega =0$. 
  Among all benchmarked quantities, however, the background screening frequency has the smallest effect.
  
\section{Discussion}

By comparing the ground state and charge-neutral excited states obtained directly from first-principles descriptions and from minimal downfolded Hamiltonians, we have provided a systematic and quantitative sensitivity analysis of the DFT+cRPA scheme for the \textit{ab initio} derivation of interacting model Hamiltonians for the description of correlation effects in the vanadocene molecule. 
We expect that the results in this study will be most applicable to similar systems, such as defects in gapped semiconductors or correlated bulk materials with well-separated correlated sub-spaces.
These systems are ideal for cRPA, as noted by some of us~\cite{PhysRevB.104.045134}, and even so, we were able to quantify errors in every step of the DFT+cRPA process.

To summarize, we find the following about the DFT+cRPA process:
 Orbital shapes are crucial. The single-particle energies for FGWF are not that different from MLWF but the model is extremely inaccurate because the two-body terms highly depend on the orbital shapes. 
 Na\"ive application of the orbital dependent double counting corrections results in very poor estimations of crystal-field excitations, because it fails to describe the one-particle levels. Small corrections do not change much within our uncertainties. 
The cRPA-screened interactions are much better than the bare ones. Monopole-only screening is clearly better than traditional. Frequency dependence, in this case, is not that big, unless one goes to very high energies. 

From this data, we can make a few recommendations.
One should be very careful about the basis orbitals, especially in cases where disentanglement of bands is necessary, since orbital shapes can completely change the results. Similarly, the functional used to choose localized orbitals might have significant effects on the interactions computed. 
Double counting corrections do not necessarily improve the accuracy of the Hamiltonian. These ``corrections'' can also be conflated with the tendency of DFT to incorrectly compute the one-particle energies. 
Even in the case of vanadocene, which is a best-case scenario for cRPA, a na\"ive application of cRPA screening obtains interactions that are better than bare, but appear to be limiting accuracy. In this case they can be improved by monopole-only screening in particular, although it remains to be seen if such a method is generally more accurate.

Looking forward, we have demonstrated that it is possible to generate reference quantum mechanical data of sufficient quality and quantity to systematically analyze heuristic (but inexpensive) methods like DFT+cRPA, at least for model systems such as vanadocene. 
The application of multiple first-principles methods allows us to evaluate the uncertainty in the reference data, partially mitigating the problem of overfitting. 
The data from this work can serve as a reference for other methods, such as $GW$-based approaches, and could be applied to a variety of materials and systems and has been made publicly available~\cite{Chang_DFT_cRPA_benchmarking_in_2024}.

\section{Methods}
\label{sec:methods}

  \subsection{DFT Calculations}

  We use Quantum Espresso (QE)~\cite{giannozzi_quantum_2009,giannozzi_advanced_2017,giannozzi_quantum_2020} and embed the vanadocene molecule in a $15\!\times\!15\!\times\!15$\AA$^3$\ supercell to minimize spurious wave function overlap and undesired screening between repeated cells. The plane-wave cut-off was set to $440\,$Ryd, and we apply the spin-restricted (spin-unpolarized) generalized gradient approximation within a PBE functional~\cite{PhysRevLett.77.3865}. In these spin-restricted calculations we find an ordering of $e_2 < a_1 < e_1$, which is different from spin-unrestricted results for both C$_{2h}$ and $D_{5h}$ VCp$_2$ where the $e_2 < a_1$ ordering is reversed between the spin channels~\cite{jackson_vanadocene_2012,sergentu_similar_2018}.
  The occupations of the three lowest $d$ orbitals are constrained to exactly $1$ to avoid partial state occupations due to the conventional smearing methods in plane-wave DFT codes. 
  This constraint does, however, not significantly affect the Kohn Sham energies compared to calculations with a large smearing, c.f. Tab.~\ref{tab:KSEnergies}. 

    \begin{table}
        \centering
        \caption{Single-particle energies (in eV) from DFT output (KS), after wannierization (MLWF and FGWF), and including double counting corrections (MLWF+$H$). DFT energies are given for constrained and smeared occupations. \label{tab:KSEnergies}} 
        \begin{ruledtabular}
            \begin{tabular}{c|ccc}
             type & $e_2$ & $a_1$ & $e_1$ \\
             \hline
                KS cons. & -0.082 & 0.0 & 2.310 \\
                KS sme.  & -0.061 & 0.0 & 2.311 \\
                \hline
                MLWF\tnote{e} & -0.082 & 0.0 & 2.310 \\
                FGWF\tnote{e} &  0.036 & 0.0 & 1.986 \\ 
                \hline
                MLWF + $H_{\text{DC}}^{\text{Har}}$ & -0.451 & 0.0 & 1.140 \\
                MLWF + $H_{\text{CFC}}^{(a_1)}$ &  0.061 & 0.0 & 2.453 \\
                MLWF + $H_{\text{CFC}}^{(e_1)}$ & -0.082 & 0.0 & 2.160
            \end{tabular}
        \end{ruledtabular}
    \end{table}

  \subsection{Wannier Basis Sets}

    \begin{figure}[h]
        \centering
        \includegraphics[width=0.99\columnwidth]{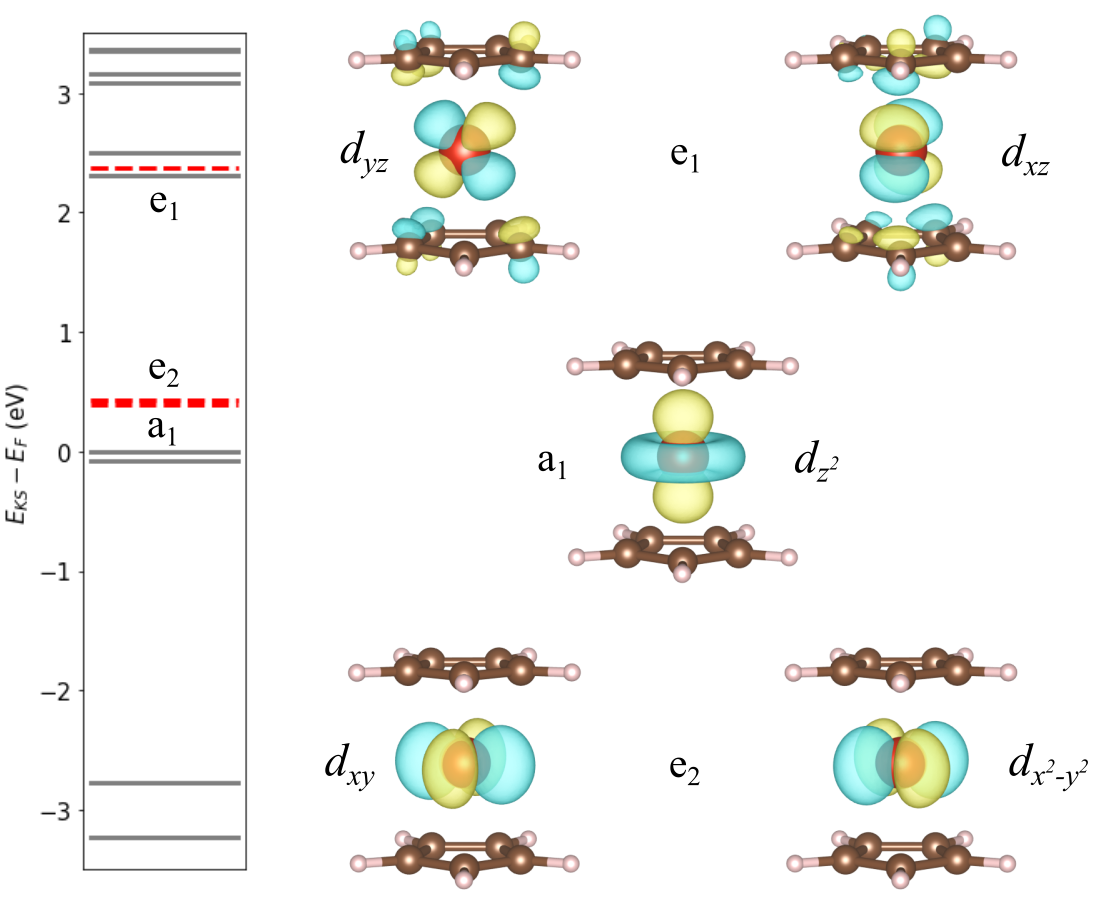}
        \caption{Single Particle Properties. Left: DFT Kohn-Sham energies. Red lines mark single-particle energies resulting from diagonalization of the downfolded Hamiltonian using first guess Wannier functions. Right: First guess Wannier functions used within the downfolding procedure.}
        \label{fig:fgwf}
    \end{figure}
  
  We use the three highest (partially) occupied and the two lowest unoccupied molecular orbitals with predominant V $d$-orbital character to construct five maximally localized Wannier functions (MLWF) $\ket{\phi_i}$ utilizing RESPACK~\cite{nakamura_respack_2021}. 
  For the construction of these MLWFs we use an inner (frozen) window to constrain the energies of the five Kohn-Sham states of interest. The resulting Wannier orbitals resemble the expected V $d$ orbital symmetries, however, with small lobes at the carbon rings as visualized in Fig.~\ref{fig:mlwf}, which is in good agreement with Ref.~\cite{sergentu_similar_2018}. These indicate a finite hybridization between the V $d$ states and the carbon rings.

  To investigate this specific choice of Wannier functions and to qualitatively compare to projector based methods, we construct a second Wannier basis without using the frozen (inner) window, utilizing a larger Wannierization (outer) window, and without performing maximal localizations. The resulting first-guess Wannier function (FGWF) single-particle energies do not perfectly reproduce the KS ones, as indicated in Fig.~\ref{fig:fgwf} and listed in Tab.~\ref{tab:KSEnergies} and cannot be generated via some uniform rotation of the MLWFs. These FGWF orbitals are nevertheless similar to the MLWFs, but do have a slightly different shape, which is reflected in the overlap matrix elements between Wannier orbitals of the same symmetry:
  \begin{align*}
      \braket{ \phi^{\text{MLWF}}_{e_2} | \phi^{\text{FGWF}}_{e_2} } \approx 0.936, \\
      \braket{ \phi^{\text{MLWF}}_{a_1} | \phi^{\text{FGWF}}_{a_1} } \approx 0.961, \\
      \braket{ \phi^{\text{MLWF}}_{e_1} | \phi^{\text{FGWF}}_{e_1} } \approx 0.987. \\
  \end{align*}

  \subsection{Model Hamiltonian Matrix Elements and cRPA Screening}
  
  With these Wannier basis sets, we evaluate the hopping matrix elements $t_{ij} = \braket{\phi_i | H_{\text{DFT}} | \phi_j}$ and the bare $v_{ijkl} = \braket{\phi_i \phi_j|v|\phi_k \phi_l}$ as well as the statically cRPA screened $U_{ijkl} = \braket{\phi_i \phi_j|U(\omega=0)|\phi_k \phi_l}$
  Coulomb matrix elements within RESPACK~\cite{nakamura_respack_2021}. The resulting single-particle energies are given in Tab.~\ref{tab:KSEnergies}, and the density-density and Hund's exchange matrix elements are listed in Tab.~\ref{tab:CME} and Tab.~\ref{tab:CME-FGWF}. 
  We see that density-density elements are screened by about $55$ to $60$\% (up to $7.8\,$eV), while Hund's elements are reduced by only $5$ to $10$\% (up to $60\,$meV). Analysing the leading and sub-leading screening contributions yields a screening of $\varepsilon_1 = \varepsilon_\text{mono} \approx 2.44$ in the predominate monopole channel and $\varepsilon_{i>1} = \varepsilon_\text{multi} \approx 1.04$ to $1.21$ in the sub-leading multipole channels. These screening channels are defined by $\varepsilon_i = v_i / U_i$ with $v_i$ being the eigenvalues of the full $v_{ijkl}$ tensor evaluated in a product basis. $U_i$ are the approximated eigenvalues of $U_{ijkl}$ obtained using the eigenbasis of the bare $v_{ijkl}$~\cite{rosner_wannier_2015}.
  The full cRPA screening is thus most dominant in the mono-pole channel, which mostly affects density-density interactions, while the screening becomes small in the multi-pole channels, which mostly affect Hund's exchange elements. 
  
  Inspired by Scott and Booth~\cite{scott_rigorous_2023} we also construct a cRPA Coulomb tensor, which is screened mostly in the leading monopole channel, such that only $\varepsilon_1 > 1$, while all other $\varepsilon_{i>1} = 1$. The resulting matrix elements are listed in Tab.~\ref{tab:CME}. As expected, the approx. ``monopole'' cRPA Coulomb matrix elements agree with the full cRPA ones in the density-density and approx. with the bare ones in the Hund's exchange channels. However, note that the monopole screened Hund's exchange elements are up to $77\,$meV larger than the fully screened ones, which is relevant for the spin-flip excitation energies.

    \begin{figure}[h]
        \centering
        \includegraphics[width=0.99\columnwidth]{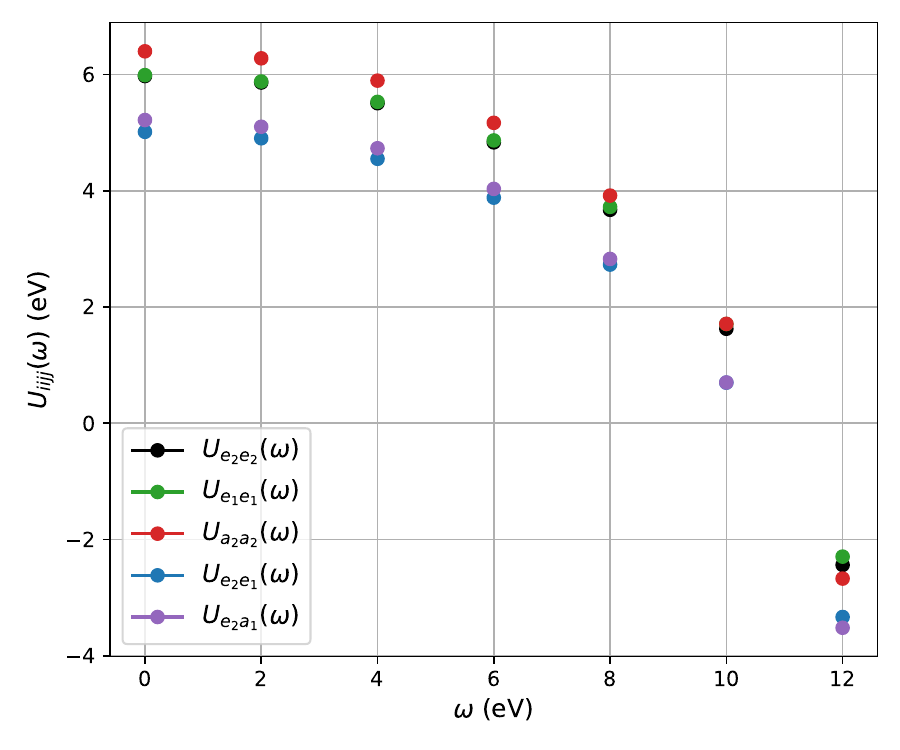}
        \caption{Approximated frequency dependence of the monopole screened cRPA Coulomb matrix elements in the density-density channel}.
        \label{fig:Uw}
    \end{figure}

  For the analysis of finite background screening frequencies, we calculate $U_{ijkl}(i\omega_n) = \braket{\phi_i \phi_j|U(i\omega_n)|\phi_k \phi_l}$ for a few Matsubara frequencies $i\omega_n$. Afterwards we fit the leading eigenvalue to a single plasmon pole model of the form $U_1(i\omega_n) = v_1 - g^2 \frac{2\omega_p}{\omega_n^2 + \omega_p^2}$ yielding $g\approx17.6\,$eV and $\omega_p \approx 16.5$eV. This allows us to adjust the real frequency of the leading eigenvalue as $U_1(\omega) \approx v_1 + g^2 \frac{2\omega_p}{\omega^2 - \omega_p^2 + i\delta}$ for which we use $\delta=2.5\,$eV. This way, we can continuously adjust the leading eigenvalue in the monopole screening model as done for Fig.~\ref{fig:EX_Freq}. The rather large $\delta=2.5\,$eV is used here to mimic the effects of screening channels beyond the single plasmon pole model. The dependence of the density-density monopole-screened Coulomb interaction matrix elements is depicted in Fig.~\ref{fig:Uw}.
    
  Although the FGWF wave functions show deviations of just up to $6.4\%$ in comparison to the MLWF ones (as measured by the overlaps given above), the resulting bare Coulomb matrix elements are different by up to $27\%$ as a result of the $\phi_i^4$ dependence of $v$ and as visible from Tab.~\ref{tab:CME-FGWF}. 
  This is a result of the enhanced atomic character of the FGWFs and their reduced hybridization with the carbon rings, yielding a smaller spread and, thus, enhanced bare Coulomb matrix elements.

\subsection{Double counting}
\label{sec:doublecounting}
  
  We choose three orbital-dependent correction schemes, which all act on the single-particle $t_{ij}$ hopping matrix elements according to the following terms, which we add to the model Hamiltonian:
    \begin{equation}
    \label{eq:DC_hartree}
        H^{\text{Har}}_{\text{DC}}=\sum_{ij,\sigma}c^\dagger_{i\sigma}c_{j\sigma} \sum_{kl}\rho_{kl} U_{iljk},
    \end{equation}
  with $\rho_{kl}$ the single-particle density-matrix in the Wannier basis obtained from the DFT calculation. This is the conventional Hartree double counting correction as discussed in detail in Refs.~\cite{bockstedte_ab_2018,ma_quantum_2020,muechler_quantum_2022}.  We also apply \textit{ad hoc} chosen crystal-field corrections of the form
    \begin{equation}
    \label{eq:CFC}
        H_{\text{CFC}}^{(a_1/e_1)}=\sum_{i,\sigma} \Delta_{i}^{(a_1/e_1)} c^\dagger_{i\sigma}c_{i\sigma}
    \end{equation}
  with only $\Delta^{(a_1)}_{i=a_1} = -0.143$\,eV or $\Delta^{(e_1)}_{i=e_1} = -0.150$\,eV non-zero elements for the respective $H_{\text{CFC}}$. These corrections thus solely shift the $a_1$ or $e_1$ states, while they preserve all symmetries.
  With these, we aim to investigate the effects of possibly wrong single-particle energies. 

\subsection{Exact Solutions and Benchmark Quantities}

    \begin{figure*}
        \centering
        \includegraphics[width=0.99\textwidth]{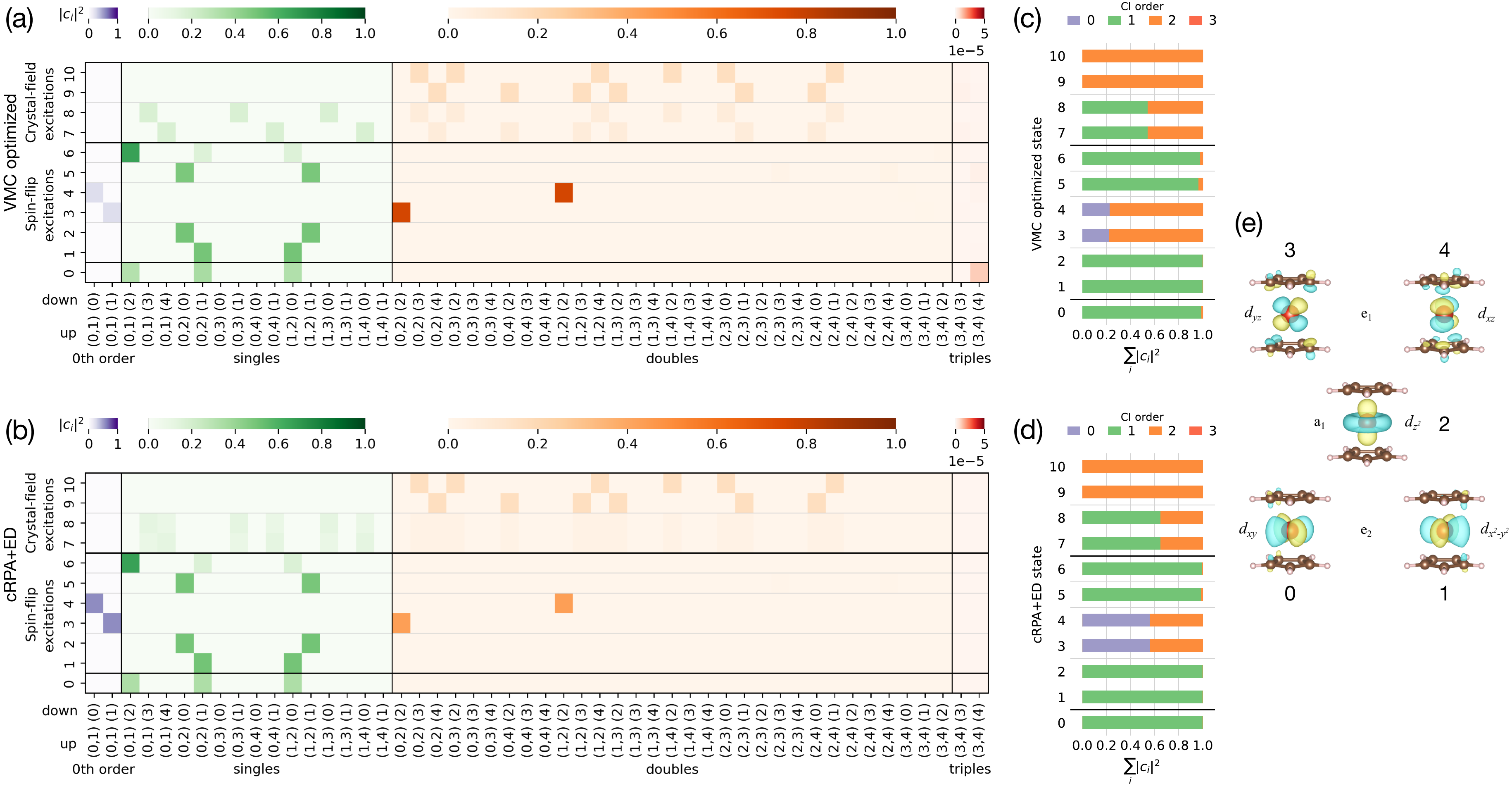}
        \caption{(a,b) The squares of configuration interaction (exact diagonalization) coefficients ($|c_i|^2$) of the optimized multi-Slater-Jastrow wave functions obtained using variational Monte Carlo and eigenstates obtained from ED using the model given by the MLWFs. The four panels correspond to the 0th order configuration interaction (purple), single excitations (green), double excitations (orange), and triple excitations (red) within the CAS(5,\,(2,1)) space.
        The tick labels on $x$ axis, $(i,j)\,(k)$, represent a Slater determinant $c^\dagger_{k\downarrow} c^{\dagger}_{j\uparrow} c^{\dagger}_{i\uparrow}\ket{\Phi_0}$ (see panel (e) for the MLWFs), where $\ket{\Phi_0}$ is the Slater determinant of all the valence electrons doubly occupying the orbitals up to the V $d$ shell.
        Panels (c) and (d) show the cumulated contributions in each order of CIs for the VMC wave functions and the model eigenstates.
        The black horizontal lines group the many-body states into ground state, spin-flip excitations, and crystal-field excitation, whereas the grey horizontal lines group the degenerate states together.
        We can see that states 3 and 4 (degenerate in energy and correspond to the 2nd excited state, labeled by blue dashed lines in Fig.~\ref{fig:EX_FP}) and all the crystal-field excitations show significant contributions from the double excitations.}
        \label{fig:CI}
    \end{figure*}

  We use the exact diagonalization (ED) routines (atom\_diag) from TRIQS~\cite{parcollet_triqs_2015} to solve the downfolded many-body Hamiltonian $H=\hat{t}_{\alpha\beta}c^\dagger_\alpha c^{\phantom{\dagger}}_\beta+\hat{U}_{\alpha\beta\gamma\delta} c^\dagger_\alpha c^\dagger_\beta c^{\phantom{\dagger}}_\gamma c^{\phantom{\dagger}}_\delta$, giving us access to both the energies and the many-body wave functions in the target space. Here, we restrict ourselves to charge-neutral excitations, i.e., those with $N=3$ electrons in total, such that the relevant many-body wave functions are of the form $\ket{\Psi}=\sum_{\alpha\beta\gamma} A_{\alpha\beta\gamma} c^\dagger_\alpha c^\dagger_\beta c^\dagger_\gamma\ket{0}$, where the sum runs over spin and orbital indices. For three electrons, the possible values of $S_z$ are $\pm3/2$ and $\pm1/2$. Note that our model has $SU(2)$ spin symmetry, so the spin $S$ is a good quantum number, and states with the same $S$ and different $m_s$ have the same energy. We can thus restrict ourselves to states with $m_s=1/2$. 

  Using the Wannier functions $\phi_\alpha(\mathbf{r})$, we calculate the electron density $\rho_i(\mathbf{r})$ in real space for a given many-body wave function $i$ as $\rho_i(\mathbf{r})=\sum_{i,\alpha} \braket{\Psi_i | c_\alpha^\dagger c_\alpha | \Psi_i} |\phi_\alpha(\mathbf{r})|^2 $ using the many body states $\Psi_i$ as obtained from the exact diagonalization of the Hamiltonian. For the spin-resolved density $\rho_i^\sigma(\mathbf{r})$ we use a similar expression yielding $\int d\mathbf{r} \rho_i^\uparrow(\mathbf{r})=2$ and $\int d\mathbf{r} \rho_i^\downarrow(\mathbf{r})=1$ in the $m_s=1/2$ channel.
  In addition to this, we also calculate the 1-RDMs $(\rho_i)_{\alpha\beta} = \braket{\Psi_i|c^\dagger_\alpha c_{\beta}|\Psi_i}$ for each wave function $\Psi_i$.

  Based on these quantities, that is, the many-body state energies, wave functions, 1-RDMs, and charge densities, we can unambiguously identify all states among all applied methods. A corresponding comparison between the many-body wave functions represented as linear combinations of Slater determinants from DFT+cRPA and the multi-Slater part of the trial wave functions for DMC is depicted in Fig.~\ref{fig:CI}. The corresponding comparisons for the 1-RDMs and the many-body charge densities are given in the Supplemental Notes 1 and 2.

\subsection{Real-space Fixed-node Diffusion Monte Carlo (DMC)}

In order to obtain approximate many-body eigenstates using fixed-node diffusion Monte Carlo (DMC), we constructed the trial wave functions in the following multi-Slater-Jastrow form,
\begin{equation}
\begin{split}
    &\Psi_{\text{T}}(\mathbf{R}) = \Psi_{\text{T}}(\mathbf{r}_1, \mathbf{r}_2, \cdots, \mathbf{r}_N) \\
    &= \text{e}^{J(\mathbf{r}_1, \mathbf{r}_2, \cdots, \mathbf{r}_N)} 
    \sum_k c_k D^{\uparrow}_k\left(\chi_{i\uparrow}(\mathbf{r}_j)\right)
    D^{\downarrow}_k\left(\chi_{i\downarrow}(\mathbf{r}_j)\right),
\end{split}
\end{equation}
where $\mathbf{r}_i$ represents the real-space coordinates of the $i$th electron and $\mathbf{R}$ encompasses the configurations of all the electrons.
The multi-Slater-determinant expansion, given by $\sum_k c_k D_k^{\uparrow}\left(\chi_{i \uparrow}\left(\mathbf{r}_j\right)\right) D_k^{\downarrow}\left(\chi_{i \downarrow}\left(\mathbf{r}_j\right)\right)$, was generated from restricted Hartree-Fock (RHF) followed by state-averaged complete active space self-consistent field (CASSCF)~\cite{ruedenberg_are_1982} using the PySCF package~\cite{pyscf2020} with correlation-consistent effective core potentials, and a corresponding triple-zeta basis set~\cite{annaberdiyev_new_2018}.
The system consists of 63 valence electrons, and in CASSCF, we chose an active space that includes two spin-up electrons, one spin-down electron, and five molecular orbitals that are V-$3d$-like determined by the atomic valence active space (AVAS) procedure~\cite{AVAS2017}.
For all the eigenstates, the parameters in the two-body Jastrow factor $J$, all the determinant coefficients $c_k$, and the molecular orbitals in the active space were fully optimized using variational Monte Carlo~\cite{foulkes_quantum_2001} using the method described in Refs.~\cite{pathak_non-orthogonal_2018, pathak_light_2020}.
The excited states were optimized with the constraint that they are orthogonal to the optimized lower-energy eigenstates~\cite{pathak_excited_2021}. 
After the trial wave functions $\Psi_{\text{T}}$ (of either the ground state or the excited states) were fully optimized, the lowest-energy wave functions that have the same nodal structure were projected out using DMC~\cite{foulkes_quantum_2001}.
We then compute the energies and reduced density matrices of the fixed-node wave functions of all the eigenstates of interest.
All the variational and diffusion Monte Carlo calculations, including the optimization of trial wave functions and DMC, were performed using the PyQMC package \cite{wheeler_pyqmc_2022} interfaced with PySCF.

\subsubsection*{Using reduced density matrices in natural orbitals to characterize many-body eigenstates}
For a wave function $\ket{\Psi}$, its one-body and two-body reduced density matrices (1(2)-RDMs) in a given single-particle basis are defined as
\begin{equation}
\begin{split}
    \rho^{\sigma}_{ij}[\Psi] &= \left\langle\Psi\left|c^\dagger_{i\sigma} c_{j\sigma}\right|\Psi\right\rangle,\\
    \rho_{i j k l}^{\sigma \sigma^{\prime}}[\Psi] & =\left\langle\Psi\left|c_{i \sigma}^{\dagger} c_{k \sigma^{\prime}}^{\dagger} c_{l \sigma^{\prime}} c_{j \sigma}\right| \Psi\right\rangle, \\   
\end{split}
\end{equation}
where $c^\dagger_{i\sigma}$ ($c_{i\sigma}$) creates (annihilates) an electron in single-particle orbital $\chi_{i\sigma}$.
Almost all the one- and two-body observables can be computed using the 1- and 2-RDMs.
For example, the spin-resolved electronic density is evaluated as
\begin{equation}
    \rho^{\sigma}(\mathbf{r}) = \sum_i \chi^*_{i\sigma}(\mathbf{r})\rho^{\sigma}_{ii}\left[\Psi\right] \chi_{i\sigma}(\mathbf{r}).
\end{equation}

To directly compare the observables that characterize the fixed-node eigenstates with those of the downfolded eigenstates, it is crucial to ensure that the RDMs are calculated using the identical set of single-particle orbitals.
Given that the downfolding and DMC calculations were performed using separated packages and employed a different basis to expand the active space (MLWFs for the downfolded results and CASSCF molecular orbitals for DMC), we rotate the RDMs obtained from each method to the same basis set to align the many-body eigenstates.

The natural orbitals are defined as the principal components, that correspond to the largest five singular values, of all the 1-RDMs up to the $m$th downfolded excited state~\cite{chang_learning_2023}, 
\begin{equation}
\begin{split}
\varrho &= U\Sigma V,\\
\varrho &= \left(\rho^{\uparrow}[\Psi_0], \cdots, \rho^\uparrow[\Psi_m], \rho^{\downarrow}[\Psi_0], \cdots, \rho^{\downarrow}[\Psi_m] \right),\
\end{split}
\end{equation}
where $\rho^\sigma[\Psi_a]$ is eigenstate $\Psi_a$'s 1-RDM written in the Wannier basis $\left\lbrace \phi_i(\mathbf{r})\right\rbrace$.
We then transform all the 1- and 2-RDMs computed from both downfolding and DMC to these natural orbitals.
This ensures that we characterize downfolded and DMC eigenstates within the same Hilbert space.
We found $\mathrm{Tr}\left(\rho_{\text{downfolded}}\left[ \Psi_i\right]\right) - \mathrm{Tr}\left(\rho_{\text{DMC}}\left[ \Psi_i\right]\right)$ to be $0.1\sim 0.2$, for all the $\Psi_i$'s of interest (up to the 2nd charge excitation).
With three electrons in the active space, this suggests that when using the MLWFs derived from Wannierization of Kohn-Sham orbitals to characterize the low-energy excitations in VCp$_2$, we get about 5\% error in terms of their 1-RDMs.

\subsection{Auxiliary-Field Quantum Monte Carlo (AFQMC)}

We computed the low-energy many-body spectrum, including spin-flip, and crystal-field excitations, of the vanadocene molecule
 using phaseless AFQMC~\cite{ZhangKrakauer2003,Motta2018,Shi2021}.
All AFQMC calculations were performed using the Flatiron Institute, Center for Computational Quantum Physics's production quality AFQMC code.
A detailed description of the AFQMC method was recently reviewed in great detail~\cite{Motta2018}.
AFQMC was recently used to compute the ionization potentials 
of several 3d-transition metal complexes, including vanadocene, 
producing results which compare favorably with experiment~\cite{Rudshteyn2020}. 
Our AFQMC procedure, especially regarding the choice of trial wave function, is guided by that benchmark.
AFQMC calculations were performed using multi-Slater determinant trial wave functions 
which were derived from CASSCF calculations.
The active space for CASSCF was chosen using the atomic valence active space (AVAS) procedure~\cite{AVAS2017} and the ANO-RCC basis to define V-$3d$ reference atomic orbitals, with an AVAS threshold of 0.1.
This yields an active space consisting of 13 active electrons and 15 active orbitals.
CASSCF orbitals were optimized for the ground state, 
and excited-state CAS-wave functions were computed in one-shot state-specific CASCI calculations using the ground-state CASSCF orbitals to define the active space.
AFQMC trial wave functions are truncated CAS-wave functions where Slater determinants with weight less than 0.0014 are discarded.
For states with $S=3/2$, including the ground state and both crystal-field excitations, the trial wave functions consisted of only about 100 Slater determinants.
For all spin-flip excitations, the trial wave functions consisted of about 150 Slater determinants.

\subsection{Equation-of-Motion Coupled Cluster (EOMCC) }

Equation-of-motion coupled-cluster calculations were performed at the single-, and double-excitation level (EOM-CCSD),
 as implemented in PySCF~\cite{pyscf2020}.
The correlation consistent effective core potential (ccECP) / pseudopotential and the corresponding ccECP-cc-pVQZ basis were used for all atoms.
The ROHF ground state determinant, with $M_S=3/2$ and $S^2=3.75$, was used as a reference for EOM-CCSD.
Due to the very large dimension of the Hilbert space in the ccECP-cc-pVQZ basis, 
we performed EOM-CCSD in a CAS-like active space with 502 active orbitals and including all electrons in the active space.
We checked that the excitation energies are converged by comparing with similar calculations performed in an active space consisting of 452 orbitals.
The average absolute deviation of the excitation energies, computed over the states in Figure~\ref{fig:EX_FP},
between EOM-CCSD(502o,63e) and EOM-CCSD(452o,63e)
is $28(15)\,$meV with a maximum absolute deviation of $53\,$meV from the first crystal-field excitation. 

\section*{Data availability}
The data that support the findings of this work are available from the corresponding author upon reasonable request. All ab initio and modelling results (AFQMC, DMC, EOM-CC, DFT, cRPA, and ED) are furthermore publicly available~\cite{Chang_DFT_cRPA_benchmarking_in_2024}.

\section*{Acknowledgement}

We thank Yusuke Nomura for support for using RESPACK and Timothy Berkelbach for helpful discussions.
YC was supported by the U.S. Department of Energy, Office of Science, Office of Basic Energy Sciences, Computational Materials Sciences Program, under Award No. DE-SC0020177 for the initial diffusion Monte Carlo calculations and analysis performed at the University of Illinois.
EvL acknowledges support from the Swedish Research Council (Vetenskapsrådet, VR) under grant 2022-03090.
CED acknowledges support from NSF Grant No.~DMR-2237674. The Flatiron Institute is a division of the Simons Foundation.
TW acknowledges support from the Deutsche Forschungsgemeinschaft (DFG, German Research Foundation) through the cluster of excellence “CUI: Advanced Imaging of Matter” of the Deutsche Forschungsgemeinschaft (DFG EXC 2056, Project ID 390715994) and research unit QUAST FOR 5249 (project ID: 449872909; project P5).
LKW was supported by a grant from the Simons Foundation as part of the Simons Collaboration on the many-electron problem. MR thanks the Flatiron Institute for hospitality and acknowledges financial support from the Dutch research program ‘Materials for the Quantum
Age’ (QuMat). This program  (Registration Number 024.005.006) is part of the Gravitation program financed by the Dutch Ministry of Education, Culture and Science (OCW).

\bibliography{references}

\providecommand{\noopsort}[1]{}\providecommand{\singleletter}[1]{#1}%
\begin{thebibliography}{80}%
\makeatletter
\providecommand \@ifxundefined [1]{%
 \@ifx{#1\undefined}
}%
\providecommand \@ifnum [1]{%
 \ifnum #1\expandafter \@firstoftwo
 \else \expandafter \@secondoftwo
 \fi
}%
\providecommand \@ifx [1]{%
 \ifx #1\expandafter \@firstoftwo
 \else \expandafter \@secondoftwo
 \fi
}%
\providecommand \natexlab [1]{#1}%
\providecommand \enquote  [1]{``#1''}%
\providecommand \bibnamefont  [1]{#1}%
\providecommand \bibfnamefont [1]{#1}%
\providecommand \citenamefont [1]{#1}%
\providecommand \href@noop [0]{\@secondoftwo}%
\providecommand \href [0]{\begingroup \@sanitize@url \@href}%
\providecommand \@href[1]{\@@startlink{#1}\@@href}%
\providecommand \@@href[1]{\endgroup#1\@@endlink}%
\providecommand \@sanitize@url [0]{\catcode `\\12\catcode `\$12\catcode
  `\&12\catcode `\#12\catcode `\^12\catcode `\_12\catcode `\%12\relax}%
\providecommand \@@startlink[1]{}%
\providecommand \@@endlink[0]{}%
\providecommand \url  [0]{\begingroup\@sanitize@url \@url }%
\providecommand \@url [1]{\endgroup\@href {#1}{\urlprefix }}%
\providecommand \urlprefix  [0]{URL }%
\providecommand \Eprint [0]{\href }%
\providecommand \doibase [0]{https://doi.org/}%
\providecommand \selectlanguage [0]{\@gobble}%
\providecommand \bibinfo  [0]{\@secondoftwo}%
\providecommand \bibfield  [0]{\@secondoftwo}%
\providecommand \translation [1]{[#1]}%
\providecommand \BibitemOpen [0]{}%
\providecommand \bibitemStop [0]{}%
\providecommand \bibitemNoStop [0]{.\EOS\space}%
\providecommand \EOS [0]{\spacefactor3000\relax}%
\providecommand \BibitemShut  [1]{\csname bibitem#1\endcsname}%
\let\auto@bib@innerbib\@empty
\bibitem [{\citenamefont {Aryasetiawan}\ and\ \citenamefont
  {Nilsson}(2022)}]{Aryasetiawan2022}%
  \BibitemOpen
  \bibfield  {author} {\bibinfo {author} {\bibfnamefont {F.}~\bibnamefont
  {Aryasetiawan}}\ and\ \bibinfo {author} {\bibfnamefont {F.}~\bibnamefont
  {Nilsson}},\ }\href {https://aip.scitation.org/doi/10.1063/9780735422490}
  {\emph {\bibinfo {title} {Downfolding {Methods} in {Many-Electron}
  {Theory}}}}\ (\bibinfo  {publisher} {AIP Publishing LLC},\ \bibinfo {year}
  {2022})\BibitemShut {NoStop}%
\bibitem [{\citenamefont {Georges}\ \emph {et~al.}(1996)\citenamefont
  {Georges}, \citenamefont {Kotliar}, \citenamefont {Krauth},\ and\
  \citenamefont {Rozenberg}}]{Georges96}%
  \BibitemOpen
  \bibfield  {author} {\bibinfo {author} {\bibfnamefont {A.}~\bibnamefont
  {Georges}}, \bibinfo {author} {\bibfnamefont {G.}~\bibnamefont {Kotliar}},
  \bibinfo {author} {\bibfnamefont {W.}~\bibnamefont {Krauth}},\ and\ \bibinfo
  {author} {\bibfnamefont {M.~J.}\ \bibnamefont {Rozenberg}},\ }\bibfield
  {title} {\bibinfo {title} {Dynamical mean-field theory of strongly correlated
  fermion systems and the limit of infinite dimensions},\ }\href
  {https://doi.org/10.1103/RevModPhys.68.13} {\bibfield  {journal} {\bibinfo
  {journal} {Rev. Mod. Phys.}\ }\textbf {\bibinfo {volume} {68}},\ \bibinfo
  {pages} {13} (\bibinfo {year} {1996})}\BibitemShut {NoStop}%
\bibitem [{\citenamefont {Rubtsov}\ \emph {et~al.}(2008)\citenamefont
  {Rubtsov}, \citenamefont {Katsnelson},\ and\ \citenamefont
  {Lichtenstein}}]{Rubtsov08}%
  \BibitemOpen
  \bibfield  {author} {\bibinfo {author} {\bibfnamefont {A.~N.}\ \bibnamefont
  {Rubtsov}}, \bibinfo {author} {\bibfnamefont {M.~I.}\ \bibnamefont
  {Katsnelson}},\ and\ \bibinfo {author} {\bibfnamefont {A.~I.}\ \bibnamefont
  {Lichtenstein}},\ }\bibfield  {title} {\bibinfo {title} {Dual fermion
  approach to nonlocal correlations in the {Hubbard} model},\ }\href
  {https://doi.org/10.1103/PhysRevB.77.033101} {\bibfield  {journal} {\bibinfo
  {journal} {Phys. Rev. B}\ }\textbf {\bibinfo {volume} {77}},\ \bibinfo
  {pages} {033101} (\bibinfo {year} {2008})}\BibitemShut {NoStop}%
\bibitem [{\citenamefont {Rubtsov}\ \emph {et~al.}(2012)\citenamefont
  {Rubtsov}, \citenamefont {Katsnelson},\ and\ \citenamefont
  {Lichtenstein}}]{Rubtsov12}%
  \BibitemOpen
  \bibfield  {author} {\bibinfo {author} {\bibfnamefont {A.}~\bibnamefont
  {Rubtsov}}, \bibinfo {author} {\bibfnamefont {M.}~\bibnamefont
  {Katsnelson}},\ and\ \bibinfo {author} {\bibfnamefont {A.}~\bibnamefont
  {Lichtenstein}},\ }\bibfield  {title} {\bibinfo {title} {Dual boson approach
  to collective excitations in correlated fermionic systems},\ }\href
  {https://doi.org/https://doi.org/10.1016/j.aop.2012.01.002} {\bibfield
  {journal} {\bibinfo  {journal} {Ann. Phys. (N.Y.)}\ }\textbf {\bibinfo
  {volume} {327}},\ \bibinfo {pages} {1320} (\bibinfo {year}
  {2012})}\BibitemShut {NoStop}%
\bibitem [{\citenamefont {Galler}\ \emph {et~al.}(2019)\citenamefont {Galler},
  \citenamefont {Thunström}, \citenamefont {Kaufmann}, \citenamefont {Pickem},
  \citenamefont {Tomczak},\ and\ \citenamefont {Held}}]{Galler19}%
  \BibitemOpen
  \bibfield  {author} {\bibinfo {author} {\bibfnamefont {A.}~\bibnamefont
  {Galler}}, \bibinfo {author} {\bibfnamefont {P.}~\bibnamefont {Thunström}},
  \bibinfo {author} {\bibfnamefont {J.}~\bibnamefont {Kaufmann}}, \bibinfo
  {author} {\bibfnamefont {M.}~\bibnamefont {Pickem}}, \bibinfo {author}
  {\bibfnamefont {J.~M.}\ \bibnamefont {Tomczak}},\ and\ \bibinfo {author}
  {\bibfnamefont {K.}~\bibnamefont {Held}},\ }\bibfield  {title} {\bibinfo
  {title} {The {A}binitio{D}{$\Gamma$}{A} {P}roject v1.0: {N}on-local
  correlations beyond and susceptibilities within dynamical mean-field
  theory},\ }\href {https://doi.org/https://doi.org/10.1016/j.cpc.2019.07.012}
  {\bibfield  {journal} {\bibinfo  {journal} {Comput. Phys. Commun.}\ }\textbf
  {\bibinfo {volume} {245}},\ \bibinfo {pages} {106847} (\bibinfo {year}
  {2019})}\BibitemShut {NoStop}%
\bibitem [{\citenamefont {Kananenka}\ \emph {et~al.}(2015)\citenamefont
  {Kananenka}, \citenamefont {Gull},\ and\ \citenamefont {Zgid}}]{Zgid15}%
  \BibitemOpen
  \bibfield  {author} {\bibinfo {author} {\bibfnamefont {A.~A.}\ \bibnamefont
  {Kananenka}}, \bibinfo {author} {\bibfnamefont {E.}~\bibnamefont {Gull}},\
  and\ \bibinfo {author} {\bibfnamefont {D.}~\bibnamefont {Zgid}},\ }\bibfield
  {title} {\bibinfo {title} {Systematically improvable multiscale solver for
  correlated electron systems},\ }\href
  {https://doi.org/10.1103/PhysRevB.91.121111} {\bibfield  {journal} {\bibinfo
  {journal} {Phys. Rev. B}\ }\textbf {\bibinfo {volume} {91}},\ \bibinfo
  {pages} {121111} (\bibinfo {year} {2015})}\BibitemShut {NoStop}%
\bibitem [{\citenamefont {Lee}\ \emph {et~al.}(2023)\citenamefont {Lee},
  \citenamefont {Poncé}, \citenamefont {Bushick}, \citenamefont {Hajinazar},
  \citenamefont {Lafuente-Bartolome}, \citenamefont {Leveillee}, \citenamefont
  {Lian}, \citenamefont {Macheda}, \citenamefont {Paudyal}, \citenamefont
  {Sio}, \citenamefont {Zacharias}, \citenamefont {Zhang}, \citenamefont
  {Bonini}, \citenamefont {Kioupakis}, \citenamefont {Margine},\ and\
  \citenamefont {Giustino}}]{lee2023electronphonon}%
  \BibitemOpen
  \bibfield  {author} {\bibinfo {author} {\bibfnamefont {H.}~\bibnamefont
  {Lee}}, \bibinfo {author} {\bibfnamefont {S.}~\bibnamefont {Poncé}},
  \bibinfo {author} {\bibfnamefont {K.}~\bibnamefont {Bushick}}, \bibinfo
  {author} {\bibfnamefont {S.}~\bibnamefont {Hajinazar}}, \bibinfo {author}
  {\bibfnamefont {J.}~\bibnamefont {Lafuente-Bartolome}}, \bibinfo {author}
  {\bibfnamefont {J.}~\bibnamefont {Leveillee}}, \bibinfo {author}
  {\bibfnamefont {C.}~\bibnamefont {Lian}}, \bibinfo {author} {\bibfnamefont
  {F.}~\bibnamefont {Macheda}}, \bibinfo {author} {\bibfnamefont
  {H.}~\bibnamefont {Paudyal}}, \bibinfo {author} {\bibfnamefont {W.~H.}\
  \bibnamefont {Sio}}, \bibinfo {author} {\bibfnamefont {M.}~\bibnamefont
  {Zacharias}}, \bibinfo {author} {\bibfnamefont {X.}~\bibnamefont {Zhang}},
  \bibinfo {author} {\bibfnamefont {N.}~\bibnamefont {Bonini}}, \bibinfo
  {author} {\bibfnamefont {E.}~\bibnamefont {Kioupakis}}, \bibinfo {author}
  {\bibfnamefont {E.~R.}\ \bibnamefont {Margine}},\ and\ \bibinfo {author}
  {\bibfnamefont {F.}~\bibnamefont {Giustino}},\ }\href@noop {} {\bibinfo
  {title} {Electron-phonon physics from first principles using the {EPW}
  code}},\ \bibinfo {howpublished} {Preprint at
  https://arxiv.org/abs/2302.08085} (\bibinfo {year} {2023})\BibitemShut
  {NoStop}%
\bibitem [{\citenamefont {Falter}\ and\ \citenamefont
  {Selmke}(1981)}]{falter_renormalization_1981}%
  \BibitemOpen
  \bibfield  {author} {\bibinfo {author} {\bibfnamefont {C.}~\bibnamefont
  {Falter}}\ and\ \bibinfo {author} {\bibfnamefont {M.}~\bibnamefont
  {Selmke}},\ }\bibfield  {title} {\bibinfo {title} {Renormalization of the
  dielectric response with applications to effective ion interactions and
  phonons},\ }\href {https://doi.org/10.1103/PhysRevB.24.586} {\bibfield
  {journal} {\bibinfo  {journal} {Phys. Rev. B}\ }\textbf {\bibinfo {volume}
  {24}},\ \bibinfo {pages} {586} (\bibinfo {year} {1981})}\BibitemShut
  {NoStop}%
\bibitem [{\citenamefont {Falter}(1988)}]{falter_unifying_1988}%
  \BibitemOpen
  \bibfield  {author} {\bibinfo {author} {\bibfnamefont {C.}~\bibnamefont
  {Falter}},\ }\bibfield  {title} {\bibinfo {title} {A unifying approach to
  lattice dynamical and electronic properties of solids},\ }\href
  {https://doi.org/10.1016/0370-1573(88)90058-0} {\bibfield  {journal}
  {\bibinfo  {journal} {Physics Reports}\ }\textbf {\bibinfo {volume} {164}},\
  \bibinfo {pages} {1} (\bibinfo {year} {1988})}\BibitemShut {NoStop}%
\bibitem [{\citenamefont {Nomura}\ and\ \citenamefont
  {Arita}(2015)}]{nomura_ab_2015}%
  \BibitemOpen
  \bibfield  {author} {\bibinfo {author} {\bibfnamefont {Y.}~\bibnamefont
  {Nomura}}\ and\ \bibinfo {author} {\bibfnamefont {R.}~\bibnamefont {Arita}},\
  }\bibfield  {title} {\bibinfo {title} {Ab initio downfolding for
  electron-phonon-coupled systems: Constrained density-functional perturbation
  theory},\ }\href {https://doi.org/10.1103/PhysRevB.92.245108} {\bibfield
  {journal} {\bibinfo  {journal} {Phys. Rev. B}\ }\textbf {\bibinfo {volume}
  {92}},\ \bibinfo {pages} {245108} (\bibinfo {year} {2015})}\BibitemShut
  {NoStop}%
\bibitem [{\citenamefont {Aryasetiawan}\ \emph {et~al.}(2004)\citenamefont
  {Aryasetiawan}, \citenamefont {Imada}, \citenamefont {Georges}, \citenamefont
  {Kotliar}, \citenamefont {Biermann},\ and\ \citenamefont
  {Lichtenstein}}]{Aryasetiawan2004}%
  \BibitemOpen
  \bibfield  {author} {\bibinfo {author} {\bibfnamefont {F.}~\bibnamefont
  {Aryasetiawan}}, \bibinfo {author} {\bibfnamefont {M.}~\bibnamefont {Imada}},
  \bibinfo {author} {\bibfnamefont {A.}~\bibnamefont {Georges}}, \bibinfo
  {author} {\bibfnamefont {G.}~\bibnamefont {Kotliar}}, \bibinfo {author}
  {\bibfnamefont {S.}~\bibnamefont {Biermann}},\ and\ \bibinfo {author}
  {\bibfnamefont {A.~I.}\ \bibnamefont {Lichtenstein}},\ }\bibfield  {title}
  {\bibinfo {title} {Frequency-dependent local interactions and low-energy
  effective models from electronic structure calculations},\ }\href
  {https://doi.org/10.1103/PhysRevB.70.195104} {\bibfield  {journal} {\bibinfo
  {journal} {Phys. Rev. B}\ }\textbf {\bibinfo {volume} {70}},\ \bibinfo
  {pages} {195104} (\bibinfo {year} {2004})}\BibitemShut {NoStop}%
\bibitem [{\citenamefont {McMahan}\ \emph
  {et~al.}(1988{\natexlab{a}})\citenamefont {McMahan}, \citenamefont {Martin},\
  and\ \citenamefont {Satpathy}}]{PhysRevB.38.6650}%
  \BibitemOpen
  \bibfield  {author} {\bibinfo {author} {\bibfnamefont {A.~K.}\ \bibnamefont
  {McMahan}}, \bibinfo {author} {\bibfnamefont {R.~M.}\ \bibnamefont
  {Martin}},\ and\ \bibinfo {author} {\bibfnamefont {S.}~\bibnamefont
  {Satpathy}},\ }\bibfield  {title} {\bibinfo {title} {Calculated effective
  {H}amiltonian for {${\mathrm{La}}_{2}$Cu${\mathrm{O}}_{4}$} and solution in
  the impurity {A}nderson approximation},\ }\href
  {https://doi.org/10.1103/PhysRevB.38.6650} {\bibfield  {journal} {\bibinfo
  {journal} {Phys. Rev. B}\ }\textbf {\bibinfo {volume} {38}},\ \bibinfo
  {pages} {6650} (\bibinfo {year} {1988}{\natexlab{a}})}\BibitemShut {NoStop}%
\bibitem [{\citenamefont {Hirayama}\ \emph {et~al.}(2013)\citenamefont
  {Hirayama}, \citenamefont {Miyake},\ and\ \citenamefont
  {Imada}}]{Hirayama13}%
  \BibitemOpen
  \bibfield  {author} {\bibinfo {author} {\bibfnamefont {M.}~\bibnamefont
  {Hirayama}}, \bibinfo {author} {\bibfnamefont {T.}~\bibnamefont {Miyake}},\
  and\ \bibinfo {author} {\bibfnamefont {M.}~\bibnamefont {Imada}},\ }\bibfield
   {title} {\bibinfo {title} {Derivation of static low-energy effective models
  by an ab initio downfolding method without double counting of coulomb
  correlations: Application to {SrVO}${}_{3}$, {FeSe}, and {FeTe}},\ }\href
  {https://doi.org/10.1103/PhysRevB.87.195144} {\bibfield  {journal} {\bibinfo
  {journal} {Phys. Rev. B}\ }\textbf {\bibinfo {volume} {87}},\ \bibinfo
  {pages} {195144} (\bibinfo {year} {2013})}\BibitemShut {NoStop}%
\bibitem [{\citenamefont {Honerkamp}(2012)}]{Honerkamp12}%
  \BibitemOpen
  \bibfield  {author} {\bibinfo {author} {\bibfnamefont {C.}~\bibnamefont
  {Honerkamp}},\ }\bibfield  {title} {\bibinfo {title} {Effective interactions
  in multiband systems from constrained summations},\ }\href
  {https://doi.org/10.1103/PhysRevB.85.195129} {\bibfield  {journal} {\bibinfo
  {journal} {Phys. Rev. B}\ }\textbf {\bibinfo {volume} {85}},\ \bibinfo
  {pages} {195129} (\bibinfo {year} {2012})}\BibitemShut {NoStop}%
\bibitem [{\citenamefont {Bartlett}\ and\ \citenamefont
  {Musia\l{}}(2007)}]{RevModPhys.79.291}%
  \BibitemOpen
  \bibfield  {author} {\bibinfo {author} {\bibfnamefont {R.~J.}\ \bibnamefont
  {Bartlett}}\ and\ \bibinfo {author} {\bibfnamefont {M.}~\bibnamefont
  {Musia\l{}}},\ }\bibfield  {title} {\bibinfo {title} {Coupled-cluster theory
  in quantum chemistry},\ }\href {https://doi.org/10.1103/RevModPhys.79.291}
  {\bibfield  {journal} {\bibinfo  {journal} {Rev. Mod. Phys.}\ }\textbf
  {\bibinfo {volume} {79}},\ \bibinfo {pages} {291} (\bibinfo {year}
  {2007})}\BibitemShut {NoStop}%
\bibitem [{\citenamefont {Lyakh}\ \emph {et~al.}(2012)\citenamefont {Lyakh},
  \citenamefont {Musiał}, \citenamefont {Lotrich},\ and\ \citenamefont
  {Bartlett}}]{lyakh_multireference_2012}%
  \BibitemOpen
  \bibfield  {author} {\bibinfo {author} {\bibfnamefont {D.~I.}\ \bibnamefont
  {Lyakh}}, \bibinfo {author} {\bibfnamefont {M.}~\bibnamefont {Musiał}},
  \bibinfo {author} {\bibfnamefont {V.~F.}\ \bibnamefont {Lotrich}},\ and\
  \bibinfo {author} {\bibfnamefont {R.~J.}\ \bibnamefont {Bartlett}},\
  }\bibfield  {title} {\bibinfo {title} {Multireference {Nature} of
  {Chemistry}: {The} {Coupled}-{Cluster} {View}},\ }\href
  {https://doi.org/10.1021/cr2001417} {\bibfield  {journal} {\bibinfo
  {journal} {Chem. Rev.}\ }\textbf {\bibinfo {volume} {112}},\ \bibinfo {pages}
  {182} (\bibinfo {year} {2012})}\BibitemShut {NoStop}%
\bibitem [{\citenamefont {Kowalski}\ and\ \citenamefont
  {Bauman}(2023)}]{PhysRevLett.131.200601}%
  \BibitemOpen
  \bibfield  {author} {\bibinfo {author} {\bibfnamefont {K.}~\bibnamefont
  {Kowalski}}\ and\ \bibinfo {author} {\bibfnamefont {N.~P.}\ \bibnamefont
  {Bauman}},\ }\bibfield  {title} {\bibinfo {title} {Quantum flow algorithms
  for simulating many-body systems on quantum computers},\ }\href
  {https://doi.org/10.1103/PhysRevLett.131.200601} {\bibfield  {journal}
  {\bibinfo  {journal} {Phys. Rev. Lett.}\ }\textbf {\bibinfo {volume} {131}},\
  \bibinfo {pages} {200601} (\bibinfo {year} {2023})}\BibitemShut {NoStop}%
\bibitem [{\citenamefont {Lichtenstein}\ and\ \citenamefont
  {Katsnelson}(1998)}]{lichtenstein_ab_1998}%
  \BibitemOpen
  \bibfield  {author} {\bibinfo {author} {\bibfnamefont {A.~I.}\ \bibnamefont
  {Lichtenstein}}\ and\ \bibinfo {author} {\bibfnamefont {M.~I.}\ \bibnamefont
  {Katsnelson}},\ }\bibfield  {title} {\bibinfo {title} {Ab initio calculations
  of quasiparticle band structure in correlated systems: {LDA}++ approach},\
  }\href {https://doi.org/10.1103/PhysRevB.57.6884} {\bibfield  {journal}
  {\bibinfo  {journal} {Phys. Rev. B}\ }\textbf {\bibinfo {volume} {57}},\
  \bibinfo {pages} {6884} (\bibinfo {year} {1998})}\BibitemShut {NoStop}%
\bibitem [{\citenamefont {Anisimov}\ \emph {et~al.}(1997)\citenamefont
  {Anisimov}, \citenamefont {Aryasetiawan},\ and\ \citenamefont
  {Lichtenstein}}]{anisimov_first-principles_1997}%
  \BibitemOpen
  \bibfield  {author} {\bibinfo {author} {\bibfnamefont {V.~I.}\ \bibnamefont
  {Anisimov}}, \bibinfo {author} {\bibfnamefont {F.}~\bibnamefont
  {Aryasetiawan}},\ and\ \bibinfo {author} {\bibfnamefont {A.~I.}\ \bibnamefont
  {Lichtenstein}},\ }\bibfield  {title} {\bibinfo {title} {First-principles
  calculations of the electronic structure and spectra of strongly correlated
  systems: the {LDA}+{U} method},\ }\href
  {https://doi.org/10.1088/0953-8984/9/4/002} {\bibfield  {journal} {\bibinfo
  {journal} {J. Phys.: Condens. Matter}\ }\textbf {\bibinfo {volume} {9}},\
  \bibinfo {pages} {767} (\bibinfo {year} {1997})}\BibitemShut {NoStop}%
\bibitem [{\citenamefont {Muechler}\ \emph {et~al.}(2022)\citenamefont
  {Muechler}, \citenamefont {Badrtdinov}, \citenamefont {Hampel}, \citenamefont
  {Cano}, \citenamefont {Rösner},\ and\ \citenamefont
  {Dreyer}}]{muechler_quantum_2022}%
  \BibitemOpen
  \bibfield  {author} {\bibinfo {author} {\bibfnamefont {L.}~\bibnamefont
  {Muechler}}, \bibinfo {author} {\bibfnamefont {D.~I.}\ \bibnamefont
  {Badrtdinov}}, \bibinfo {author} {\bibfnamefont {A.}~\bibnamefont {Hampel}},
  \bibinfo {author} {\bibfnamefont {J.}~\bibnamefont {Cano}}, \bibinfo {author}
  {\bibfnamefont {M.}~\bibnamefont {Rösner}},\ and\ \bibinfo {author}
  {\bibfnamefont {C.~E.}\ \bibnamefont {Dreyer}},\ }\bibfield  {title}
  {\bibinfo {title} {Quantum embedding methods for correlated excited states of
  point defects: {Case} studies and challenges},\ }\href
  {https://doi.org/10.1103/PhysRevB.105.235104} {\bibfield  {journal} {\bibinfo
   {journal} {Phys. Rev. B}\ }\textbf {\bibinfo {volume} {105}},\ \bibinfo
  {pages} {235104} (\bibinfo {year} {2022})}\BibitemShut {NoStop}%
\bibitem [{\citenamefont {Sheng}\ \emph {et~al.}(2022)\citenamefont {Sheng},
  \citenamefont {Vorwerk}, \citenamefont {Govoni},\ and\ \citenamefont
  {Galli}}]{sheng_greens_2022}%
  \BibitemOpen
  \bibfield  {author} {\bibinfo {author} {\bibfnamefont {N.}~\bibnamefont
  {Sheng}}, \bibinfo {author} {\bibfnamefont {C.}~\bibnamefont {Vorwerk}},
  \bibinfo {author} {\bibfnamefont {M.}~\bibnamefont {Govoni}},\ and\ \bibinfo
  {author} {\bibfnamefont {G.}~\bibnamefont {Galli}},\ }\bibfield  {title}
  {\bibinfo {title} {Green’s {F}unction {F}ormulation of {Q}uantum {D}efect
  {E}mbedding {T}heory},\ }\href {https://doi.org/10.1021/acs.jctc.2c00240}
  {\bibfield  {journal} {\bibinfo  {journal} {J. Chem. Theory Comput.}\
  }\textbf {\bibinfo {volume} {18}},\ \bibinfo {pages} {3512} (\bibinfo {year}
  {2022})}\BibitemShut {NoStop}%
\bibitem [{\citenamefont {Haule}(2015)}]{haule_exact_2015}%
  \BibitemOpen
  \bibfield  {author} {\bibinfo {author} {\bibfnamefont {K.}~\bibnamefont
  {Haule}},\ }\bibfield  {title} {\bibinfo {title} {Exact {D}ouble {C}ounting
  in {C}ombining the {D}ynamical {M}ean {F}ield {T}heory and the {D}ensity
  {F}unctional {T}heory},\ }\href
  {https://doi.org/10.1103/PhysRevLett.115.196403} {\bibfield  {journal}
  {\bibinfo  {journal} {Phys. Rev. Lett.}\ }\textbf {\bibinfo {volume} {115}},\
  \bibinfo {pages} {196403} (\bibinfo {year} {2015})}\BibitemShut {NoStop}%
\bibitem [{\citenamefont {Kristanovski}\ \emph {et~al.}(2018)\citenamefont
  {Kristanovski}, \citenamefont {Shick}, \citenamefont {Lechermann},\ and\
  \citenamefont {Lichtenstein}}]{kristanovski_role_2018}%
  \BibitemOpen
  \bibfield  {author} {\bibinfo {author} {\bibfnamefont {O.}~\bibnamefont
  {Kristanovski}}, \bibinfo {author} {\bibfnamefont {A.~B.}\ \bibnamefont
  {Shick}}, \bibinfo {author} {\bibfnamefont {F.}~\bibnamefont {Lechermann}},\
  and\ \bibinfo {author} {\bibfnamefont {A.~I.}\ \bibnamefont {Lichtenstein}},\
  }\bibfield  {title} {\bibinfo {title} {Role of nonspherical double counting
  in {DFT}+{DMFT}: {Total} energy and structural optimization of pnictide
  superconductors},\ }\href {https://doi.org/10.1103/PhysRevB.97.201116}
  {\bibfield  {journal} {\bibinfo  {journal} {Phys. Rev. B}\ }\textbf {\bibinfo
  {volume} {97}},\ \bibinfo {pages} {201116} (\bibinfo {year}
  {2018})}\BibitemShut {NoStop}%
\bibitem [{\citenamefont {Kotliar}\ \emph {et~al.}(2006)\citenamefont
  {Kotliar}, \citenamefont {Savrasov}, \citenamefont {Haule}, \citenamefont
  {Oudovenko}, \citenamefont {Parcollet},\ and\ \citenamefont
  {Marianetti}}]{kotliar_electronic_2006}%
  \BibitemOpen
  \bibfield  {author} {\bibinfo {author} {\bibfnamefont {G.}~\bibnamefont
  {Kotliar}}, \bibinfo {author} {\bibfnamefont {S.~Y.}\ \bibnamefont
  {Savrasov}}, \bibinfo {author} {\bibfnamefont {K.}~\bibnamefont {Haule}},
  \bibinfo {author} {\bibfnamefont {V.~S.}\ \bibnamefont {Oudovenko}}, \bibinfo
  {author} {\bibfnamefont {O.}~\bibnamefont {Parcollet}},\ and\ \bibinfo
  {author} {\bibfnamefont {C.~A.}\ \bibnamefont {Marianetti}},\ }\bibfield
  {title} {\bibinfo {title} {Electronic structure calculations with dynamical
  mean-field theory},\ }\href {https://doi.org/10.1103/RevModPhys.78.865}
  {\bibfield  {journal} {\bibinfo  {journal} {Rev. Mod. Phys.}\ }\textbf
  {\bibinfo {volume} {78}},\ \bibinfo {pages} {865} (\bibinfo {year}
  {2006})}\BibitemShut {NoStop}%
\bibitem [{\citenamefont {Dederichs}\ \emph {et~al.}(1984)\citenamefont
  {Dederichs}, \citenamefont {Bl\"ugel}, \citenamefont {Zeller},\ and\
  \citenamefont {Akai}}]{Dederichs1984}%
  \BibitemOpen
  \bibfield  {author} {\bibinfo {author} {\bibfnamefont {P.~H.}\ \bibnamefont
  {Dederichs}}, \bibinfo {author} {\bibfnamefont {S.}~\bibnamefont {Bl\"ugel}},
  \bibinfo {author} {\bibfnamefont {R.}~\bibnamefont {Zeller}},\ and\ \bibinfo
  {author} {\bibfnamefont {H.}~\bibnamefont {Akai}},\ }\bibfield  {title}
  {\bibinfo {title} {{Ground States of Constrained Systems: Application to
  Cerium Impurities}},\ }\href {https://doi.org/10.1103/PhysRevLett.53.2512}
  {\bibfield  {journal} {\bibinfo  {journal} {Phys. Rev. Lett.}\ }\textbf
  {\bibinfo {volume} {53}},\ \bibinfo {pages} {2512} (\bibinfo {year}
  {1984})}\BibitemShut {NoStop}%
\bibitem [{\citenamefont {McMahan}\ \emph
  {et~al.}(1988{\natexlab{b}})\citenamefont {McMahan}, \citenamefont {Martin},\
  and\ \citenamefont {Satpathy}}]{McMahan1988}%
  \BibitemOpen
  \bibfield  {author} {\bibinfo {author} {\bibfnamefont {A.~K.}\ \bibnamefont
  {McMahan}}, \bibinfo {author} {\bibfnamefont {R.~M.}\ \bibnamefont
  {Martin}},\ and\ \bibinfo {author} {\bibfnamefont {S.}~\bibnamefont
  {Satpathy}},\ }\bibfield  {title} {\bibinfo {title} {{Calculated effective
  Hamiltonian for ${\mathrm{La}}_{2}$Cu${\mathrm{O}}_{4}$ and solution in the
  impurity Anderson approximation}},\ }\href
  {https://doi.org/10.1103/PhysRevB.38.6650} {\bibfield  {journal} {\bibinfo
  {journal} {Phys. Rev. B}\ }\textbf {\bibinfo {volume} {38}},\ \bibinfo
  {pages} {6650} (\bibinfo {year} {1988}{\natexlab{b}})}\BibitemShut {NoStop}%
\bibitem [{\citenamefont {Zhang}\ and\ \citenamefont
  {Satpathy}(1991)}]{Zhang91}%
  \BibitemOpen
  \bibfield  {author} {\bibinfo {author} {\bibfnamefont {Z.}~\bibnamefont
  {Zhang}}\ and\ \bibinfo {author} {\bibfnamefont {S.}~\bibnamefont
  {Satpathy}},\ }\bibfield  {title} {\bibinfo {title} {Electron states,
  magnetism, and the {Verwey} transition in magnetite},\ }\href
  {https://doi.org/10.1103/PhysRevB.44.13319} {\bibfield  {journal} {\bibinfo
  {journal} {Phys. Rev. B}\ }\textbf {\bibinfo {volume} {44}},\ \bibinfo
  {pages} {13319} (\bibinfo {year} {1991})}\BibitemShut {NoStop}%
\bibitem [{\citenamefont {Koch}\ and\ \citenamefont
  {Jo/rgensen}(1990)}]{koch_coupled_1990}%
  \BibitemOpen
  \bibfield  {author} {\bibinfo {author} {\bibfnamefont {H.}~\bibnamefont
  {Koch}}\ and\ \bibinfo {author} {\bibfnamefont {P.}~\bibnamefont
  {Jo/rgensen}},\ }\bibfield  {title} {\bibinfo {title} {Coupled cluster
  response functions},\ }\href {https://doi.org/10.1063/1.458814} {\bibfield
  {journal} {\bibinfo  {journal} {J. Chem. Phys.}\ }\textbf {\bibinfo {volume}
  {93}},\ \bibinfo {pages} {3333} (\bibinfo {year} {1990})}\BibitemShut
  {NoStop}%
\bibitem [{\citenamefont {Stanton}\ and\ \citenamefont
  {Bartlett}(1993)}]{stanton_equation_1993}%
  \BibitemOpen
  \bibfield  {author} {\bibinfo {author} {\bibfnamefont {J.~F.}\ \bibnamefont
  {Stanton}}\ and\ \bibinfo {author} {\bibfnamefont {R.~J.}\ \bibnamefont
  {Bartlett}},\ }\bibfield  {title} {\bibinfo {title} {The equation of motion
  coupled‐cluster method. {A} systematic biorthogonal approach to molecular
  excitation energies, transition probabilities, and excited state
  properties},\ }\href {https://doi.org/10.1063/1.464746} {\bibfield  {journal}
  {\bibinfo  {journal} {J. Chem. Phys.}\ }\textbf {\bibinfo {volume} {98}},\
  \bibinfo {pages} {7029} (\bibinfo {year} {1993})}\BibitemShut {NoStop}%
\bibitem [{\citenamefont {Zhang}\ and\ \citenamefont
  {Krakauer}(2003)}]{ZhangKrakauer2003}%
  \BibitemOpen
  \bibfield  {author} {\bibinfo {author} {\bibfnamefont {S.}~\bibnamefont
  {Zhang}}\ and\ \bibinfo {author} {\bibfnamefont {H.}~\bibnamefont
  {Krakauer}},\ }\bibfield  {title} {\bibinfo {title} {Quantum {Monte} {Carlo}
  {Method} {Using} {Phase}-free {Random} {Walks} with {Slater}
  {Determinants}},\ }\href {https://doi.org/10.1103/PhysRevLett.90.136401}
  {\bibfield  {journal} {\bibinfo  {journal} {Phys. Rev. Lett.}\ }\textbf
  {\bibinfo {volume} {90}},\ \bibinfo {pages} {136401} (\bibinfo {year}
  {2003})}\BibitemShut {NoStop}%
\bibitem [{\citenamefont {Foulkes}\ \emph {et~al.}(2001)\citenamefont
  {Foulkes}, \citenamefont {Mitas}, \citenamefont {Needs},\ and\ \citenamefont
  {Rajagopal}}]{foulkes_quantum_2001}%
  \BibitemOpen
  \bibfield  {author} {\bibinfo {author} {\bibfnamefont {W.~M.~C.}\
  \bibnamefont {Foulkes}}, \bibinfo {author} {\bibfnamefont {L.}~\bibnamefont
  {Mitas}}, \bibinfo {author} {\bibfnamefont {R.~J.}\ \bibnamefont {Needs}},\
  and\ \bibinfo {author} {\bibfnamefont {G.}~\bibnamefont {Rajagopal}},\
  }\bibfield  {title} {\bibinfo {title} {Quantum {Monte} {Carlo} simulations of
  solids},\ }\href {https://doi.org/10.1103/RevModPhys.73.33} {\bibfield
  {journal} {\bibinfo  {journal} {Rev. Mod. Phys.}\ }\textbf {\bibinfo {volume}
  {73}},\ \bibinfo {pages} {33} (\bibinfo {year} {2001})}\BibitemShut {NoStop}%
\bibitem [{\citenamefont {Balabanov}\ and\ \citenamefont
  {Boggs}(2000)}]{balabanov_ab_2000}%
  \BibitemOpen
  \bibfield  {author} {\bibinfo {author} {\bibfnamefont {N.~B.}\ \bibnamefont
  {Balabanov}}\ and\ \bibinfo {author} {\bibfnamefont {J.~E.}\ \bibnamefont
  {Boggs}},\ }\bibfield  {title} {\bibinfo {title} {Ab {Initio} {Study} of
  {Molecular} and {Electronic} {Structures} of {Early} {Transition} {Metal}
  {Trihydrides} {MH}$_{\textrm{3}}$ ({M} = {Sc}, {Ti}, {V}, {Fe})},\ }\href
  {https://doi.org/10.1021/jp9933774} {\bibfield  {journal} {\bibinfo
  {journal} {J. Phys. Chem. A}\ }\textbf {\bibinfo {volume} {104}},\ \bibinfo
  {pages} {1597} (\bibinfo {year} {2000})}\BibitemShut {NoStop}%
\bibitem [{\citenamefont {Pathak}\ \emph {et~al.}(2021)\citenamefont {Pathak},
  \citenamefont {Busemeyer}, \citenamefont {Rodrigues},\ and\ \citenamefont
  {Wagner}}]{pathak_excited_2021}%
  \BibitemOpen
  \bibfield  {author} {\bibinfo {author} {\bibfnamefont {S.}~\bibnamefont
  {Pathak}}, \bibinfo {author} {\bibfnamefont {B.}~\bibnamefont {Busemeyer}},
  \bibinfo {author} {\bibfnamefont {J.~N.~B.}\ \bibnamefont {Rodrigues}},\ and\
  \bibinfo {author} {\bibfnamefont {L.~K.}\ \bibnamefont {Wagner}},\ }\bibfield
   {title} {\bibinfo {title} {Excited states in variational {Monte} {Carlo}
  using a penalty method},\ }\href {https://doi.org/10.1063/5.0030949}
  {\bibfield  {journal} {\bibinfo  {journal} {J. Chem. Phys.}\ }\textbf
  {\bibinfo {volume} {154}},\ \bibinfo {pages} {034101} (\bibinfo {year}
  {2021})}\BibitemShut {NoStop}%
\bibitem [{\citenamefont {Rudshteyn}\ \emph {et~al.}(2022)\citenamefont
  {Rudshteyn}, \citenamefont {Weber}, \citenamefont {Coskun}, \citenamefont
  {Devlaminck}, \citenamefont {Zhang}, \citenamefont {Reichman}, \citenamefont
  {Shee},\ and\ \citenamefont {Friesner}}]{Rudshteyn2020}%
  \BibitemOpen
  \bibfield  {author} {\bibinfo {author} {\bibfnamefont {B.}~\bibnamefont
  {Rudshteyn}}, \bibinfo {author} {\bibfnamefont {J.~L.}\ \bibnamefont
  {Weber}}, \bibinfo {author} {\bibfnamefont {D.}~\bibnamefont {Coskun}},
  \bibinfo {author} {\bibfnamefont {P.~A.}\ \bibnamefont {Devlaminck}},
  \bibinfo {author} {\bibfnamefont {S.}~\bibnamefont {Zhang}}, \bibinfo
  {author} {\bibfnamefont {D.~R.}\ \bibnamefont {Reichman}}, \bibinfo {author}
  {\bibfnamefont {J.}~\bibnamefont {Shee}},\ and\ \bibinfo {author}
  {\bibfnamefont {R.~A.}\ \bibnamefont {Friesner}},\ }\bibfield  {title}
  {\bibinfo {title} {{Calculation of Metallocene Ionization Potentials via
  Auxiliary Field Quantum Monte Carlo: Toward Benchmark Quantum Chemistry for
  Transition Metals}},\ }\href {https://doi.org/10.1021/acs.jctc.1c01071}
  {\bibfield  {journal} {\bibinfo  {journal} {J. Chem. Theory Comput.}\
  }\textbf {\bibinfo {volume} {18}},\ \bibinfo {pages} {2845} (\bibinfo {year}
  {2022})}\BibitemShut {NoStop}%
\bibitem [{\citenamefont {Shee}\ \emph {et~al.}(2023)\citenamefont {Shee},
  \citenamefont {Weber}, \citenamefont {Reichman}, \citenamefont {Friesner},\
  and\ \citenamefont {Zhang}}]{shee_potentially_2023}%
  \BibitemOpen
  \bibfield  {author} {\bibinfo {author} {\bibfnamefont {J.}~\bibnamefont
  {Shee}}, \bibinfo {author} {\bibfnamefont {J.~L.}\ \bibnamefont {Weber}},
  \bibinfo {author} {\bibfnamefont {D.~R.}\ \bibnamefont {Reichman}}, \bibinfo
  {author} {\bibfnamefont {R.~A.}\ \bibnamefont {Friesner}},\ and\ \bibinfo
  {author} {\bibfnamefont {S.}~\bibnamefont {Zhang}},\ }\bibfield  {title}
  {\bibinfo {title} {On the potentially transformative role of auxiliary-field
  quantum {Monte} {Carlo} in quantum chemistry: {A} highly accurate method for
  transition metals and beyond},\ }\href {https://doi.org/10.1063/5.0134009}
  {\bibfield  {journal} {\bibinfo  {journal} {J. Chem. Phys.}\ }\textbf
  {\bibinfo {volume} {158}},\ \bibinfo {pages} {140901} (\bibinfo {year}
  {2023})}\BibitemShut {NoStop}%
\bibitem [{\citenamefont {Prins}\ and\ \citenamefont
  {Voorst}(1968{\natexlab{a}})}]{prins_bonding_1968}%
  \BibitemOpen
  \bibfield  {author} {\bibinfo {author} {\bibfnamefont {R.}~\bibnamefont
  {Prins}}\ and\ \bibinfo {author} {\bibfnamefont {J.~D. W.~V.}\ \bibnamefont
  {Voorst}},\ }\bibfield  {title} {\bibinfo {title} {Bonding in {Sandwich}
  {Compounds}},\ }\href {https://doi.org/10.1063/1.1669928} {\bibfield
  {journal} {\bibinfo  {journal} {J. Chem. Phys.}\ }\textbf {\bibinfo {volume}
  {49}},\ \bibinfo {pages} {4665} (\bibinfo {year}
  {1968}{\natexlab{a}})}\BibitemShut {NoStop}%
\bibitem [{\citenamefont {Xu}\ \emph {et~al.}(2003)\citenamefont {Xu},
  \citenamefont {Xie}, \citenamefont {Feng},\ and\ \citenamefont
  {Schaefer}}]{xu2003}%
  \BibitemOpen
  \bibfield  {author} {\bibinfo {author} {\bibfnamefont {Z.-F.}\ \bibnamefont
  {Xu}}, \bibinfo {author} {\bibfnamefont {Y.}~\bibnamefont {Xie}}, \bibinfo
  {author} {\bibfnamefont {W.-L.}\ \bibnamefont {Feng}},\ and\ \bibinfo
  {author} {\bibfnamefont {H.~F.}\ \bibnamefont {Schaefer}},\ }\bibfield
  {title} {\bibinfo {title} {Systematic {Investigation} of {Electronic} and
  {Molecular} {Structures} for the {First} {Transition} {Metal} {Series}
  {Metallocenes} {M(C$_5$H$_5$)$_2$} ({M = V, Cr, Mn, Fe, Co, and Ni})},\
  }\href {https://doi.org/10.1021/jp0219855} {\bibfield  {journal} {\bibinfo
  {journal} {J. Phys. Chem. A}\ }\textbf {\bibinfo {volume} {107}},\ \bibinfo
  {pages} {2716} (\bibinfo {year} {2003})}\BibitemShut {NoStop}%
\bibitem [{\citenamefont {Gard}\ \emph {et~al.}(1975)\citenamefont {Gard},
  \citenamefont {Haaland}, \citenamefont {Novak},\ and\ \citenamefont
  {Seip}}]{GARD1975181}%
  \BibitemOpen
  \bibfield  {author} {\bibinfo {author} {\bibfnamefont {E.}~\bibnamefont
  {Gard}}, \bibinfo {author} {\bibfnamefont {A.}~\bibnamefont {Haaland}},
  \bibinfo {author} {\bibfnamefont {D.~P.}\ \bibnamefont {Novak}},\ and\
  \bibinfo {author} {\bibfnamefont {R.}~\bibnamefont {Seip}},\ }\bibfield
  {title} {\bibinfo {title} {The molecular structures of
  dicyclopentadienylvanadium, {(C$_5$H$_5$)$_2$V}, and
  dicyclopentadienylchromium, {(C$_5$H$_5$)$_2$Cr}, determined by gas phase
  electron diffraction},\ }\href
  {https://doi.org/https://doi.org/10.1016/S0022-328X(00)91459-1} {\bibfield
  {journal} {\bibinfo  {journal} {J. Organomet. Chem}\ }\textbf {\bibinfo
  {volume} {88}},\ \bibinfo {pages} {181} (\bibinfo {year} {1975})}\BibitemShut
  {NoStop}%
\bibitem [{\citenamefont {van Loon}\ \emph {et~al.}(2021)\citenamefont {van
  Loon}, \citenamefont {R\"osner}, \citenamefont {Katsnelson},\ and\
  \citenamefont {Wehling}}]{PhysRevB.104.045134}%
  \BibitemOpen
  \bibfield  {author} {\bibinfo {author} {\bibfnamefont {E.~G. C.~P.}\
  \bibnamefont {van Loon}}, \bibinfo {author} {\bibfnamefont {M.}~\bibnamefont
  {R\"osner}}, \bibinfo {author} {\bibfnamefont {M.~I.}\ \bibnamefont
  {Katsnelson}},\ and\ \bibinfo {author} {\bibfnamefont {T.~O.}\ \bibnamefont
  {Wehling}},\ }\bibfield  {title} {\bibinfo {title} {Random phase
  approximation for gapped systems: {Role} of vertex corrections and
  applicability of the constrained random phase approximation},\ }\href
  {https://doi.org/10.1103/PhysRevB.104.045134} {\bibfield  {journal} {\bibinfo
   {journal} {Phys. Rev. B}\ }\textbf {\bibinfo {volume} {104}},\ \bibinfo
  {pages} {045134} (\bibinfo {year} {2021})}\BibitemShut {NoStop}%
\bibitem [{\citenamefont {Annaberdiyev}\ \emph {et~al.}(2018)\citenamefont
  {Annaberdiyev}, \citenamefont {Wang}, \citenamefont {Melton}, \citenamefont
  {Chandler~Bennett}, \citenamefont {Shulenburger},\ and\ \citenamefont
  {Mitas}}]{annaberdiyev_new_2018}%
  \BibitemOpen
  \bibfield  {author} {\bibinfo {author} {\bibfnamefont {A.}~\bibnamefont
  {Annaberdiyev}}, \bibinfo {author} {\bibfnamefont {G.}~\bibnamefont {Wang}},
  \bibinfo {author} {\bibfnamefont {C.~A.}\ \bibnamefont {Melton}}, \bibinfo
  {author} {\bibfnamefont {M.}~\bibnamefont {Chandler~Bennett}}, \bibinfo
  {author} {\bibfnamefont {L.}~\bibnamefont {Shulenburger}},\ and\ \bibinfo
  {author} {\bibfnamefont {L.}~\bibnamefont {Mitas}},\ }\bibfield  {title}
  {\bibinfo {title} {A new generation of effective core potentials from
  correlated calculations: 3d transition metal series},\ }\href
  {https://doi.org/10.1063/1.5040472} {\bibfield  {journal} {\bibinfo
  {journal} {J. Chem. Phys.}\ }\textbf {\bibinfo {volume} {149}},\ \bibinfo
  {pages} {134108} (\bibinfo {year} {2018})}\BibitemShut {NoStop}%
\bibitem [{\citenamefont {Aryasetiawan}\ \emph {et~al.}(2006)\citenamefont
  {Aryasetiawan}, \citenamefont {Karlsson}, \citenamefont {Jepsen},\ and\
  \citenamefont {Sch\"onberger}}]{Aryasetiawan2006}%
  \BibitemOpen
  \bibfield  {author} {\bibinfo {author} {\bibfnamefont {F.}~\bibnamefont
  {Aryasetiawan}}, \bibinfo {author} {\bibfnamefont {K.}~\bibnamefont
  {Karlsson}}, \bibinfo {author} {\bibfnamefont {O.}~\bibnamefont {Jepsen}},\
  and\ \bibinfo {author} {\bibfnamefont {U.}~\bibnamefont {Sch\"onberger}},\
  }\bibfield  {title} {\bibinfo {title} {Calculations of {Hubbard} {$U$} from
  first-principles},\ }\href {https://doi.org/10.1103/PhysRevB.74.125106}
  {\bibfield  {journal} {\bibinfo  {journal} {Phys. Rev. B}\ }\textbf {\bibinfo
  {volume} {74}},\ \bibinfo {pages} {125106} (\bibinfo {year}
  {2006})}\BibitemShut {NoStop}%
\bibitem [{\citenamefont {Nomura}\ \emph {et~al.}(2012)\citenamefont {Nomura},
  \citenamefont {Kaltak}, \citenamefont {Nakamura}, \citenamefont {Taranto},
  \citenamefont {Sakai}, \citenamefont {Toschi}, \citenamefont {Arita},
  \citenamefont {Held}, \citenamefont {Kresse},\ and\ \citenamefont
  {Imada}}]{Nomura12}%
  \BibitemOpen
  \bibfield  {author} {\bibinfo {author} {\bibfnamefont {Y.}~\bibnamefont
  {Nomura}}, \bibinfo {author} {\bibfnamefont {M.}~\bibnamefont {Kaltak}},
  \bibinfo {author} {\bibfnamefont {K.}~\bibnamefont {Nakamura}}, \bibinfo
  {author} {\bibfnamefont {C.}~\bibnamefont {Taranto}}, \bibinfo {author}
  {\bibfnamefont {S.}~\bibnamefont {Sakai}}, \bibinfo {author} {\bibfnamefont
  {A.}~\bibnamefont {Toschi}}, \bibinfo {author} {\bibfnamefont
  {R.}~\bibnamefont {Arita}}, \bibinfo {author} {\bibfnamefont
  {K.}~\bibnamefont {Held}}, \bibinfo {author} {\bibfnamefont {G.}~\bibnamefont
  {Kresse}},\ and\ \bibinfo {author} {\bibfnamefont {M.}~\bibnamefont
  {Imada}},\ }\bibfield  {title} {\bibinfo {title} {Effective on-site
  interaction for dynamical mean-field theory},\ }\href
  {https://doi.org/10.1103/PhysRevB.86.085117} {\bibfield  {journal} {\bibinfo
  {journal} {Phys. Rev. B}\ }\textbf {\bibinfo {volume} {86}},\ \bibinfo
  {pages} {085117} (\bibinfo {year} {2012})}\BibitemShut {NoStop}%
\bibitem [{\citenamefont {Martins}\ \emph {et~al.}(2011)\citenamefont
  {Martins}, \citenamefont {Aichhorn}, \citenamefont {Vaugier},\ and\
  \citenamefont {Biermann}}]{Biermann11}%
  \BibitemOpen
  \bibfield  {author} {\bibinfo {author} {\bibfnamefont {C.}~\bibnamefont
  {Martins}}, \bibinfo {author} {\bibfnamefont {M.}~\bibnamefont {Aichhorn}},
  \bibinfo {author} {\bibfnamefont {L.}~\bibnamefont {Vaugier}},\ and\ \bibinfo
  {author} {\bibfnamefont {S.}~\bibnamefont {Biermann}},\ }\bibfield  {title}
  {\bibinfo {title} {Reduced {E}ffective {S}pin-{O}rbital {D}egeneracy and
  {S}pin-{O}rbital {O}rdering in {P}aramagnetic {T}ransition-{M}etal {O}xides:
  {Sr}$_2${IrO}$_4$ versus {Sr}$_2${Rh}{O}$_4$},\ }\href
  {https://doi.org/10.1103/PhysRevLett.107.266404} {\bibfield  {journal}
  {\bibinfo  {journal} {Phys. Rev. Lett.}\ }\textbf {\bibinfo {volume} {107}},\
  \bibinfo {pages} {266404} (\bibinfo {year} {2011})}\BibitemShut {NoStop}%
\bibitem [{\citenamefont {Werner}\ \emph {et~al.}(2015)\citenamefont {Werner},
  \citenamefont {Sakuma}, \citenamefont {Nilsson},\ and\ \citenamefont
  {Aryasetiawan}}]{Werner15}%
  \BibitemOpen
  \bibfield  {author} {\bibinfo {author} {\bibfnamefont {P.}~\bibnamefont
  {Werner}}, \bibinfo {author} {\bibfnamefont {R.}~\bibnamefont {Sakuma}},
  \bibinfo {author} {\bibfnamefont {F.}~\bibnamefont {Nilsson}},\ and\ \bibinfo
  {author} {\bibfnamefont {F.}~\bibnamefont {Aryasetiawan}},\ }\bibfield
  {title} {\bibinfo {title} {Dynamical screening in
  {${\text{La}}_{2}{\text{CuO}}_{4}$}},\ }\href
  {https://doi.org/10.1103/PhysRevB.91.125142} {\bibfield  {journal} {\bibinfo
  {journal} {Phys. Rev. B}\ }\textbf {\bibinfo {volume} {91}},\ \bibinfo
  {pages} {125142} (\bibinfo {year} {2015})}\BibitemShut {NoStop}%
\bibitem [{\citenamefont {Prins}\ and\ \citenamefont
  {Voorst}(1968{\natexlab{b}})}]{prins_bonding_2003}%
  \BibitemOpen
  \bibfield  {author} {\bibinfo {author} {\bibfnamefont {R.}~\bibnamefont
  {Prins}}\ and\ \bibinfo {author} {\bibfnamefont {J.~D. W.~v.}\ \bibnamefont
  {Voorst}},\ }\bibfield  {title} {\bibinfo {title} {Bonding in {S}andwich
  {C}ompounds},\ }\href {https://doi.org/10.1063/1.1669928} {\bibfield
  {journal} {\bibinfo  {journal} {J. Chem. Phys.}\ }\textbf {\bibinfo {volume}
  {49}},\ \bibinfo {pages} {4665} (\bibinfo {year}
  {1968}{\natexlab{b}})}\BibitemShut {NoStop}%
\bibitem [{\citenamefont {Jackson}\ \emph {et~al.}(2012)\citenamefont
  {Jackson}, \citenamefont {Krzystek}, \citenamefont {Ozarowski}, \citenamefont
  {Wijeratne}, \citenamefont {Wicker}, \citenamefont {Mindiola},\ and\
  \citenamefont {Telser}}]{jackson_vanadocene_2012}%
  \BibitemOpen
  \bibfield  {author} {\bibinfo {author} {\bibfnamefont {T.~A.}\ \bibnamefont
  {Jackson}}, \bibinfo {author} {\bibfnamefont {J.}~\bibnamefont {Krzystek}},
  \bibinfo {author} {\bibfnamefont {A.}~\bibnamefont {Ozarowski}}, \bibinfo
  {author} {\bibfnamefont {G.~B.}\ \bibnamefont {Wijeratne}}, \bibinfo {author}
  {\bibfnamefont {B.~F.}\ \bibnamefont {Wicker}}, \bibinfo {author}
  {\bibfnamefont {D.~J.}\ \bibnamefont {Mindiola}},\ and\ \bibinfo {author}
  {\bibfnamefont {J.}~\bibnamefont {Telser}},\ }\bibfield  {title} {\bibinfo
  {title} {Vanadocene de {N}ovo: {S}pectroscopic and {C}omputational {A}nalysis
  of {B}is($\eta^5$-cyclopentadienyl)vanadium({II})},\ }\href
  {https://doi.org/10.1021/om300892y} {\bibfield  {journal} {\bibinfo
  {journal} {Organometallics}\ }\textbf {\bibinfo {volume} {31}},\ \bibinfo
  {pages} {8265} (\bibinfo {year} {2012})}\BibitemShut {NoStop}%
\bibitem [{\citenamefont {Phung}\ \emph {et~al.}(2012)\citenamefont {Phung},
  \citenamefont {Vancoillie},\ and\ \citenamefont {Pierloot}}]{Phung2012}%
  \BibitemOpen
  \bibfield  {author} {\bibinfo {author} {\bibfnamefont {Q.~M.}\ \bibnamefont
  {Phung}}, \bibinfo {author} {\bibfnamefont {S.}~\bibnamefont {Vancoillie}},\
  and\ \bibinfo {author} {\bibfnamefont {K.}~\bibnamefont {Pierloot}},\
  }\bibfield  {title} {\bibinfo {title} {A {Multiconfigurational}
  {Perturbation} {Theory} and {Density} {Functional} {Theory} {Study} on the
  {Heterolytic} {Dissociation} {Enthalpy} of {First-Row} {Metallocenes}},\
  }\href {https://doi.org/10.1021/ct200875m} {\bibfield  {journal} {\bibinfo
  {journal} {J. Chem. Theory Comput.}\ }\textbf {\bibinfo {volume} {8}},\
  \bibinfo {pages} {883} (\bibinfo {year} {2012})}\BibitemShut {NoStop}%
\bibitem [{\citenamefont {Nain}\ \emph {et~al.}(2022)\citenamefont {Nain},
  \citenamefont {Khurana},\ and\ \citenamefont {Ali}}]{nain_harnessing_2022}%
  \BibitemOpen
  \bibfield  {author} {\bibinfo {author} {\bibfnamefont {S.}~\bibnamefont
  {Nain}}, \bibinfo {author} {\bibfnamefont {R.}~\bibnamefont {Khurana}},\ and\
  \bibinfo {author} {\bibfnamefont {M.~E.}\ \bibnamefont {Ali}},\ }\bibfield
  {title} {\bibinfo {title} {Harnessing {C}olossal {M}agnetic {A}nisotropy in
  {S}andwiched 3d$^2$-{M}etallocenes},\ }\href
  {https://doi.org/10.1021/acs.jpca.2c01605} {\bibfield  {journal} {\bibinfo
  {journal} {J. Phys. Chem. A}\ }\textbf {\bibinfo {volume} {126}},\ \bibinfo
  {pages} {2811} (\bibinfo {year} {2022})}\BibitemShut {NoStop}%
\bibitem [{\citenamefont {Marzari}\ and\ \citenamefont
  {Vanderbilt}(1997)}]{Marzari97}%
  \BibitemOpen
  \bibfield  {author} {\bibinfo {author} {\bibfnamefont {N.}~\bibnamefont
  {Marzari}}\ and\ \bibinfo {author} {\bibfnamefont {D.}~\bibnamefont
  {Vanderbilt}},\ }\bibfield  {title} {\bibinfo {title} {Maximally localized
  generalized {Wannier} functions for composite energy bands},\ }\href
  {https://doi.org/10.1103/PhysRevB.56.12847} {\bibfield  {journal} {\bibinfo
  {journal} {Phys. Rev. B}\ }\textbf {\bibinfo {volume} {56}},\ \bibinfo
  {pages} {12847} (\bibinfo {year} {1997})}\BibitemShut {NoStop}%
\bibitem [{\citenamefont {Amadon}\ \emph {et~al.}(2008)\citenamefont {Amadon},
  \citenamefont {Lechermann}, \citenamefont {Georges}, \citenamefont {Jollet},
  \citenamefont {Wehling},\ and\ \citenamefont
  {Lichtenstein}}]{amadon_plane-wave_2008}%
  \BibitemOpen
  \bibfield  {author} {\bibinfo {author} {\bibfnamefont {B.}~\bibnamefont
  {Amadon}}, \bibinfo {author} {\bibfnamefont {F.}~\bibnamefont {Lechermann}},
  \bibinfo {author} {\bibfnamefont {A.}~\bibnamefont {Georges}}, \bibinfo
  {author} {\bibfnamefont {F.}~\bibnamefont {Jollet}}, \bibinfo {author}
  {\bibfnamefont {T.~O.}\ \bibnamefont {Wehling}},\ and\ \bibinfo {author}
  {\bibfnamefont {A.~I.}\ \bibnamefont {Lichtenstein}},\ }\bibfield  {title}
  {\bibinfo {title} {Plane-wave based electronic structure calculations for
  correlated materials using dynamical mean-field theory and projected local
  orbitals},\ }\href {https://doi.org/10.1103/PhysRevB.77.205112} {\bibfield
  {journal} {\bibinfo  {journal} {Phys. Rev. B}\ }\textbf {\bibinfo {volume}
  {77}},\ \bibinfo {pages} {205112} (\bibinfo {year} {2008})}\BibitemShut
  {NoStop}%
\bibitem [{\citenamefont {Haule}\ \emph {et~al.}(2010)\citenamefont {Haule},
  \citenamefont {Yee},\ and\ \citenamefont {Kim}}]{haule_dynamical_2010}%
  \BibitemOpen
  \bibfield  {author} {\bibinfo {author} {\bibfnamefont {K.}~\bibnamefont
  {Haule}}, \bibinfo {author} {\bibfnamefont {C.-H.}\ \bibnamefont {Yee}},\
  and\ \bibinfo {author} {\bibfnamefont {K.}~\bibnamefont {Kim}},\ }\bibfield
  {title} {\bibinfo {title} {Dynamical mean-field theory within the
  full-potential methods: {E}lectronic structure of $\mathrm{CeIrIn}_5$,
  $\mathrm{CeCoIn}_5$, and $\mathrm{CeRhIn}_5$},\ }\href
  {https://doi.org/10.1103/PhysRevB.81.195107} {\bibfield  {journal} {\bibinfo
  {journal} {Phys. Rev. B}\ }\textbf {\bibinfo {volume} {81}},\ \bibinfo
  {pages} {195107} (\bibinfo {year} {2010})}\BibitemShut {NoStop}%
\bibitem [{\citenamefont {Karolak}\ \emph {et~al.}(2011)\citenamefont
  {Karolak}, \citenamefont {Wehling}, \citenamefont {Lechermann},\ and\
  \citenamefont {Lichtenstein}}]{karolak_general_2011}%
  \BibitemOpen
  \bibfield  {author} {\bibinfo {author} {\bibfnamefont {M.}~\bibnamefont
  {Karolak}}, \bibinfo {author} {\bibfnamefont {T.~O.}\ \bibnamefont
  {Wehling}}, \bibinfo {author} {\bibfnamefont {F.}~\bibnamefont
  {Lechermann}},\ and\ \bibinfo {author} {\bibfnamefont {A.~I.}\ \bibnamefont
  {Lichtenstein}},\ }\bibfield  {title} {\bibinfo {title} {General {DFT++}
  method implemented with projector augmented waves: electronic structure of
  {SrVO}$_3$ and the mott transition in {Ca}$_{2 -x}${Sr}$_x${RuO}$_4$},\
  }\href {https://doi.org/10.1088/0953-8984/23/8/085601} {\bibfield  {journal}
  {\bibinfo  {journal} {J. Phys.: Condens. Matter}\ }\textbf {\bibinfo {volume}
  {23}},\ \bibinfo {pages} {085601} (\bibinfo {year} {2011})}\BibitemShut
  {NoStop}%
\bibitem [{\citenamefont {Schüler}\ \emph {et~al.}(2018)\citenamefont
  {Schüler}, \citenamefont {Peil}, \citenamefont {Kraberger}, \citenamefont
  {Pordzik}, \citenamefont {Marsman}, \citenamefont {Kresse}, \citenamefont
  {Wehling},\ and\ \citenamefont {Aichhorn}}]{schuler_charge_2018}%
  \BibitemOpen
  \bibfield  {author} {\bibinfo {author} {\bibfnamefont {M.}~\bibnamefont
  {Schüler}}, \bibinfo {author} {\bibfnamefont {O.~E.}\ \bibnamefont {Peil}},
  \bibinfo {author} {\bibfnamefont {G.~J.}\ \bibnamefont {Kraberger}}, \bibinfo
  {author} {\bibfnamefont {R.}~\bibnamefont {Pordzik}}, \bibinfo {author}
  {\bibfnamefont {M.}~\bibnamefont {Marsman}}, \bibinfo {author} {\bibfnamefont
  {G.}~\bibnamefont {Kresse}}, \bibinfo {author} {\bibfnamefont {T.~O.}\
  \bibnamefont {Wehling}},\ and\ \bibinfo {author} {\bibfnamefont
  {M.}~\bibnamefont {Aichhorn}},\ }\bibfield  {title} {\bibinfo {title} {Charge
  self-consistent many-body corrections using optimized projected localized
  orbitals},\ }\href {https://doi.org/10.1088/1361-648X/aae80a} {\bibfield
  {journal} {\bibinfo  {journal} {J. Phys.: Condens. Matter}\ }\textbf
  {\bibinfo {volume} {30}},\ \bibinfo {pages} {475901} (\bibinfo {year}
  {2018})}\BibitemShut {NoStop}%
\bibitem [{\citenamefont {Perdew}\ and\ \citenamefont {Levy}(1983)}]{Perdew83}%
  \BibitemOpen
  \bibfield  {author} {\bibinfo {author} {\bibfnamefont {J.~P.}\ \bibnamefont
  {Perdew}}\ and\ \bibinfo {author} {\bibfnamefont {M.}~\bibnamefont {Levy}},\
  }\bibfield  {title} {\bibinfo {title} {Physical {Content} of the {Exact}
  {Kohn-Sham} {Orbital} {Energies}: {Band} {Gaps} and {Derivative}
  {Discontinuities}},\ }\href {https://doi.org/10.1103/PhysRevLett.51.1884}
  {\bibfield  {journal} {\bibinfo  {journal} {Phys. Rev. Lett.}\ }\textbf
  {\bibinfo {volume} {51}},\ \bibinfo {pages} {1884} (\bibinfo {year}
  {1983})}\BibitemShut {NoStop}%
\bibitem [{\citenamefont {Sham}\ and\ \citenamefont
  {Schl\"uter}(1985)}]{Sham85}%
  \BibitemOpen
  \bibfield  {author} {\bibinfo {author} {\bibfnamefont {L.~J.}\ \bibnamefont
  {Sham}}\ and\ \bibinfo {author} {\bibfnamefont {M.}~\bibnamefont
  {Schl\"uter}},\ }\bibfield  {title} {\bibinfo {title} {Density-functional
  theory of the band gap},\ }\href {https://doi.org/10.1103/PhysRevB.32.3883}
  {\bibfield  {journal} {\bibinfo  {journal} {Phys. Rev. B}\ }\textbf {\bibinfo
  {volume} {32}},\ \bibinfo {pages} {3883} (\bibinfo {year}
  {1985})}\BibitemShut {NoStop}%
\bibitem [{\citenamefont {Bockstedte}\ \emph {et~al.}(2018)\citenamefont
  {Bockstedte}, \citenamefont {Sch{\"u}tz}, \citenamefont {Garratt},
  \citenamefont {Ivády},\ and\ \citenamefont {Gali}}]{bockstedte_ab_2018}%
  \BibitemOpen
  \bibfield  {author} {\bibinfo {author} {\bibfnamefont {M.}~\bibnamefont
  {Bockstedte}}, \bibinfo {author} {\bibfnamefont {F.}~\bibnamefont
  {Sch{\"u}tz}}, \bibinfo {author} {\bibfnamefont {T.}~\bibnamefont {Garratt}},
  \bibinfo {author} {\bibfnamefont {V.}~\bibnamefont {Ivády}},\ and\ \bibinfo
  {author} {\bibfnamefont {A.}~\bibnamefont {Gali}},\ }\bibfield  {title}
  {\bibinfo {title} {Ab initio description of highly correlated states in
  defects for realizing quantum bits},\ }\href
  {https://doi.org/10.1038/s41535-018-0103-6} {\bibfield  {journal} {\bibinfo
  {journal} {npj Quant. Mater.}\ }\textbf {\bibinfo {volume} {3}},\ \bibinfo
  {pages} {1} (\bibinfo {year} {2018})}\BibitemShut {NoStop}%
\bibitem [{\citenamefont {Ma}\ \emph {et~al.}(2020)\citenamefont {Ma},
  \citenamefont {Govoni},\ and\ \citenamefont {Galli}}]{ma_quantum_2020}%
  \BibitemOpen
  \bibfield  {author} {\bibinfo {author} {\bibfnamefont {H.}~\bibnamefont
  {Ma}}, \bibinfo {author} {\bibfnamefont {M.}~\bibnamefont {Govoni}},\ and\
  \bibinfo {author} {\bibfnamefont {G.}~\bibnamefont {Galli}},\ }\bibfield
  {title} {\bibinfo {title} {Quantum simulations of materials on near-term
  quantum computers},\ }\href {https://doi.org/10.1038/s41524-020-00353-z}
  {\bibfield  {journal} {\bibinfo  {journal} {npj Comput. Mater.}\ }\textbf
  {\bibinfo {volume} {6}},\ \bibinfo {pages} {1} (\bibinfo {year}
  {2020})}\BibitemShut {NoStop}%
\bibitem [{\citenamefont {Solovyev}\ \emph {et~al.}(1994)\citenamefont
  {Solovyev}, \citenamefont {Dederichs},\ and\ \citenamefont
  {Anisimov}}]{solovyev_corrected_1994}%
  \BibitemOpen
  \bibfield  {author} {\bibinfo {author} {\bibfnamefont {I.~V.}\ \bibnamefont
  {Solovyev}}, \bibinfo {author} {\bibfnamefont {P.~H.}\ \bibnamefont
  {Dederichs}},\ and\ \bibinfo {author} {\bibfnamefont {V.~I.}\ \bibnamefont
  {Anisimov}},\ }\bibfield  {title} {\bibinfo {title} {Corrected atomic limit
  in the local-density approximation and the electronic structure of $d$
  impurities in {Rb}},\ }\href {https://doi.org/10.1103/PhysRevB.50.16861}
  {\bibfield  {journal} {\bibinfo  {journal} {Phys. Rev. B}\ }\textbf {\bibinfo
  {volume} {50}},\ \bibinfo {pages} {16861} (\bibinfo {year}
  {1994})}\BibitemShut {NoStop}%
\bibitem [{\citenamefont {Czy\ifmmode~\dot{z}\else \.{z}\fi{}yk}\ and\
  \citenamefont {Sawatzky}(1994)}]{czyzyk_local-density_1994}%
  \BibitemOpen
  \bibfield  {author} {\bibinfo {author} {\bibfnamefont {M.~T.}\ \bibnamefont
  {Czy\ifmmode~\dot{z}\else \.{z}\fi{}yk}}\ and\ \bibinfo {author}
  {\bibfnamefont {G.~A.}\ \bibnamefont {Sawatzky}},\ }\bibfield  {title}
  {\bibinfo {title} {Local-density functional and on-site correlations: The
  electronic structure of {${\mathrm{La}}_{2}$${\mathrm{CuO}}_{4}$} and
  {${\mathrm{LaCuO}}_{3}$}},\ }\href
  {https://doi.org/10.1103/PhysRevB.49.14211} {\bibfield  {journal} {\bibinfo
  {journal} {Phys. Rev. B}\ }\textbf {\bibinfo {volume} {49}},\ \bibinfo
  {pages} {14211} (\bibinfo {year} {1994})}\BibitemShut {NoStop}%
\bibitem [{\citenamefont {Anisimov}\ \emph {et~al.}(1991)\citenamefont
  {Anisimov}, \citenamefont {Zaanen},\ and\ \citenamefont
  {Andersen}}]{anisimov_band_1991}%
  \BibitemOpen
  \bibfield  {author} {\bibinfo {author} {\bibfnamefont {V.~I.}\ \bibnamefont
  {Anisimov}}, \bibinfo {author} {\bibfnamefont {J.}~\bibnamefont {Zaanen}},\
  and\ \bibinfo {author} {\bibfnamefont {O.~K.}\ \bibnamefont {Andersen}},\
  }\bibfield  {title} {\bibinfo {title} {Band theory and {M}ott insulators:
  {H}ubbard {$U$} instead of {S}toner {$I$}},\ }\href
  {https://doi.org/10.1103/PhysRevB.44.943} {\bibfield  {journal} {\bibinfo
  {journal} {Phys. Rev. B}\ }\textbf {\bibinfo {volume} {44}},\ \bibinfo
  {pages} {943} (\bibinfo {year} {1991})}\BibitemShut {NoStop}%
\bibitem [{\citenamefont {Liechtenstein}\ \emph {et~al.}(1995)\citenamefont
  {Liechtenstein}, \citenamefont {Anisimov},\ and\ \citenamefont
  {Zaanen}}]{liechtenstein_density-functional_1995}%
  \BibitemOpen
  \bibfield  {author} {\bibinfo {author} {\bibfnamefont {A.~I.}\ \bibnamefont
  {Liechtenstein}}, \bibinfo {author} {\bibfnamefont {V.~I.}\ \bibnamefont
  {Anisimov}},\ and\ \bibinfo {author} {\bibfnamefont {J.}~\bibnamefont
  {Zaanen}},\ }\bibfield  {title} {\bibinfo {title} {Density-functional theory
  and strong interactions: {O}rbital ordering in {M}ott-{H}ubbard insulators},\
  }\href {https://doi.org/10.1103/PhysRevB.52.R5467} {\bibfield  {journal}
  {\bibinfo  {journal} {Phys. Rev. B}\ }\textbf {\bibinfo {volume} {52}},\
  \bibinfo {pages} {R5467} (\bibinfo {year} {1995})}\BibitemShut {NoStop}%
\bibitem [{\citenamefont {Scott}\ and\ \citenamefont
  {Booth}(2023)}]{scott_rigorous_2023}%
  \BibitemOpen
  \bibfield  {author} {\bibinfo {author} {\bibfnamefont {C.~J.~C.}\
  \bibnamefont {Scott}}\ and\ \bibinfo {author} {\bibfnamefont {G.~H.}\
  \bibnamefont {Booth}},\ }\href@noop {} {\bibinfo {title} {Rigorous screened
  interactions for realistic correlated electron systems}} (\bibinfo {year}
  {2023}),\ \Eprint {https://arxiv.org/abs/2307.13584} {arXiv:2307.13584
  [cond-mat.str-el]} \BibitemShut {NoStop}%
\bibitem [{\citenamefont {Chang}\ \emph {et~al.}(2024)\citenamefont {Chang},
  \citenamefont {van Loon}, \citenamefont {Eskridge},\ and\ \citenamefont
  {Rösner}}]{Chang_DFT_cRPA_benchmarking_in_2024}%
  \BibitemOpen
  \bibfield  {author} {\bibinfo {author} {\bibfnamefont {Y.}~\bibnamefont
  {Chang}}, \bibinfo {author} {\bibfnamefont {E.~G. C.~P.}\ \bibnamefont {van
  Loon}}, \bibinfo {author} {\bibfnamefont {B.}~\bibnamefont {Eskridge}},\ and\
  \bibinfo {author} {\bibfnamefont {M.}~\bibnamefont {Rösner}},\ }\href
  {https://github.com/YueqingChang/Downfolding_benchmark_vanadocene/tree/main}
  {\bibinfo {title} {{DFT+cRPA benchmarking in vanadocene molecule}, {GitHub
  repository}}} (\bibinfo {year} {2024})\BibitemShut {NoStop}%
\bibitem [{\citenamefont {Giannozzi}\ \emph {et~al.}(2009)\citenamefont
  {Giannozzi}, \citenamefont {Baroni}, \citenamefont {Bonini}, \citenamefont
  {Calandra}, \citenamefont {Car}, \citenamefont {Cavazzoni}, \citenamefont
  {Ceresoli}, \citenamefont {Chiarotti}, \citenamefont {Cococcioni},
  \citenamefont {Dabo}, \citenamefont {Corso}, \citenamefont {Gironcoli},
  \citenamefont {Fabris}, \citenamefont {Fratesi}, \citenamefont {Gebauer},
  \citenamefont {Gerstmann}, \citenamefont {Gougoussis}, \citenamefont
  {Kokalj}, \citenamefont {Lazzeri}, \citenamefont {Martin-Samos},
  \citenamefont {Marzari}, \citenamefont {Mauri}, \citenamefont {Mazzarello},
  \citenamefont {Paolini}, \citenamefont {Pasquarello}, \citenamefont
  {Paulatto}, \citenamefont {Sbraccia}, \citenamefont {Scandolo}, \citenamefont
  {Sclauzero}, \citenamefont {Seitsonen}, \citenamefont {Smogunov},
  \citenamefont {Umari},\ and\ \citenamefont
  {Wentzcovitch}}]{giannozzi_quantum_2009}%
  \BibitemOpen
  \bibfield  {author} {\bibinfo {author} {\bibfnamefont {P.}~\bibnamefont
  {Giannozzi}}, \bibinfo {author} {\bibfnamefont {S.}~\bibnamefont {Baroni}},
  \bibinfo {author} {\bibfnamefont {N.}~\bibnamefont {Bonini}}, \bibinfo
  {author} {\bibfnamefont {M.}~\bibnamefont {Calandra}}, \bibinfo {author}
  {\bibfnamefont {R.}~\bibnamefont {Car}}, \bibinfo {author} {\bibfnamefont
  {C.}~\bibnamefont {Cavazzoni}}, \bibinfo {author} {\bibfnamefont
  {D.}~\bibnamefont {Ceresoli}}, \bibinfo {author} {\bibfnamefont {G.~L.}\
  \bibnamefont {Chiarotti}}, \bibinfo {author} {\bibfnamefont {M.}~\bibnamefont
  {Cococcioni}}, \bibinfo {author} {\bibfnamefont {I.}~\bibnamefont {Dabo}},
  \bibinfo {author} {\bibfnamefont {A.~D.}\ \bibnamefont {Corso}}, \bibinfo
  {author} {\bibfnamefont {S.~d.}\ \bibnamefont {Gironcoli}}, \bibinfo {author}
  {\bibfnamefont {S.}~\bibnamefont {Fabris}}, \bibinfo {author} {\bibfnamefont
  {G.}~\bibnamefont {Fratesi}}, \bibinfo {author} {\bibfnamefont
  {R.}~\bibnamefont {Gebauer}}, \bibinfo {author} {\bibfnamefont
  {U.}~\bibnamefont {Gerstmann}}, \bibinfo {author} {\bibfnamefont
  {C.}~\bibnamefont {Gougoussis}}, \bibinfo {author} {\bibfnamefont
  {A.}~\bibnamefont {Kokalj}}, \bibinfo {author} {\bibfnamefont
  {M.}~\bibnamefont {Lazzeri}}, \bibinfo {author} {\bibfnamefont
  {L.}~\bibnamefont {Martin-Samos}}, \bibinfo {author} {\bibfnamefont
  {N.}~\bibnamefont {Marzari}}, \bibinfo {author} {\bibfnamefont
  {F.}~\bibnamefont {Mauri}}, \bibinfo {author} {\bibfnamefont
  {R.}~\bibnamefont {Mazzarello}}, \bibinfo {author} {\bibfnamefont
  {S.}~\bibnamefont {Paolini}}, \bibinfo {author} {\bibfnamefont
  {A.}~\bibnamefont {Pasquarello}}, \bibinfo {author} {\bibfnamefont
  {L.}~\bibnamefont {Paulatto}}, \bibinfo {author} {\bibfnamefont
  {C.}~\bibnamefont {Sbraccia}}, \bibinfo {author} {\bibfnamefont
  {S.}~\bibnamefont {Scandolo}}, \bibinfo {author} {\bibfnamefont
  {G.}~\bibnamefont {Sclauzero}}, \bibinfo {author} {\bibfnamefont {A.~P.}\
  \bibnamefont {Seitsonen}}, \bibinfo {author} {\bibfnamefont {A.}~\bibnamefont
  {Smogunov}}, \bibinfo {author} {\bibfnamefont {P.}~\bibnamefont {Umari}},\
  and\ \bibinfo {author} {\bibfnamefont {R.~M.}\ \bibnamefont {Wentzcovitch}},\
  }\bibfield  {title} {\bibinfo {title} {{QUANTUM} {ESPRESSO}: a modular and
  open-source software project for quantum simulations of materials},\ }\href
  {https://doi.org/10.1088/0953-8984/21/39/395502} {\bibfield  {journal}
  {\bibinfo  {journal} {J. Phys.: Condens. Matter}\ }\textbf {\bibinfo {volume}
  {21}},\ \bibinfo {pages} {395502} (\bibinfo {year} {2009})}\BibitemShut
  {NoStop}%
\bibitem [{\citenamefont {Giannozzi}\ \emph {et~al.}(2017)\citenamefont
  {Giannozzi}, \citenamefont {Andreussi}, \citenamefont {Brumme}, \citenamefont
  {Bunau}, \citenamefont {Nardelli}, \citenamefont {Calandra}, \citenamefont
  {Car}, \citenamefont {Cavazzoni}, \citenamefont {Ceresoli}, \citenamefont
  {Cococcioni}, \citenamefont {Colonna}, \citenamefont {Carnimeo},
  \citenamefont {Corso}, \citenamefont {Gironcoli}, \citenamefont {Delugas},
  \citenamefont {{DiStasio}}, \citenamefont {Ferretti}, \citenamefont {Floris},
  \citenamefont {Fratesi}, \citenamefont {Fugallo}, \citenamefont {Gebauer},
  \citenamefont {Gerstmann}, \citenamefont {Giustino}, \citenamefont {Gorni},
  \citenamefont {Jia}, \citenamefont {Kawamura}, \citenamefont {Ko},
  \citenamefont {Kokalj}, \citenamefont {K\"uc\"ukbenli}, \citenamefont
  {Lazzeri}, \citenamefont {Marsili}, \citenamefont {Marzari}, \citenamefont
  {Mauri}, \citenamefont {Nguyen}, \citenamefont {Nguyen}, \citenamefont
  {Otero-de-la Roza}, \citenamefont {Paulatto}, \citenamefont {Ponc\'e},
  \citenamefont {Rocca}, \citenamefont {Sabatini}, \citenamefont {Santra},
  \citenamefont {Schlipf}, \citenamefont {Seitsonen}, \citenamefont {Smogunov},
  \citenamefont {Timrov}, \citenamefont {Thonhauser}, \citenamefont {Umari},
  \citenamefont {Vast}, \citenamefont {Wu},\ and\ \citenamefont
  {Baroni}}]{giannozzi_advanced_2017}%
  \BibitemOpen
  \bibfield  {author} {\bibinfo {author} {\bibfnamefont {P.}~\bibnamefont
  {Giannozzi}}, \bibinfo {author} {\bibfnamefont {O.}~\bibnamefont
  {Andreussi}}, \bibinfo {author} {\bibfnamefont {T.}~\bibnamefont {Brumme}},
  \bibinfo {author} {\bibfnamefont {O.}~\bibnamefont {Bunau}}, \bibinfo
  {author} {\bibfnamefont {M.~B.}\ \bibnamefont {Nardelli}}, \bibinfo {author}
  {\bibfnamefont {M.}~\bibnamefont {Calandra}}, \bibinfo {author}
  {\bibfnamefont {R.}~\bibnamefont {Car}}, \bibinfo {author} {\bibfnamefont
  {C.}~\bibnamefont {Cavazzoni}}, \bibinfo {author} {\bibfnamefont
  {D.}~\bibnamefont {Ceresoli}}, \bibinfo {author} {\bibfnamefont
  {M.}~\bibnamefont {Cococcioni}}, \bibinfo {author} {\bibfnamefont
  {N.}~\bibnamefont {Colonna}}, \bibinfo {author} {\bibfnamefont
  {I.}~\bibnamefont {Carnimeo}}, \bibinfo {author} {\bibfnamefont {A.~D.}\
  \bibnamefont {Corso}}, \bibinfo {author} {\bibfnamefont {S.~d.}\ \bibnamefont
  {Gironcoli}}, \bibinfo {author} {\bibfnamefont {P.}~\bibnamefont {Delugas}},
  \bibinfo {author} {\bibfnamefont {R.~A.}\ \bibnamefont {{DiStasio}}},
  \bibinfo {author} {\bibfnamefont {A.}~\bibnamefont {Ferretti}}, \bibinfo
  {author} {\bibfnamefont {A.}~\bibnamefont {Floris}}, \bibinfo {author}
  {\bibfnamefont {G.}~\bibnamefont {Fratesi}}, \bibinfo {author} {\bibfnamefont
  {G.}~\bibnamefont {Fugallo}}, \bibinfo {author} {\bibfnamefont
  {R.}~\bibnamefont {Gebauer}}, \bibinfo {author} {\bibfnamefont
  {U.}~\bibnamefont {Gerstmann}}, \bibinfo {author} {\bibfnamefont
  {F.}~\bibnamefont {Giustino}}, \bibinfo {author} {\bibfnamefont
  {T.}~\bibnamefont {Gorni}}, \bibinfo {author} {\bibfnamefont
  {J.}~\bibnamefont {Jia}}, \bibinfo {author} {\bibfnamefont {M.}~\bibnamefont
  {Kawamura}}, \bibinfo {author} {\bibfnamefont {H.-Y.}\ \bibnamefont {Ko}},
  \bibinfo {author} {\bibfnamefont {A.}~\bibnamefont {Kokalj}}, \bibinfo
  {author} {\bibfnamefont {E.}~\bibnamefont {K\"uc\"ukbenli}}, \bibinfo
  {author} {\bibfnamefont {M.}~\bibnamefont {Lazzeri}}, \bibinfo {author}
  {\bibfnamefont {M.}~\bibnamefont {Marsili}}, \bibinfo {author} {\bibfnamefont
  {N.}~\bibnamefont {Marzari}}, \bibinfo {author} {\bibfnamefont
  {F.}~\bibnamefont {Mauri}}, \bibinfo {author} {\bibfnamefont {N.~L.}\
  \bibnamefont {Nguyen}}, \bibinfo {author} {\bibfnamefont {H.-V.}\
  \bibnamefont {Nguyen}}, \bibinfo {author} {\bibfnamefont {A.}~\bibnamefont
  {Otero-de-la Roza}}, \bibinfo {author} {\bibfnamefont {L.}~\bibnamefont
  {Paulatto}}, \bibinfo {author} {\bibfnamefont {S.}~\bibnamefont {Ponc\'e}},
  \bibinfo {author} {\bibfnamefont {D.}~\bibnamefont {Rocca}}, \bibinfo
  {author} {\bibfnamefont {R.}~\bibnamefont {Sabatini}}, \bibinfo {author}
  {\bibfnamefont {B.}~\bibnamefont {Santra}}, \bibinfo {author} {\bibfnamefont
  {M.}~\bibnamefont {Schlipf}}, \bibinfo {author} {\bibfnamefont {A.~P.}\
  \bibnamefont {Seitsonen}}, \bibinfo {author} {\bibfnamefont {A.}~\bibnamefont
  {Smogunov}}, \bibinfo {author} {\bibfnamefont {I.}~\bibnamefont {Timrov}},
  \bibinfo {author} {\bibfnamefont {T.}~\bibnamefont {Thonhauser}}, \bibinfo
  {author} {\bibfnamefont {P.}~\bibnamefont {Umari}}, \bibinfo {author}
  {\bibfnamefont {N.}~\bibnamefont {Vast}}, \bibinfo {author} {\bibfnamefont
  {X.}~\bibnamefont {Wu}},\ and\ \bibinfo {author} {\bibfnamefont
  {S.}~\bibnamefont {Baroni}},\ }\bibfield  {title} {\bibinfo {title} {Advanced
  capabilities for materials modelling with {Q}uantum {ESPRESSO}},\ }\href
  {https://doi.org/10.1088/1361-648X/aa8f79} {\bibfield  {journal} {\bibinfo
  {journal} {J. Phys.: Condens. Matter}\ }\textbf {\bibinfo {volume} {29}},\
  \bibinfo {pages} {465901} (\bibinfo {year} {2017})}\BibitemShut {NoStop}%
\bibitem [{\citenamefont {Giannozzi}\ \emph {et~al.}(2020)\citenamefont
  {Giannozzi}, \citenamefont {Baseggio}, \citenamefont {Bonf\`a}, \citenamefont
  {Brunato}, \citenamefont {Car}, \citenamefont {Carnimeo}, \citenamefont
  {Cavazzoni}, \citenamefont {de~Gironcoli}, \citenamefont {Delugas},
  \citenamefont {Ferrari~Ruffino}, \citenamefont {Ferretti}, \citenamefont
  {Marzari}, \citenamefont {Timrov}, \citenamefont {Urru},\ and\ \citenamefont
  {Baroni}}]{giannozzi_quantum_2020}%
  \BibitemOpen
  \bibfield  {author} {\bibinfo {author} {\bibfnamefont {P.}~\bibnamefont
  {Giannozzi}}, \bibinfo {author} {\bibfnamefont {O.}~\bibnamefont {Baseggio}},
  \bibinfo {author} {\bibfnamefont {P.}~\bibnamefont {Bonf\`a}}, \bibinfo
  {author} {\bibfnamefont {D.}~\bibnamefont {Brunato}}, \bibinfo {author}
  {\bibfnamefont {R.}~\bibnamefont {Car}}, \bibinfo {author} {\bibfnamefont
  {I.}~\bibnamefont {Carnimeo}}, \bibinfo {author} {\bibfnamefont
  {C.}~\bibnamefont {Cavazzoni}}, \bibinfo {author} {\bibfnamefont
  {S.}~\bibnamefont {de~Gironcoli}}, \bibinfo {author} {\bibfnamefont
  {P.}~\bibnamefont {Delugas}}, \bibinfo {author} {\bibfnamefont
  {F.}~\bibnamefont {Ferrari~Ruffino}}, \bibinfo {author} {\bibfnamefont
  {A.}~\bibnamefont {Ferretti}}, \bibinfo {author} {\bibfnamefont
  {N.}~\bibnamefont {Marzari}}, \bibinfo {author} {\bibfnamefont
  {I.}~\bibnamefont {Timrov}}, \bibinfo {author} {\bibfnamefont
  {A.}~\bibnamefont {Urru}},\ and\ \bibinfo {author} {\bibfnamefont
  {S.}~\bibnamefont {Baroni}},\ }\bibfield  {title} {\bibinfo {title} {Quantum
  {ESPRESSO} toward the exascale},\ }\href {https://doi.org/10.1063/5.0005082}
  {\bibfield  {journal} {\bibinfo  {journal} {J. Chem. Phys.}\ }\textbf
  {\bibinfo {volume} {152}},\ \bibinfo {pages} {154105} (\bibinfo {year}
  {2020})}\BibitemShut {NoStop}%
\bibitem [{\citenamefont {Perdew}\ \emph {et~al.}(1996)\citenamefont {Perdew},
  \citenamefont {Burke},\ and\ \citenamefont
  {Ernzerhof}}]{PhysRevLett.77.3865}%
  \BibitemOpen
  \bibfield  {author} {\bibinfo {author} {\bibfnamefont {J.~P.}\ \bibnamefont
  {Perdew}}, \bibinfo {author} {\bibfnamefont {K.}~\bibnamefont {Burke}},\ and\
  \bibinfo {author} {\bibfnamefont {M.}~\bibnamefont {Ernzerhof}},\ }\bibfield
  {title} {\bibinfo {title} {Generalized {G}radient {A}pproximation {M}ade
  {S}imple},\ }\href {https://doi.org/10.1103/PhysRevLett.77.3865} {\bibfield
  {journal} {\bibinfo  {journal} {Phys. Rev. Lett.}\ }\textbf {\bibinfo
  {volume} {77}},\ \bibinfo {pages} {3865} (\bibinfo {year}
  {1996})}\BibitemShut {NoStop}%
\bibitem [{\citenamefont {Sergentu}\ \emph {et~al.}(2018)\citenamefont
  {Sergentu}, \citenamefont {Gendron},\ and\ \citenamefont
  {Autschbach}}]{sergentu_similar_2018}%
  \BibitemOpen
  \bibfield  {author} {\bibinfo {author} {\bibfnamefont {D.-C.}\ \bibnamefont
  {Sergentu}}, \bibinfo {author} {\bibfnamefont {F.}~\bibnamefont {Gendron}},\
  and\ \bibinfo {author} {\bibfnamefont {J.}~\bibnamefont {Autschbach}},\
  }\bibfield  {title} {\bibinfo {title} {Similar ligand–metal bonding for
  transition metals and actinides? 5f$^1$ {U}({C}$_7${H}$_7$)$_2-$ versus
  3d$^n$ metallocenes},\ }\href {https://doi.org/10.1039/C7SC05373H} {\bibfield
   {journal} {\bibinfo  {journal} {Chem. Sci.}\ }\textbf {\bibinfo {volume}
  {9}},\ \bibinfo {pages} {6292} (\bibinfo {year} {2018})}\BibitemShut
  {NoStop}%
\bibitem [{\citenamefont {Nakamura}\ \emph {et~al.}(2021)\citenamefont
  {Nakamura}, \citenamefont {Yoshimoto}, \citenamefont {Nomura}, \citenamefont
  {Tadano}, \citenamefont {Kawamura}, \citenamefont {Kosugi}, \citenamefont
  {Yoshimi}, \citenamefont {Misawa},\ and\ \citenamefont
  {Motoyama}}]{nakamura_respack_2021}%
  \BibitemOpen
  \bibfield  {author} {\bibinfo {author} {\bibfnamefont {K.}~\bibnamefont
  {Nakamura}}, \bibinfo {author} {\bibfnamefont {Y.}~\bibnamefont {Yoshimoto}},
  \bibinfo {author} {\bibfnamefont {Y.}~\bibnamefont {Nomura}}, \bibinfo
  {author} {\bibfnamefont {T.}~\bibnamefont {Tadano}}, \bibinfo {author}
  {\bibfnamefont {M.}~\bibnamefont {Kawamura}}, \bibinfo {author}
  {\bibfnamefont {T.}~\bibnamefont {Kosugi}}, \bibinfo {author} {\bibfnamefont
  {K.}~\bibnamefont {Yoshimi}}, \bibinfo {author} {\bibfnamefont
  {T.}~\bibnamefont {Misawa}},\ and\ \bibinfo {author} {\bibfnamefont
  {Y.}~\bibnamefont {Motoyama}},\ }\bibfield  {title} {\bibinfo {title}
  {{RESPACK}: {A}n ab initio tool for derivation of effective low-energy model
  of material},\ }\href {https://doi.org/10.1016/j.cpc.2020.107781} {\bibfield
  {journal} {\bibinfo  {journal} {Comput. Phys. Commun.}\ }\textbf {\bibinfo
  {volume} {261}},\ \bibinfo {pages} {107781} (\bibinfo {year}
  {2021})}\BibitemShut {NoStop}%
\bibitem [{\citenamefont {R{\"o}sner}\ \emph {et~al.}(2015)\citenamefont
  {R{\"o}sner}, \citenamefont {Şaşıoğlu}, \citenamefont {Friedrich},
  \citenamefont {Bl{\"u}gel},\ and\ \citenamefont
  {Wehling}}]{rosner_wannier_2015}%
  \BibitemOpen
  \bibfield  {author} {\bibinfo {author} {\bibfnamefont {M.}~\bibnamefont
  {R{\"o}sner}}, \bibinfo {author} {\bibfnamefont {E.}~\bibnamefont
  {Şaşıoğlu}}, \bibinfo {author} {\bibfnamefont {C.}~\bibnamefont
  {Friedrich}}, \bibinfo {author} {\bibfnamefont {S.}~\bibnamefont
  {Bl{\"u}gel}},\ and\ \bibinfo {author} {\bibfnamefont {T.~O.}\ \bibnamefont
  {Wehling}},\ }\bibfield  {title} {\bibinfo {title} {Wannier function approach
  to realistic {C}oulomb interactions in layered materials and
  heterostructures},\ }\href {https://doi.org/10.1103/PhysRevB.92.085102}
  {\bibfield  {journal} {\bibinfo  {journal} {Phys. Rev. B}\ }\textbf {\bibinfo
  {volume} {92}},\ \bibinfo {pages} {085102} (\bibinfo {year}
  {2015})}\BibitemShut {NoStop}%
\bibitem [{\citenamefont {Parcollet}\ \emph {et~al.}(2015)\citenamefont
  {Parcollet}, \citenamefont {Ferrero}, \citenamefont {Ayral}, \citenamefont
  {Hafermann}, \citenamefont {Krivenko}, \citenamefont {Messio},\ and\
  \citenamefont {Seth}}]{parcollet_triqs_2015}%
  \BibitemOpen
  \bibfield  {author} {\bibinfo {author} {\bibfnamefont {O.}~\bibnamefont
  {Parcollet}}, \bibinfo {author} {\bibfnamefont {M.}~\bibnamefont {Ferrero}},
  \bibinfo {author} {\bibfnamefont {T.}~\bibnamefont {Ayral}}, \bibinfo
  {author} {\bibfnamefont {H.}~\bibnamefont {Hafermann}}, \bibinfo {author}
  {\bibfnamefont {I.}~\bibnamefont {Krivenko}}, \bibinfo {author}
  {\bibfnamefont {L.}~\bibnamefont {Messio}},\ and\ \bibinfo {author}
  {\bibfnamefont {P.}~\bibnamefont {Seth}},\ }\bibfield  {title} {\bibinfo
  {title} {{TRIQS}: {A} toolbox for research on interacting quantum systems},\
  }\href {https://doi.org/10.1016/j.cpc.2015.04.023} {\bibfield  {journal}
  {\bibinfo  {journal} {Comput. Phys. Commun.}\ }\textbf {\bibinfo {volume}
  {196}},\ \bibinfo {pages} {398} (\bibinfo {year} {2015})}\BibitemShut
  {NoStop}%
\bibitem [{\citenamefont {Ruedenberg}\ \emph {et~al.}(1982)\citenamefont
  {Ruedenberg}, \citenamefont {Schmidt}, \citenamefont {Gilbert},\ and\
  \citenamefont {Elbert}}]{ruedenberg_are_1982}%
  \BibitemOpen
  \bibfield  {author} {\bibinfo {author} {\bibfnamefont {K.}~\bibnamefont
  {Ruedenberg}}, \bibinfo {author} {\bibfnamefont {M.~W.}\ \bibnamefont
  {Schmidt}}, \bibinfo {author} {\bibfnamefont {M.~M.}\ \bibnamefont
  {Gilbert}},\ and\ \bibinfo {author} {\bibfnamefont {S.}~\bibnamefont
  {Elbert}},\ }\bibfield  {title} {\bibinfo {title} {Are atoms intrinsic to
  molecular electronic wavefunctions? {I}. {T}he {FORS} model},\ }\href
  {https://doi.org/https://doi.org/10.1016/0301-0104(82)87004-3} {\bibfield
  {journal} {\bibinfo  {journal} {Chem. Phys.}\ }\textbf {\bibinfo {volume}
  {71}},\ \bibinfo {pages} {41} (\bibinfo {year} {1982})}\BibitemShut {NoStop}%
\bibitem [{\citenamefont {Sun}\ \emph {et~al.}(2020)\citenamefont {Sun},
  \citenamefont {Zhang}, \citenamefont {Banerjee}, \citenamefont {Bao},
  \citenamefont {Barbry}, \citenamefont {Blunt}, \citenamefont {Bogdanov},
  \citenamefont {Booth}, \citenamefont {Chen}, \citenamefont {Cui},
  \citenamefont {Eriksen}, \citenamefont {Gao}, \citenamefont {Guo},
  \citenamefont {Hermann}, \citenamefont {Hermes}, \citenamefont {Koh},
  \citenamefont {Koval}, \citenamefont {Lehtola}, \citenamefont {Li},
  \citenamefont {Liu}, \citenamefont {Mardirossian}, \citenamefont {McClain},
  \citenamefont {Motta}, \citenamefont {Mussard}, \citenamefont {Pham},
  \citenamefont {Pulkin}, \citenamefont {Purwanto}, \citenamefont {Robinson},
  \citenamefont {Ronca}, \citenamefont {Sayfutyarova}, \citenamefont
  {Scheurer}, \citenamefont {Schurkus}, \citenamefont {Smith}, \citenamefont
  {Sun}, \citenamefont {Sun}, \citenamefont {Upadhyay}, \citenamefont {Wagner},
  \citenamefont {Wang}, \citenamefont {White}, \citenamefont {Whitfield},
  \citenamefont {Williamson}, \citenamefont {Wouters}, \citenamefont {Yang},
  \citenamefont {Yu}, \citenamefont {Zhu}, \citenamefont {Berkelbach},
  \citenamefont {Sharma}, \citenamefont {Sokolov},\ and\ \citenamefont
  {Chan}}]{pyscf2020}%
  \BibitemOpen
  \bibfield  {author} {\bibinfo {author} {\bibfnamefont {Q.}~\bibnamefont
  {Sun}}, \bibinfo {author} {\bibfnamefont {X.}~\bibnamefont {Zhang}}, \bibinfo
  {author} {\bibfnamefont {S.}~\bibnamefont {Banerjee}}, \bibinfo {author}
  {\bibfnamefont {P.}~\bibnamefont {Bao}}, \bibinfo {author} {\bibfnamefont
  {M.}~\bibnamefont {Barbry}}, \bibinfo {author} {\bibfnamefont {N.~S.}\
  \bibnamefont {Blunt}}, \bibinfo {author} {\bibfnamefont {N.~A.}\ \bibnamefont
  {Bogdanov}}, \bibinfo {author} {\bibfnamefont {G.~H.}\ \bibnamefont {Booth}},
  \bibinfo {author} {\bibfnamefont {J.}~\bibnamefont {Chen}}, \bibinfo {author}
  {\bibfnamefont {Z.-H.}\ \bibnamefont {Cui}}, \bibinfo {author} {\bibfnamefont
  {J.~J.}\ \bibnamefont {Eriksen}}, \bibinfo {author} {\bibfnamefont
  {Y.}~\bibnamefont {Gao}}, \bibinfo {author} {\bibfnamefont {S.}~\bibnamefont
  {Guo}}, \bibinfo {author} {\bibfnamefont {J.}~\bibnamefont {Hermann}},
  \bibinfo {author} {\bibfnamefont {M.~R.}\ \bibnamefont {Hermes}}, \bibinfo
  {author} {\bibfnamefont {K.}~\bibnamefont {Koh}}, \bibinfo {author}
  {\bibfnamefont {P.}~\bibnamefont {Koval}}, \bibinfo {author} {\bibfnamefont
  {S.}~\bibnamefont {Lehtola}}, \bibinfo {author} {\bibfnamefont
  {Z.}~\bibnamefont {Li}}, \bibinfo {author} {\bibfnamefont {J.}~\bibnamefont
  {Liu}}, \bibinfo {author} {\bibfnamefont {N.}~\bibnamefont {Mardirossian}},
  \bibinfo {author} {\bibfnamefont {J.~D.}\ \bibnamefont {McClain}}, \bibinfo
  {author} {\bibfnamefont {M.}~\bibnamefont {Motta}}, \bibinfo {author}
  {\bibfnamefont {B.}~\bibnamefont {Mussard}}, \bibinfo {author} {\bibfnamefont
  {H.~Q.}\ \bibnamefont {Pham}}, \bibinfo {author} {\bibfnamefont
  {A.}~\bibnamefont {Pulkin}}, \bibinfo {author} {\bibfnamefont
  {W.}~\bibnamefont {Purwanto}}, \bibinfo {author} {\bibfnamefont {P.~J.}\
  \bibnamefont {Robinson}}, \bibinfo {author} {\bibfnamefont {E.}~\bibnamefont
  {Ronca}}, \bibinfo {author} {\bibfnamefont {E.~R.}\ \bibnamefont
  {Sayfutyarova}}, \bibinfo {author} {\bibfnamefont {M.}~\bibnamefont
  {Scheurer}}, \bibinfo {author} {\bibfnamefont {H.~F.}\ \bibnamefont
  {Schurkus}}, \bibinfo {author} {\bibfnamefont {J.~E.~T.}\ \bibnamefont
  {Smith}}, \bibinfo {author} {\bibfnamefont {C.}~\bibnamefont {Sun}}, \bibinfo
  {author} {\bibfnamefont {S.-N.}\ \bibnamefont {Sun}}, \bibinfo {author}
  {\bibfnamefont {S.}~\bibnamefont {Upadhyay}}, \bibinfo {author}
  {\bibfnamefont {L.~K.}\ \bibnamefont {Wagner}}, \bibinfo {author}
  {\bibfnamefont {X.}~\bibnamefont {Wang}}, \bibinfo {author} {\bibfnamefont
  {A.}~\bibnamefont {White}}, \bibinfo {author} {\bibfnamefont {J.~D.}\
  \bibnamefont {Whitfield}}, \bibinfo {author} {\bibfnamefont {M.~J.}\
  \bibnamefont {Williamson}}, \bibinfo {author} {\bibfnamefont
  {S.}~\bibnamefont {Wouters}}, \bibinfo {author} {\bibfnamefont
  {J.}~\bibnamefont {Yang}}, \bibinfo {author} {\bibfnamefont {J.~M.}\
  \bibnamefont {Yu}}, \bibinfo {author} {\bibfnamefont {T.}~\bibnamefont
  {Zhu}}, \bibinfo {author} {\bibfnamefont {T.~C.}\ \bibnamefont {Berkelbach}},
  \bibinfo {author} {\bibfnamefont {S.}~\bibnamefont {Sharma}}, \bibinfo
  {author} {\bibfnamefont {A.~Y.}\ \bibnamefont {Sokolov}},\ and\ \bibinfo
  {author} {\bibfnamefont {G.~K.-L.}\ \bibnamefont {Chan}},\ }\bibfield
  {title} {\bibinfo {title} {{Recent developments in the PySCF program
  package}},\ }\href {https://doi.org/10.1063/5.0006074} {\bibfield  {journal}
  {\bibinfo  {journal} {J. Chem. Phys.}\ }\textbf {\bibinfo {volume} {153}},\
  \bibinfo {pages} {024109} (\bibinfo {year} {2020})}\BibitemShut {NoStop}%
\bibitem [{\citenamefont {Sayfutyarova}\ \emph {et~al.}(2017)\citenamefont
  {Sayfutyarova}, \citenamefont {Sun}, \citenamefont {Chan},\ and\
  \citenamefont {Knizia}}]{AVAS2017}%
  \BibitemOpen
  \bibfield  {author} {\bibinfo {author} {\bibfnamefont {E.~R.}\ \bibnamefont
  {Sayfutyarova}}, \bibinfo {author} {\bibfnamefont {Q.}~\bibnamefont {Sun}},
  \bibinfo {author} {\bibfnamefont {G.~K.-L.}\ \bibnamefont {Chan}},\ and\
  \bibinfo {author} {\bibfnamefont {G.}~\bibnamefont {Knizia}},\ }\bibfield
  {title} {\bibinfo {title} {{Automated Construction of Molecular Active Spaces
  from Atomic Valence Orbitals}},\ }\href
  {https://doi.org/10.1021/acs.jctc.7b00128} {\bibfield  {journal} {\bibinfo
  {journal} {J. Chem. Theory Comput.}\ }\textbf {\bibinfo {volume} {13}},\
  \bibinfo {pages} {4063} (\bibinfo {year} {2017})}\BibitemShut {NoStop}%
\bibitem [{\citenamefont {Pathak}\ and\ \citenamefont
  {Wagner}(2018)}]{pathak_non-orthogonal_2018}%
  \BibitemOpen
  \bibfield  {author} {\bibinfo {author} {\bibfnamefont {S.}~\bibnamefont
  {Pathak}}\ and\ \bibinfo {author} {\bibfnamefont {L.~K.}\ \bibnamefont
  {Wagner}},\ }\bibfield  {title} {\bibinfo {title} {Non-orthogonal
  determinants in multi-{Slater}-{Jastrow} trial wave functions for fixed-node
  diffusion {Monte} {Carlo}},\ }\href {https://doi.org/10.1063/1.5052906}
  {\bibfield  {journal} {\bibinfo  {journal} {J. Chem. Phys.}\ }\textbf
  {\bibinfo {volume} {149}},\ \bibinfo {pages} {234104} (\bibinfo {year}
  {2018})}\BibitemShut {NoStop}%
\bibitem [{\citenamefont {Pathak}\ and\ \citenamefont
  {Wagner}(2020)}]{pathak_light_2020}%
  \BibitemOpen
  \bibfield  {author} {\bibinfo {author} {\bibfnamefont {S.}~\bibnamefont
  {Pathak}}\ and\ \bibinfo {author} {\bibfnamefont {L.~K.}\ \bibnamefont
  {Wagner}},\ }\bibfield  {title} {\bibinfo {title} {A light weight
  regularization for wave function parameter gradients in quantum {Monte}
  {Carlo}},\ }\href {https://doi.org/10.1063/5.0004008} {\bibfield  {journal}
  {\bibinfo  {journal} {AIP Adv.}\ }\textbf {\bibinfo {volume} {10}},\ \bibinfo
  {pages} {085213} (\bibinfo {year} {2020})}\BibitemShut {NoStop}%
\bibitem [{\citenamefont {Wheeler}\ \emph {et~al.}(2023)\citenamefont
  {Wheeler}, \citenamefont {Pathak}, \citenamefont {Kleiner}, \citenamefont
  {Yuan}, \citenamefont {Rodrigues}, \citenamefont {Lorsung}, \citenamefont
  {Krongchon}, \citenamefont {Chang}, \citenamefont {Zhou}, \citenamefont
  {Busemeyer}, \citenamefont {Williams}, \citenamefont {Mu\~{n}oz},
  \citenamefont {Chow},\ and\ \citenamefont {Wagner}}]{wheeler_pyqmc_2022}%
  \BibitemOpen
  \bibfield  {author} {\bibinfo {author} {\bibfnamefont {W.~A.}\ \bibnamefont
  {Wheeler}}, \bibinfo {author} {\bibfnamefont {S.}~\bibnamefont {Pathak}},
  \bibinfo {author} {\bibfnamefont {K.~G.}\ \bibnamefont {Kleiner}}, \bibinfo
  {author} {\bibfnamefont {S.}~\bibnamefont {Yuan}}, \bibinfo {author}
  {\bibfnamefont {J.~N.~B.}\ \bibnamefont {Rodrigues}}, \bibinfo {author}
  {\bibfnamefont {C.}~\bibnamefont {Lorsung}}, \bibinfo {author} {\bibfnamefont
  {K.}~\bibnamefont {Krongchon}}, \bibinfo {author} {\bibfnamefont
  {Y.}~\bibnamefont {Chang}}, \bibinfo {author} {\bibfnamefont
  {Y.}~\bibnamefont {Zhou}}, \bibinfo {author} {\bibfnamefont {B.}~\bibnamefont
  {Busemeyer}}, \bibinfo {author} {\bibfnamefont {K.~T.}\ \bibnamefont
  {Williams}}, \bibinfo {author} {\bibfnamefont {A.}~\bibnamefont {Mu\~{n}oz}},
  \bibinfo {author} {\bibfnamefont {C.~Y.}\ \bibnamefont {Chow}},\ and\
  \bibinfo {author} {\bibfnamefont {L.~K.}\ \bibnamefont {Wagner}},\ }\bibfield
   {title} {\bibinfo {title} {{{PyQMC}: {An} all-{Python} real-space quantum
  {Monte} {Carlo} module in {PySCF}}},\ }\href
  {https://doi.org/10.1063/5.0139024} {\bibfield  {journal} {\bibinfo
  {journal} {J. Chem. Phys.}\ }\textbf {\bibinfo {volume} {158}},\ \bibinfo
  {pages} {114801} (\bibinfo {year} {2023})}\BibitemShut {NoStop}%
\bibitem [{\citenamefont {Chang}\ \emph {et~al.}(2023)\citenamefont {Chang},
  \citenamefont {Joshi},\ and\ \citenamefont {Wagner}}]{chang_learning_2023}%
  \BibitemOpen
  \bibfield  {author} {\bibinfo {author} {\bibfnamefont {Y.}~\bibnamefont
  {Chang}}, \bibinfo {author} {\bibfnamefont {S.}~\bibnamefont {Joshi}},\ and\
  \bibinfo {author} {\bibfnamefont {L.~K.}\ \bibnamefont {Wagner}},\
  }\href@noop {} {\bibinfo {title} {Learning emergent models from \textit{ab
  initio} many-body calculations}},\ \bibinfo {howpublished} {Preprint at
  https://arxiv.org/abs/2302.02899} (\bibinfo {year} {2023})\BibitemShut
  {NoStop}%
\bibitem [{\citenamefont {Motta}\ and\ \citenamefont
  {Zhang}(2018)}]{Motta2018}%
  \BibitemOpen
  \bibfield  {author} {\bibinfo {author} {\bibfnamefont {M.}~\bibnamefont
  {Motta}}\ and\ \bibinfo {author} {\bibfnamefont {S.}~\bibnamefont {Zhang}},\
  }\bibfield  {title} {\bibinfo {title} {Ab initio computations of molecular
  systems by the auxiliary-field quantum {Monte} {Carlo} method},\ }\href
  {https://doi.org/https://doi.org/10.1002/wcms.1364} {\bibfield  {journal}
  {\bibinfo  {journal} {WIREs Comput. Mol. Sci.}\ }\textbf {\bibinfo {volume}
  {8}},\ \bibinfo {pages} {e1364} (\bibinfo {year} {2018})}\BibitemShut
  {NoStop}%
\bibitem [{\citenamefont {Shi}\ and\ \citenamefont {Zhang}(2021)}]{Shi2021}%
  \BibitemOpen
  \bibfield  {author} {\bibinfo {author} {\bibfnamefont {H.}~\bibnamefont
  {Shi}}\ and\ \bibinfo {author} {\bibfnamefont {S.}~\bibnamefont {Zhang}},\
  }\bibfield  {title} {\bibinfo {title} {{Some recent developments in
  auxiliary-field quantum Monte Carlo for real materials}},\ }\href
  {https://doi.org/10.1063/5.0031024} {\bibfield  {journal} {\bibinfo
  {journal} {J. Chem. Phys.}\ }\textbf {\bibinfo {volume} {154}},\ \bibinfo
  {pages} {024107} (\bibinfo {year} {2021})}\BibitemShut {NoStop}%
\end{thebibliography}%

\onecolumngrid
\newpage
\newpage
\section{Supplementary Information}
\appendix

\section*{Supplementary Note 1: 1-RDM Comparisons}
\label{sup:1-RDMs}

In this work, the benchmark of all the DFT+cRPA downfolded models is based on comparing their eigenstates with the many-body wave functions from highly accurate first-principles many-body calculations (DMC). 
On one hand, the real-space DMC wave functions contain information on the probability density of all valence electrons in vanadocene (the core electrons are ignored in the ECPs), living in a Hilbert space of 63 electrons.
On the other hand, the ED eigenstates are defined only on a 3-electron Hilbert space spanned by the 5 MLWFs.
Therefore, to establish the one-to-one correspondence besides the energies and $S^2$ values, we compare the 1-RDMs computed using the same single-particle basis, which is chosen to be MLWFs here.

In Supplementary Figure~\ref{fig:1rdm_comparison}, we show the state-wise comparison of the diagonal terms of the one-particle reduced density matrices (1-RDMs) in the spin-up channel for all the DFT+cRPA ED eigenstates and the DMC eigenstates. 
The mixed estimator error in the DMC 1-RDMs have been corrected using the extrapolated estimator, $\rho_{\text{DMC}} = \rho_{\text{mixed}} +\rho_{\text{mixed}}^\dagger - \rho_{\text{VMC}}$.
The single-particle orbitals, labeled from 0 to 4, correspond to the five MLWFs shown in Fig.~1 of the main text.
States with zero/finite occupancies in the $e_1$ manifold are categorized as spin-flip (SF)/crystal-field (CF) excitations. 
Note that this information is crucial for the benchmark, especially in the cases when the states can not be matched easily based on their energies, see, for example, Fig.~5 of the main text. 

\begin{figure*}
    \centering
    \includegraphics[width=0.5\textwidth]{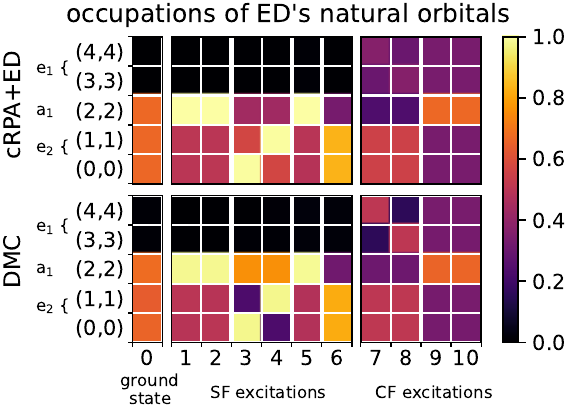}
    \caption{the state-wise comparison of the diagonal terms of the one-particle reduced density matrices (1-RDMs) in the spin-up channel for all the DFT+cRPA ED eigenstates and the DMC eigenstates. 
    The single-particle orbitals, labeled from 0 to 4, correspond to the five MLWFs shown in Supplementary Figure~\ref{fig:mlwf}.
    States with zero/finite occupancies in the $e_1$ manifold are categorized as spin-flip (SF)/crystal-field (CF) excitations.}
    \label{fig:1rdm_comparison}
\end{figure*}

\newpage

\section*{Supplementary Note 2: Charge Density Comparisons}
\label{sup:charge_density}

  In Supplementary Figures~\ref{fig:EXFirst} to \ref{fig:EXLast} we show the one-particle density matrices together with the charge densities and their differences to the ground states ($\Delta^{\uparrow/\downarrow}$) for all states discussed in the main text and as obtained from our DMC calculations (after projecting to the natural basis of the downfolded results) and from downfolding using the static cRPA Coulomb interaction. To also illustrate the spin and charge excitation characters of these states, we show the difference $\Delta^{\uparrow} \pm  \Delta^{\downarrow}$. For the spin-flip excitation, we expect $\Delta^{\uparrow} + \Delta^{\downarrow}$ to be vanishingly small, while the change in the spin polarization, i.e., $\Delta^{\uparrow} - \Delta^{\downarrow}$ should be non-zero. For crystal-field excitations we expect finite $\Delta^{\uparrow} + \Delta^{\downarrow}$.

  The overall good qualitative agreement between the DMC and downfolded results shows that we can indeed map the downfolded states 1:1 to the corresponding reference states from DMC, which allows for more detailed discussions of their differences.

  As an example, we briefly discuss the first spin-flip and crystal-field excitations shown in Figs.~\ref{fig:EXSF} and \ref{fig:EXCF}. For both excitations, all depicted quantities obtained via DMC and downfolding are in good qualitative and quantitative agreement. For the spin-flip excitation, we do not find any significant difference between the two methods: diagonal and off-diagonal one-particle matrix elements are nearly the same, and the resulting charge densities look very much alike. In the case of the crystal-field excitation, the agreement is also qualitatively good, while there are some quantitative differences. In both DMC and the downfolded model, the higher $e_1$ is occupied by approx. $1$ electron with approx. $2/3$ in the up and $1/3$ in the spin down channel. The relative occupations in the lower $e_2$ and $a_1$ orbitals, however, differ between the methods. The downfolded results underestimate the $a_1$ occupation and overestimate the $e_2$ occupations compared to the DMC. 
  
  This level of agreement holds throughout nearly all investigated excitations. Only the second spin-flip excitation shows different characteristics in $\Delta^{\uparrow/\downarrow}$ and $\Delta^{\uparrow} \pm  \Delta^{\downarrow}$. From the comparison of the one-particle density matrices, we find here again that for this excitation, the $a_1$ ($e_2$) occupations are underestimated (overestimated) using the downfolded Hamiltonian in comparison to DMC. This might be attributed to an overestimation of the $e_2$-$a_1$ splitting in the spin-restricted DFT calculation. In fact, upon adding a crystal-field correction ($H_{\text{CFC}}^{a_1}$), which moves the $a_1$ state below the $e_2$, this qualitative difference between the DMC and downfolded result can be lifted. Whether this discrepancy results here from a shortcoming in the DFT starting point, from missing double counting corrections, from the chosen model Hamiltonian, or from less accurate DMC reference data cannot be judged at the moment.

\begin{figure*}
    \centering
    \includegraphics[width=0.85\textwidth]{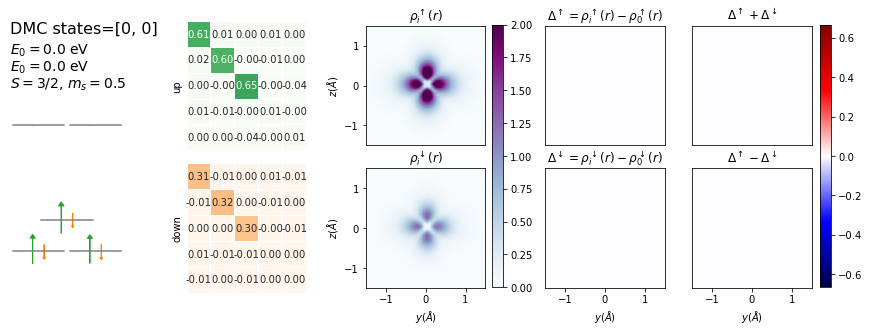}
    \includegraphics[width=0.85\textwidth]{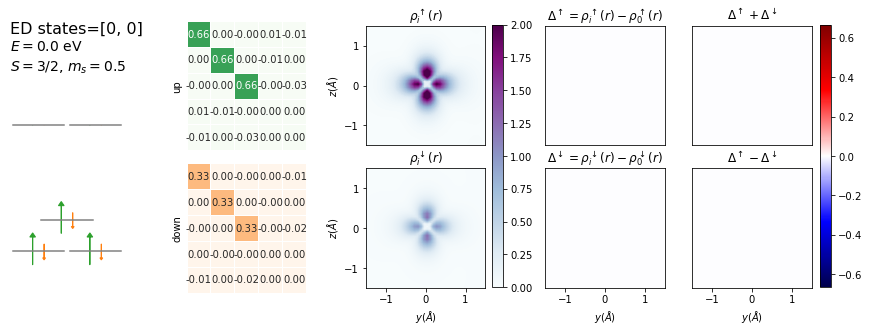}
    \caption{Ground state within DMC (top) and as obtained from downfolding using the MLWF basis, full static cRPA Coulomb matrix elements, and no double counting (bottom). We show sketches of the many-body state, spin-channel resolved one-particle reduced density matrices (ordered as $e_2$, $a_1$, $e_1$), and visualizations of the corresponding many-body charge densities. \label{fig:EXFirst}}
\end{figure*}

    \begin{figure*}
        \centering
        \includegraphics[width=0.85\textwidth]{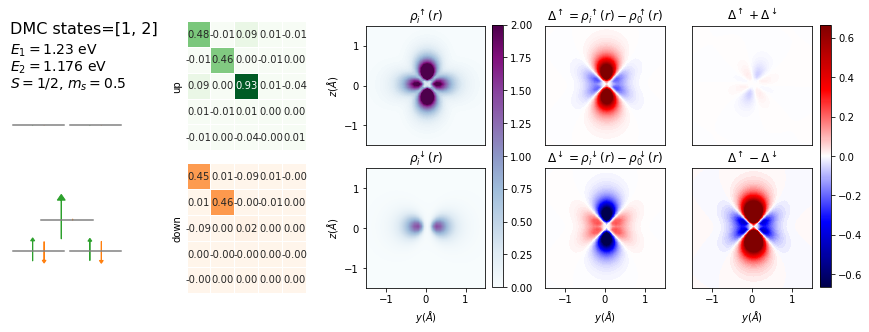}
        \includegraphics[width=0.85\textwidth]{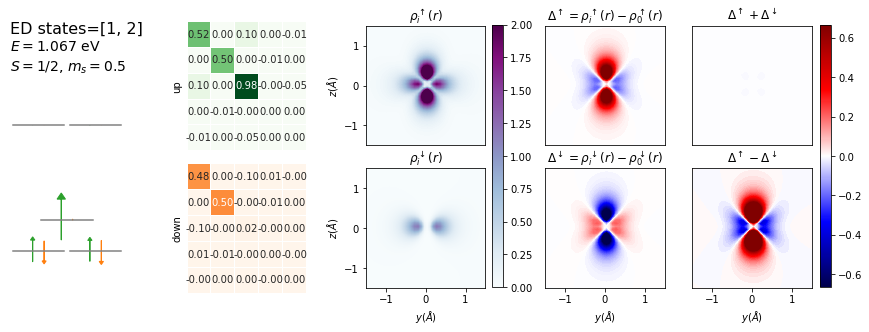}
        \caption{First spin-flip excitations within DMC (top) and as obtained from downfolding using the MLWF basis, full static cRPA Coulomb matrix elements, and no double counting (bottom). We show sketches of the many-body state, spin-channel resolved one-particle reduced density matrices (ordered as $e_2$, $a_1$, $e_1$), and visualizations of the corresponding many-body charge densities.}
        \label{fig:EXSF}
    \end{figure*}

\begin{figure*}
    \centering
    \includegraphics[width=0.85\textwidth]{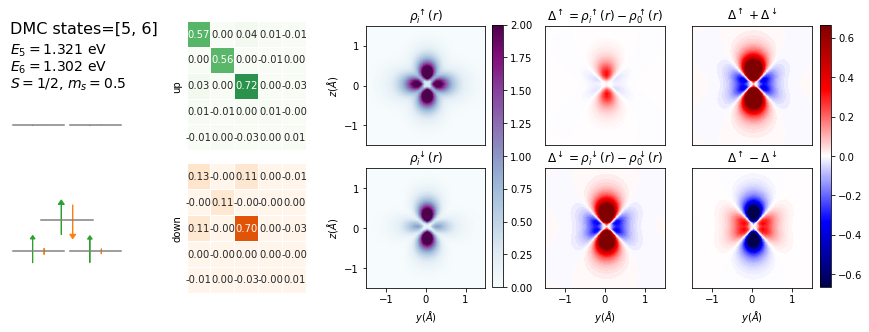}
    \includegraphics[width=0.85\textwidth]{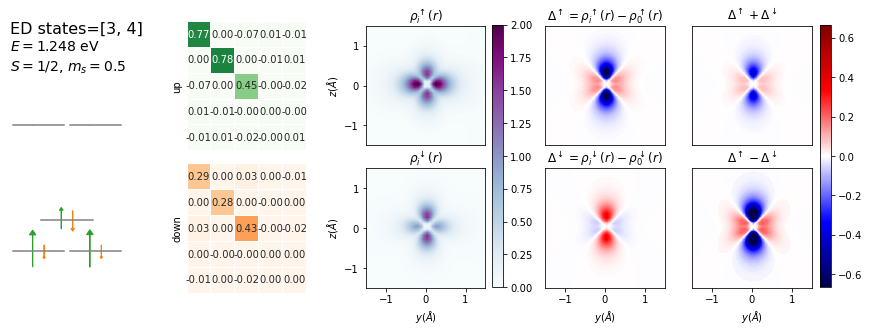}
    \caption{Second spin-flip excitations within DMC (top) and as obtained from downfolding using the MLWF basis, full static cRPA Coulomb matrix elements, and no double counting (bottom). We show sketches of the many-body state, spin-channel resolved one-particle density matrices (ordered as $e_2$, $a_1$, $e_1$), and visualizations of the corresponding many-body charge densities.}
\end{figure*}

\begin{figure*}
    \centering
    \includegraphics[width=0.85\textwidth]{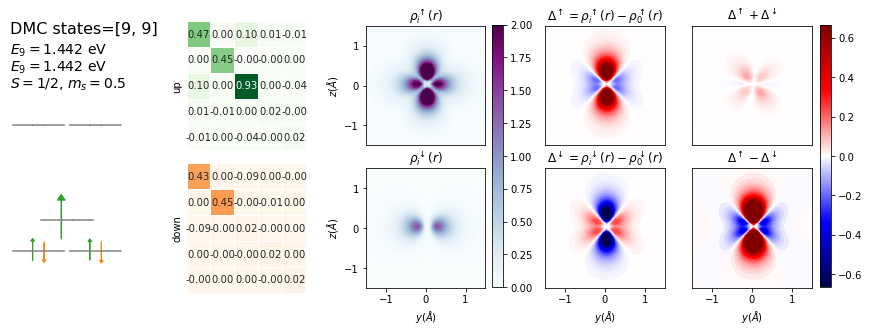}
    \includegraphics[width=0.85\textwidth]{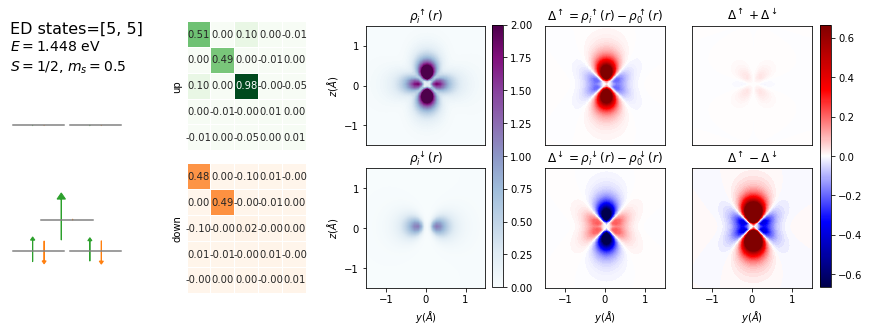}
    \caption{Third spin-flip excitations within DMC (top) and as obtained from downfolding using the MLWF basis, full static cRPA Coulomb matrix elements, and no double counting (bottom). We show sketches of the many-body state, spin-channel resolved one-particle reduced density matrices (ordered as $e_2$, $a_1$, $e_1$), and visualizations of the corresponding many-body charge densities.}
\end{figure*}

\begin{figure*}
    \centering
    \includegraphics[width=0.85\textwidth]{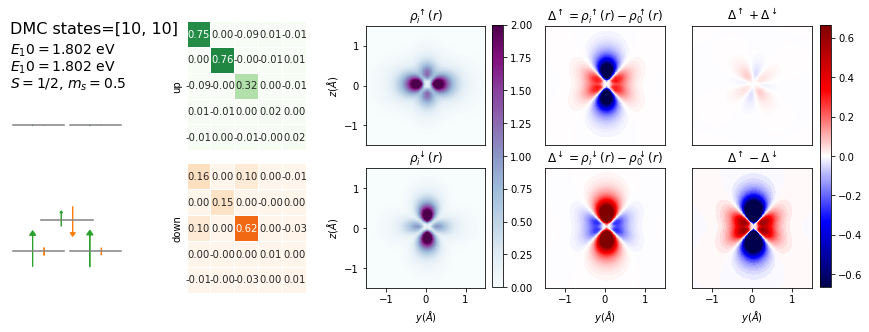}
    \includegraphics[width=0.85\textwidth]{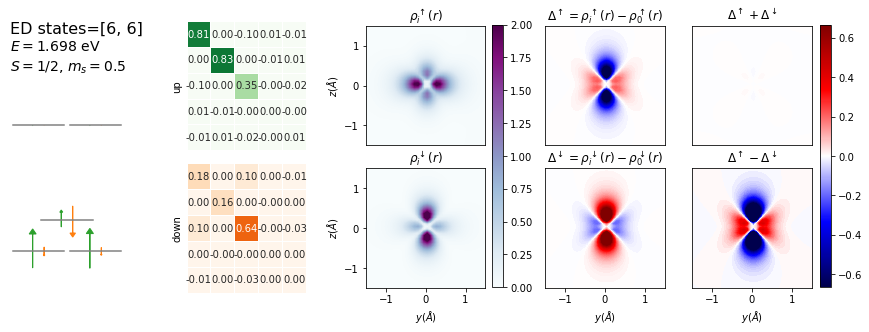}
    \caption{Fourth spin-flip excitations within DMC (top) and as obtained from downfolding using the MLWF basis, full static cRPA Coulomb matrix elements, and no double counting (bottom). We show sketches of the many-body state, spin-channel resolved one-particle reduced density matrices (ordered as $e_2$, $a_1$, $e_1$), and visualizations of the corresponding many-body charge densities.}
\end{figure*}

    \begin{figure*}
        \centering
        \includegraphics[width=0.85\textwidth]{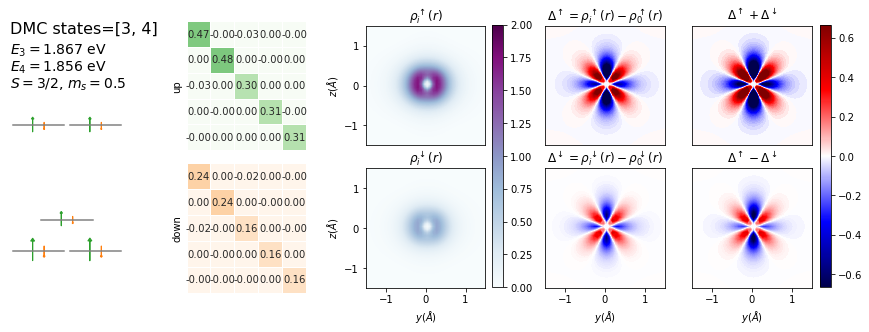}
        \includegraphics[width=0.85\textwidth]{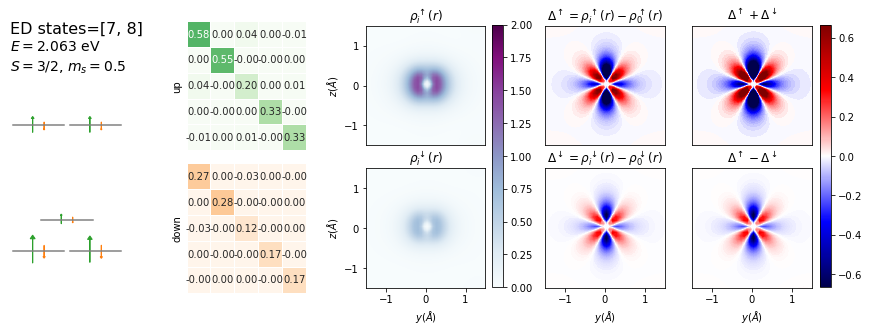}
        \caption{First crystal-field excitations within DMC (top) and as obtained from downfolding using the MLWF basis, full static cRPA Coulomb matrix elements, and no double counting (bottom). We show sketches of the many-body state, spin-channel resolved one-particle reduced density matrices (ordered as $e_2$, $a_1$, $e_1$), and visualizations of the corresponding many-body charge densities.}
        \label{fig:EXCF}
    \end{figure*}

\begin{figure*}
    \centering
    \includegraphics[width=0.85\textwidth]{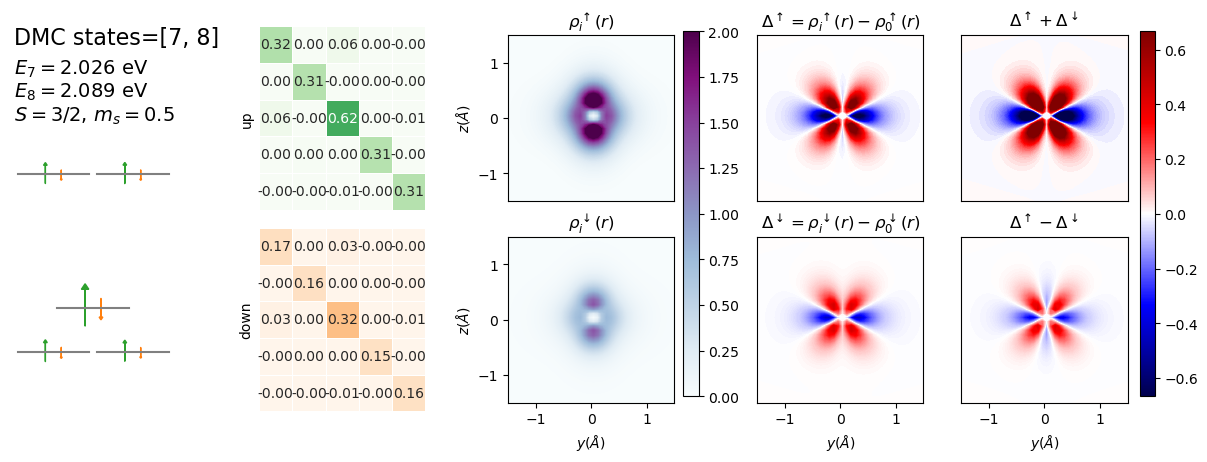}
    \includegraphics[width=0.85\textwidth]{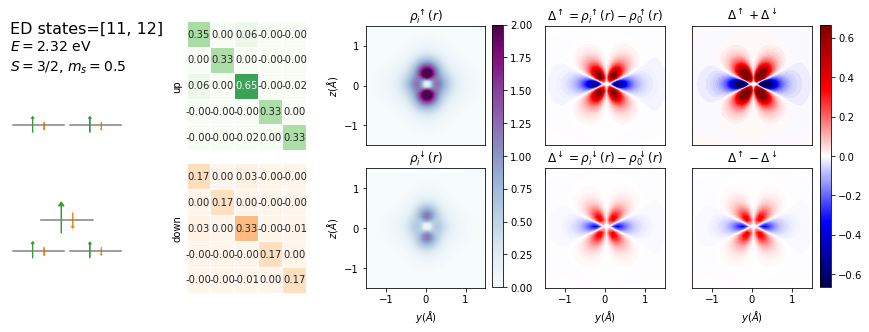}
    \caption{Second crystal-field excitations within DMC (top) and as obtained from downfolding using the MLWF basis, full static cRPA Coulomb matrix elements, and no double counting (bottom). We show sketches of the many-body state, spin-channel resolved one-particle reduced density matrices (ordered as $e_2$, $a_1$, $e_1$), and visualizations of the corresponding many-body charge densities. \label{fig:EXLast}}
\end{figure*}

\end{document}